\documentclass[journal=jctcce,manuscript=article]{achemso}
\setkeys{acs}{articletitle = true}
\pdfoutput=1
\usepackage[version=3]{mhchem} 
\usepackage[T1]{fontenc}       
\usepackage{color}

\usepackage{fp} 
\usepackage{bm} 
\usepackage{amssymb} 

\newcommand*{\vF}{\boldsymbol{F}}
\newcommand*{\vC}{\boldsymbol{C}}

\newcommand*{\vT}{\boldsymbol{T}}

\renewcommand*{\Pi}{{\varPi}}

\newcommand*{\vv}{\boldsymbol{v}}
\newcommand*{\vV}{\boldsymbol{V}}
\newcommand*{\vP}{\boldsymbol{P}}
\newcommand*{\vU}{\boldsymbol{U}}
\newcommand*{\vJ}{\boldsymbol{J}}

\newcommand*{\vD}{\boldsymbol{D}}

\usepackage{amsbsy} 
\usepackage{graphicx} 
\usepackage{algorithm,algorithmic}
\usepackage{bm}
\usepackage{array}
\usepackage{mathtools}
\usepackage{subcaption}
\usepackage[mathscr]{eucal}
\usepackage[scaled]{beramono}
\usepackage{listings}
\usepackage{lineno}

\author{Matteo De Santis}
\email{matteo.des89@gmail.com}
\affiliation[Universit\`a degli Studi di Perugia]
{Dipartimento di Chimica, Biologia e Biotecnologie, Universit\`a degli Studi di Perugia,
Via Elce di Sotto 8, 06123
Perugia, Italy}
\alsoaffiliation{Istituto di Scienze e Tecnologie Chimiche (SCITEC), Consiglio Nazionale delle Ricerche
 c/o
Dipartimento di Chimica, Biologia e Biotecnologie, Universit\`a degli Studi di Perugia,
Via Elce di Sotto 8, 06123
Perugia, Italy}

\author{Leonardo Belpassi}
\affiliation[Consiglio Nazionale delle Ricerche]
{Istituto di Scienze e Tecnologie Chimiche (SCITEC), Consiglio Nazionale delle Ricerche
 c/o
Dipartimento di Chimica, Biologia e Biotecnologie, Universit\`a degli Studi di Perugia,
Via Elce di Sotto 8, 06123
Perugia, Italy}

\author{Christoph R. Jacob}
\affiliation[TU]
{Institute of Physical and Theoretical Chemistry, Technische Universit\"at Braunschweig, Gau{\ss}str.~17, 38106 Braunschweig, Germany}

\author{Andr\'{e} Severo Pereira Gomes}
\affiliation[Univ. Lille]
{Univ. Lille, CNRS, UMR 8523-PhLAM-Physique des Lasers Atomes et Mol\'{e}cules, F-59000 Lille, France}

\author{Francesco Tarantelli}
\affiliation[Universit\`a degli Studi di Perugia]
{Dipartimento di Chimica, Biologia e Biotecnologie, Universit\`a degli Studi di Perugia,
Via Elce di Sotto 8, 06123
Perugia, Italy}

\author{Lucas Visscher}
\affiliation[VU]
{Theoretical Chemistry, Faculty of Science, Vrije Universiteit Amsterdam, De Boelelaan 1083, NL-1081HV Amsterdam, Netherlands}

\author{Loriano Storchi}
\affiliation[Universit\`a degli Studi `G. D'Annunzio']
{Dipartimento di Farmacia, Universit\`a degli Studi `G. D'Annunzio',
Via dei Vestini 31, 66100
Chieti, Italy}
\alsoaffiliation{Istituto di Scienze e Tecnologie Chimiche (SCITEC), Consiglio Nazionale delle Ricerche
 c/o
Dipartimento di Chimica, Biologia e Biotecnologie, Universit\`a degli Studi di Perugia,
Via Elce di Sotto 8, 06123
Perugia, Italy}

\title{Environmental effects with Frozen Density Embedding 
in Real-Time Time-Dependent Density Functional Theory using localized basis functions}

\begin{document}

\maketitle

\begin{abstract}
Frozen Density Embedding (FDE) represents a versatile embedding
scheme to  describe the environmental effect on the 
electron dynamics in molecular systems. The extension of the general theory of FDE to the
real-time time-dependent Kohn-Sham method has previously  been presented 
and implemented in plane-waves and periodic boundary conditions
(Pavanello et al. J. Chem. Phys. 142, 154116, 2015).

In the current paper, we extend our recent formulation
of real-time time-dependent Kohn-Sham method based on  localized basis
set functions and
developed within the Psi4NumPy framework to the FDE scheme. The latter
 has been implemented in its
``uncoupled'' flavor (in which the time evolution is only carried out for the
active subsystem, while the environment subsystems remain at their ground state), 
using and adapting
the FDE implementation already available in the PyEmbed module of the scripting framework PyADF.
The implementation was facilitated by the fact that both Psi4NumPy
and PyADF, being  native Python API, provided an ideal framework of
development using the Python advantages in terms of code readability and
reusability.  We employed this new implementation to investigate
the stability of the time propagation procedure, which is based an
efficient predictor/corrector  second-order midpoint Magnus propagator
employing an exact diagonalization, in combination with
the FDE scheme.  We demonstrate that the inclusion of the FDE potential
does not introduce any numerical instability in time propagation of the
density matrix of the active subsystem and in the limit of weak external
field, the numerical results for low-lying transition energies are
consistent with those obtained using the reference FDE calculations based on
the linear response TDDFT. The method is found to give stable numerical
results also in the presence of strong external field inducing non-linear
effects. Preliminary results are reported for high harmonic generation
(HHG) of a water molecule embedded in a small water cluster.  The effect of
the embedding potential is evident in the HHG spectrum reducing the number
of the well resolved high harmonics at high energy with respect to the
free water. This is consistent with a shift towards lower ionization
energy passing from an isolated water molecule to a small water cluster.
The computational burden for the propagation step increases approximately linearly
with the size of the surrounding frozen environment. Furthermore, we
have also shown that the updating frequency of the embedding potential
may be significantly reduced, much less  that one per time step, without
jeopardising the accuracy of the transition energies.
\end{abstract}
\section{Introduction}
\label{sec:introduction}
The last decade has seen a growing interest in the electron
dynamics taking place in molecules  subjected to an external
electromagnetic field.  Matter-radiation interaction
is involved in many different phenomena ranging from  weak-field
processes, i.e.,  photo-excitation, absorption and scattering, light
harvesting in dye sensitized solar cells~\cite{dssc1,dssc2}
and photo-ionization, to strong-field processes encompassing
high harmonic generation~\cite{prl_hhg,atto_hhg},
optical rectification~\cite{prl_or,kadlec_or}, multiphoton
ionization~\cite{Keldysh_2017} and above threshold ionization~\cite{ati}.
Furthermore, the emergence of new Free Electron Lasers (FEL) and
attosecond methodologies~\cite{attosecond_rev1,attosecond_rev2} opened
an area of research in which experiments can probe electron dynamics
and chemical reactions in real-time  and the movement of electrons in
molecules may be controlled.
These experiments can provide direct insights into bond
breaking~\cite{Attar54,Wolf2019,Adam2019}/forming\cite{Kim2015} and
ionization~\cite{sharifi07ionization,zigo2017ionization} by directly
probing nuclear and electron dynamics.

Real-time time-dependent electronic structure theory, in which 
the equation of motion is directly solved in the time domain, is clearly the most promising 
for investigating time-dependent molecular response and electronic dynamics.
The recent progress in the development of these  methodologies is impressive
(see, for instance, a recent review by Li et al.\cite{goings2018real}).
Among different approaches, because of its compromise between accuracy and efficiency,
the real-time time-dependent density functional theory 
(rt-TDDFT) is becoming very popular.
The main obstacle to implementing the rt-TDDFT method involves the algorithmic design of
a numerically stable and computationally efficient time evolution propagator.
This typically requires the repeated evaluation of the effective Hamiltonian
matrix representation (Kohn-Sham matrix), at each time step.
Despite the difficulties to realizing a stable time propagator scheme,
there are very appealing features in  a real-time approach to TDDFT,
such as the absence of explicit exchange-correlation
kernel derivatives ~\cite{ruud2010} or divergence problems
appearing in response theory and 
one has  the possibility to obtain all frequency excitations at the same cost.
Furthermore, the method is suitable to treat complex non-linear phenomena
and external fields with an explicit shape, which is a key ingredient for the
quantum optimal control theory\cite{cammi:2019}.

Several implementations have been presented~\cite{sun2007,schlegel2005,bkurti_manby2008},
after the pioneering work of Theilhaber~\cite{theilhaber} and Yabana and Bertsch~\cite{yabana_bertsch}.
Many of them rely on the real space grid methodology~\cite{yabana_bertsch} with Siesta and Octopus 
as the most recent ones~\cite{takimoto2007real, andrade2015real}. Alternative approaches employ plane waves 
such as in Qbox~\cite{schleife2012plane} or QUANTUM ESPRESSO\cite{quantumespresso:2020,Pavanello:2017} and analytic atom centered Gaussian basis implementations
(i.e., Gaussian~\cite{liang2011,morzan2014}, NWChem~\cite{nwchem}, Q-Chem~\cite{qchem1,qchem2}) 
also have gained popularity. The scheme has been also extended to include relativistic effects
at the highest level.
Repisky et al. proposed the first application and implementation
of relativistic TDDFT to atomic and molecular systems
\cite{repisky2015excitation} based on the four-component Dirac hamiltonian
and almost simultaneously Goings et al.~\cite{goings2016rtx2c} published the
development of X2C Hamiltonian-based electron dynamics and its application
to the evaluation of UV/vis spectra.
Very recently, some of us presented a rt-TDDFT implementation\cite{pybertha,pyberthagit}
based on state-of-the-art software engineering approaches (i.e. including interlanguage communication 
between High-level Languages such as Python, C, FORTRAN and prototyping techniques).  
The method, based on the design of an efficient
propagation scheme   within the Psi4NumPy~\cite{psi4numpy}
 framework, was also  extended to the relativistic four-component framework based on
the BERTHA code\cite{belpassi11_12368,belpassi06_124104,storchi10_384}, (more specifically based on the 
recently developed PyBERTHA\cite{pybertha,pyberthagit,paperelectronic,parcopaper}, that is the Python API of BERTHA).

The applications of the rt-TDDFT approach encompass
studies of linear~\cite{lopata2012} and non-linear optical
response properties~\cite{ding,takimoto2007real}, molecular
conductance~\cite{voohris}, singlet-triplet transitions~\cite{isborn},
plasmonic resonances magnetic circular dichroism~\cite{gauss_jjgoings},
core excitation, photoinduced electric current,
spin-magnetization dynamics~\cite{mag_dyn} and Ehrenfest
dynamics~\cite{rt_ehrenfestdyn,grigory2016}. Moreover, many studies
in the relativistic and quasi relativistic framework appeared,
ranging from X-ray near-edge absorption~\cite{kadek2015}, to
nonlinear optical properties ~\cite{konecny2016nlo}, to chiroptical
spectroscopy~\cite{konecnyresolution-of-identity}.

Most part of initial applications of real-time methodology to chemical systems were largely focused on the
electron dynamics and optical properties of the isolated target systems. However,
it is widely recognized that these phenomena are extremely sensitive to the polarization induced by the environment,
such that the simulation on an isolated molecule is usually not sufficient even 
for a qualitative description.
A number of studies aiming at including the effect of a chemical
environment within rt-TDDFT have appeared in the literature.  They are
based on the coupling of rt-TDDFT with the QM/MM approach which includes the
molecular environment explicitly and at a reduced cost using classical
mechanical description \cite{PhysRevLett.90.258101,morzan2014} or in a
polarizable continuous medium (PCM), where the solvent
degrees of freedom are replaced by an effective
 classical dielectric.\cite{Li:2012,Li_b:2012,PIPOLO2014112}
One of the challenges, however, in the dynamical description 
of the environment is that the response
of the solvent is not instantaneous, thus these  approaches
have been extended to include the non-equilibrium solvent response\cite{Corni_jpca:2015,mennucci2015,mennucci2017,demon2krt}.
A recent extension considers also non-equilibrium
cavity field polarization effects for molecules embedded in an homogeneous
dielectric\cite{Corni:2019}. 

Going beyond a classical description for the environment, 
very recently, Koh et al.\cite{Koh:2017} have combined
the rt-TDDFT method  with block-orthogonalized Manby-Miller theory\cite{lee_projection-based_2019} to
accelerate the rt-TDDFT  simulations, the approach is also suitable
for cheaply accounting the solvation effect on the molecular response.
Another fully quantum mechanical approach to include environment effects 
in the molecular response property is based on the frozen-density embedding (FDE) 
scheme \cite{gomes_quantum-chemical_2012,jacob_subsystem_2014,wesolowski_frozen-density_2015}.
FDE is a DFT-in-DFT embedding method that allows to partition a larger 
Kohn-Sham system into a set of smaller, coupled Kohn-Sham subsystems.
Additional to the computational advantage, FDE provides physical insight into the properties of embedded systems and the coupling interactions between them.\cite{pavanello2015}

For electronic ground states, the theory and methodology were introduced 
by Weso\l{}owski and Warshel~\cite{wesolowski93}, based on the approach originally
proposed by Senatore and Subbaswamy~\cite{Senatore86}, and later Cortona~\cite{Cortona92},
for solid-state calculations. It has been further 
generalized~\cite{iannuzzi2006,jacob2008flexible} and directed to the simultaneous
optimization of the subsystem electronic densities. 
Within the linear-response formalism Casida and Weso\l{}owski put forward a formal TDDFT generalization~\cite{doi:10.1002/qua.10744} of the FDE scheme.
Neugebauer~\cite{coupl_neugebauer2007,neugebauer2009} then introduced coupled FDE,
a subsystem TDDFT formulation which removed some of the approximations made in the initial 
TDDFT-FDE implementations. Recently, 
the approach has been further extended~\cite{toelle_1,toelle_2} to account for charge-transfer excitations,
 taking advantage of an exact FDE scheme~\cite{fux_exemb,goodpaster_exemb,goodpaster_exemb2,huang_exemb,nafziger_exemb}.

A DFT subsystem formulation of the real-time methodology has been presented 
in a seminal work by Pavanello and coworkers\cite{pavanello2015} together with its formulation 
within the FDE framework. They showed that the extension of FDE 
to rt-TDDFT can be done straightforwardly by updating the embedding potential 
between the systems at every time step and evolving the 
Kohn-Sham subsystems in time simultaneously.
Its actual implementation, based on the use of plane-waves
and ultrasoft pseudopotentials~\cite{pavanello2015,Pavanello:2017},
showed that
the updating of the embedding potentials during the 
time evolution of the electron density  
does not affect the numerical stability of the propagator.
The approach may be approximated and devised in the so called ``uncoupled'' scheme
where the density response to the external field is limited to one active subsystem 
while keeping the densities of the other subsystems frozen in time. 
Note that also in this uncoupled version the embedding potential 
is time-dependent and needs to be recomputed and updated during the
time propagation. However, the propagation scheme is restricted to the active subsystem
and the approach is promising to include environmental
effects in the real-time simulation. 
Numerous applications within the context of the linear response 
TDDFT showed that an uncoupled FDE is sufficient for reproducing supermolecular 
results with good accuracy even in the presence of hydrogen bonds 
as long as there are no couplings in the excitations between the systems.

In this work we extend rt-TDDFT based on localized basis functions to
the FDE scheme in its uncoupled version (uFDE-rt-TDDFT),
taking advantage of modern software engineering and code
reusability offered by the Python programming language.
We devised an unified framework based on Python in which
the high interoperability allowed the concerted and efficient use of the
recent rt-TDDFT procedure, which some of us have implemented in the framework
of the Psi4Numpy API\cite{pybertha,pyberthagit}, and the PyADF API\cite{pyadf}. 
The rt-TDDFT procedure has served
as the main interface where the PyADF methods, which gave direct access to
the key quantities necessary to devise the FDE scheme, can be accessed
within a unified framework. 
Since in this work we
introduced a new flavor of the rt-TDDFT Psi4Numpy-based program,
to avoid confusions, from now on, we will refer to the aforementioned
rt-TDDFT based on Psi4Numpy as Psi4-rt, while its extension to the FDE
subsystem framework will be referred as Psi4-rt-PyEmbed.

In Section 2 we review the fundamentals of FDE and its extension to rt-TDDFT methodology.
In Section 3 computational details are given with a specific focus on the interoperability of the various codes we merged and used: Psi4Numpy\cite{psi4numpy}, XCFun\cite{xcfun:2019} 
and PyADF\cite{pyadf}, including the PyEmbed module recently developed by some of us.
In Section 4 we report and comment the results of the calculations we performed on excitation transitions
for different molecular systems, including: a water-ammonia complex, a water cluster and a more extend 
acetone-in-water cluster case.
Finally, we give some preliminary results 
about the applicability and numerical stability of the method in presence of intense
 external field inducing strong non-linear effects as High Harmonic Generation  (HHG)
in the active system. Concluding remarks and perspectives are finally given in Section 5.
\section{Theory}
\label{sec:theory}
In this section we briefly review the theoretical foundations 
of the FDE scheme and its extension to the rt-TDDFT methodology.
As mentioned above, a previous implementation
was presented by Pavanello et al.~\cite{pavanello2015} using plane waves and ultrasoft pseudopotentials. 
We refer the interested reader to this seminal work for a general theoretical background,
and for additional details of the FDE-rt-TDDFT formal derivation.

\subsection{Subsystem DFT and Frozen Density Embedding formulation}

In the subsystem formulation of DFT the entire system is partitioned into N subsystems,
and the total density $\rho_\text{tot}(\bm{r})$ is represented as the sum of electron densities of the 
various subsystems [i.e., $\rho_a(\bm{r})$ ($a = 1,..,N$)]. 
Focusing on a single subsystem, we can consider the total density as partitioned in only two 
contributions as
\begin{equation}
	\rho_\text{tot}(\bm{r}) =\rho_\text{I}(\bm{r}) + \rho_\text{II}(\bm{r}).
\end{equation}
The total energy of the system can then be written as 
\begin{equation}
	E_\text{tot}[\rho_\text{I},\rho_\text{II}] = E_\text{I}[\rho_\text{I}] + E_\text{II}[\rho_\text{II}] + E_\text{int}[\rho_\text{I},\rho_\text{II}]
\label{eq:etot}
\end{equation}
with the energy of each subsystem ($E_i[\rho_i]$, with $i=\text{I},\text{II}$) given according to the usual 
definition in DFT as
\begin{equation}
\begin{aligned}
E_i[\rho_i]   &= \int\rho_i(\bm{r})v_\text{nuc}^{i}(\bm{r}) {\rm d}^3r 
                          + \frac{1}{2}\iint\frac{\rho_i(\bm{r})\rho_{i}(\bm{r}')}{|\bm{r}-\bm{r}'|}{\rm d}^3r {\rm d}^3r' + \\
              &+ E_\text{xc}[\rho_i] + T_s[\rho_i] + E_\text{nuc}^{i}.
\end{aligned}
\end{equation}
In the above expression, $v_\text{nuc}^{i}(\bm{r})$ is the nuclear potential due to the set of atoms which defines the 
subsystem and $E_\text{nuc}^{i}$ is the related nuclear repulsion energy.
$T_s[\rho_i]$ is the kinetic energy of the auxiliary non-interacting system,
which is, within the Kohn-Sham (KS) approach, commonly evaluated using the KS orbitals.
The interaction energy is given by the expression:
\begin{equation}
\begin{aligned}
  \label{eq:eint}
	E_\text{int}[\rho_\text{I},\rho_\text{II}] &= \int\rho_\text{I}(\bm{r})v_\text{nuc}^\text{II}(\bm{r}){\rm d}^3r 
	            +\int\rho_\text{II}(\bm{r})v_\text{nuc}^\text{I}(\bm{r}) {\rm d}^3r + E^\text{I,II}_\text{nuc} \\
	&+\iint\frac{\rho_\text{I}(\bm{r})\rho_\text{II}(\bm{r}')}{|\bm{r}-\bm{r}'|} {\rm d}^3r {\rm d}^3r'
	           +E^\text{nadd}_\text{xc}[\rho_\text{I},\rho_\text{II}] + T^\text{nadd}_s[\rho_\text{I},\rho_\text{II}]
\end{aligned}
\end{equation}
with $v_\text{nuc}^\text{I}$ and $v_\text{nuc}^\text{II}$ the nuclear potentials due to the set of atoms associated with the subsystem $\text{I}$ 
and $\text{II}$,
respectively. The repulsion energy for nuclei belonging to different subsystems is described by the $E^\text{I,II}_\text{nuc}$ term.
The non-additive contributions are defined as:
\begin{equation}
	X^\text{nadd}[\rho_\text{I},\rho_\text{II}] = X[\rho_\text{I}+\rho_\text{II}] - X[\rho_\text{I}] - X[\rho_\text{II}]
\end{equation}
with $X=E_\text{xc}, T_s$. These terms arise because both 
exchange-correlation and kinetic energy, in contrast to the Coulomb interaction, 
are not linear functionals of the density.

The electron density of a given fragment ($\rho_\text{I}$ or $\rho_\text{II}$ in this case) 
can be determined by minimizing the total energy
functional (Eq.\ref{eq:etot}) with respect to the density of the fragment while
keeping the density of the other subsystem frozen. This procedure is the essence of the
FDE scheme and leads to a set of Kohn-Sham-like equations (one for  each subsystem)
\begin{equation}\label{eq:act_opt}
        \Big[ -\frac{\nabla^2}{2} + v^\text{KS}_\text{eff}[\rho_\text{I}](\bm{r}) 
                + v_\text{emb}^\text{I}[\rho_\text{I},\rho_\text{II}](\bm{r})\Big]\phi_k^\text{I}(\bm{r}) = \varepsilon_k^\text{I}\phi_k^\text{I}(\bm{r})
\end{equation}
which are coupled by the embedding potential term $v^\text{I}_\text{emb}(\bm{r})$, 
which carries all dependence on the other fragment's density.
In this equation, $v^\text{KS}_\text{eff}[\rho_\text{I}](\bm{r})$ is the KS potential calculated on basis
of the density of subsystem $\text{I}$ only, whereas the embedding potential takes into account the effect of the other subsystem 
(which we consider here as the complete environment).
In the framework of FDE theory, $v^\text{I}_\text{emb}(\bm{r})$ is explicitly given by 
\begin{equation}\label{eq:vemb}
\begin{aligned}
	v^\text{I}_\text{emb}[\rho_\text{I},\rho_\text{II}](\bm{r}) 
	= \frac{\delta E_\text{int}[\rho_\text{I},\rho_\text{II}]}{\delta\rho_\text{I}(\bm{r})}=& 
	    \ v_\text{nuc}^\text{II}(\bm{r})+\int\frac{\rho_\text{II}(\bm{r}')}{|\bm{r}-\bm{r}'|} {\rm d}^3r' 
	  + \frac{\delta E_\text{xc}^\text{nadd}[\rho_\text{I}, \rho_\text{II}]}{\delta\rho_\text{I}(\bm{r})}
	  + \frac{\delta T^\text{nadd}_s[\rho_\text{I}, \rho_\text{II}]}{\delta\rho_\text{I}(\bm{r})},
\end{aligned}
\end{equation}
where the non-additive exchange-correlation and kinetic energy contributions 
are defined as the difference between the associated exchange-correlation and kinetic
potentials defined using $\rho_\text{tot}(\bm{r})$ and $\rho_\text{I}(\bm{r})$. For both potentials,
one needs to account for the fact that only the density is known for the total system so that
potentials that require input in the form of KS orbitals are prohibited.
For the exchange-correlation potential, one may make use of accurate density functional 
approximations and its quality is therefore similar to that of ordinary KS. The potential for the non-additive kinetic term 
($\frac{\delta T^\text{nadd}_s[\rho]}{\delta\rho_\text{I}(\bm{r})}$,
in Eq.\ref{eq:vemb}) is more problematic as less accurate orbital-free kinetic energy density functionals (KEDFs)
are available for this purpose.
Examples of popular functional approximations applied in this context are the Thomas-Fermi (TF) 
kinetic energy functional\cite{thomas_1927} or 
the GGA functional  PW91k\cite{PhysRevA.50.5328}. 
These functionals have shown to be accurate in the 
case of  weakly interacting systems including hydrogen bond systems, whereas 
their use for subsystems interacting with a larger covalent character 
is problematic (see Ref.\cite{fux_exemb} and references therein).
The research for more accurate KEDFs is a key aspect for the 
applicability of the FDE scheme as a general scheme, including the partitioning 
of the system also breaking covalent bonds.\cite{Pavanello:2020}

In general, the set of coupled equations that arise in the FDE scheme for the
 subsystems  have to be solved iteratively.
Typically, one may employ a procedure of ``freeze-and-thaw'' where the
electron density of the active subsystem is determined keeping frozen
the electron density of the others subsystems, which is then frozen
when the electron density of the other subsystems is worked out.
This procedure may be repeated many times until all subsystems'
densities are converged.  In this case the FDE scheme can be seen as
an alternative formulation of the conventional KS-DFT approach for
large systems (by construction it scales linearly with the number of
subsystems). The update of the density for (part of) the environment can
be important when trial densities obtained from isolated subsystems are
not are not a very good starting point, as is the case for ionic species
\cite{acetone_w_cluster,Bouchafra:2019,Halbert:2020}.

The implementation of FDE is relatively straightforward, 
in that the $v^\text{I}_\text{emb}(\bm{r})$ potential is a one-electron operator that
needs to be added to the usual KS hamiltonian. 
When using localized basis functions, the matrix representation of the embedding potential
(${\bf V^\text{emb}}$) may be evaluated using numerical integration grids 
similar to those used for the exchange-correlation term in the KS method. 
This contribution is then added  to the KS matrix and the eigenvalue problem 
is solved in the usual self-consistent field  manner.

We note that, irrespective of whether one or many subsystem densities are
optimized, the matrix ${\bf V^\text{emb}}$ needs to be updated during SCF procedure because it also depends
on the density of the active subsystem (see Eq.\ref{eq:vemb}).

Going beyond the ground state is necessary to access many interesting
properties, which for DFT are expressed via response 
theory\cite{doi:10.1002/qua.10744,coupl_neugebauer2007,neugebauer2009,visscher12,Olejniczak:2017},
such as electronic absorption\cite{Johannes:2005} or NMR shielding\cite{Bulo:2008,Halbert:2020},
and for which FDE has been shown to work properly since these are quite often relatively 
local. In a response formulation, the embedding potential as well as its derivatives
enter the equations and, if more than one subsystem is allowed to react to the external
perturbations~\cite{coupl_neugebauer2007,neugebauer2009,visscher12,Olejniczak:2017}, the
derivatives of the embedding potential introduce the coupling in the subsystems' response
(as the embedding potential introduces the coupling of the subsystems' electronic structure
in the ground state).

While such couplings in response may be very important in certain situations, such as
for strongly interacting systems\cite{toelle_1,toelle_2} or for extensive properties\cite{neugebauer2009}, 
disregarding them still can provide a very accurate picture, notably for localized excited 
states\cite{Johannes:2005,acetone_w_cluster}.
In this simplified ``uncoupled'' framework, one  considers only the response of the subsystem of interest 
(and thus the embedding potential and its derivative
with respect to this subsystem's density). While neglecting environment response may seem a drastic approximation,
good performance relative to supermolecular reference data has been obtained for excitation energies of a chromophore 
in a solvent or a crystal environment, even when only retaining
the embedding potential~\cite{acetone_w_cluster}. We will therefore employ this framework in the following.

\subsection{The Real-Time Time-Dependent Kohn-Sham method and its
extension to FDE}

The time-dependent equation for the Kohn-Sham method can be
conveniently formulated in terms of the Liouville-von Neumann (LvN)
equation.  In an orthonormal basis set	the LvN equation reads:
\begin{equation}\label{eq:lvn_eq}
	i \frac{\partial\vD(t)}{\partial t} = \vF(t)\vD(t)
	-\vD(t)\vF(t)
\end{equation} where $i$ is the imaginary unit and $\vD(t)$
and $\vF(t)$ are the one-electron density matrix and time-dependent
Kohn-Sham matrix, respectively.	The above equation holds in both the
non-relativistic and relativistic four-component 
formulations\cite{belpassi11_12368,repisky2015excitation}.

In a non-relativistic framework, the Kohn-Sham matrix ($\vF(t)$) is defined as 
\begin{equation} \vF(t) = \vT + \vv_\text{nuc} + \vV_\text{xc}[\rho(t)] +
\vJ[\rho(t)] + \vv_\text{ext}(t) ,
\end{equation}
where $\vT$ and $\vv_\text{nuc}$ are the one-electron non-relativistic kinetic energy
and intramolecular nuclear attraction terms, respectively.
The explicit time dependence of $\vF(t)$ is due to the time-dependent external potential $\vv_\text{ext}(t)$, which accounts
for the interaction of the molecular system with an applied external electric field. Even in the absence of an
external field the Fock operator is implicitly dependent on time through
the density matrix $\vD(t)$
in the Coulomb ($\vJ[\rho(t)]$) and exchange-correlation terms ($\vV_\text{xc}[\rho(t)]$).

The propagation in time of the density matrix can be expressed as
\begin{equation}
\vD(t)= \vU(t,t_0)\,\vD(t_0)\,\vU(t,t_0)^{\dagger}
\end{equation}
where $\vU(t,t_0)$ is the matrix representation of the time-evolution operator.

If we start (initial condition, i.e. initial time $t_{0}$)
with the electronic ground state density matrix and use as orthonormal basis
the ground state molecular orbitals,
$\vD(t_0)$ assumes the form
\begin{equation}\label{eq:D}
        \vD(t_0) = \begin{pmatrix} \bm{1}_{oo} & \bm{0}_{ov}\\ \bm{0}_{vo} & \bm{0}_{vv}  
\end{pmatrix} \nonumber \\
        \quad    ,
\end{equation}
where $\bm{1}_{oo}$ is the identity matrix over the occupied orbital space of 
size $n_{occ}$ (total number of electrons). The $\vD$ matrix has the dimension of 
$n_\text{tot}$ ($n_\text{tot}=n_{occ}+n_{virt}$) that is  total number of the basis functions.

In our implementation, which uses a basis set of atomic centered (AO) Gaussian-type functions,
the ground-state molecular orbitals are conveniently used as the reference
orthonormal basis and at the time $t$ the Fock and density matrices are
related to their AO basis representation simply by:
\begin{equation}
\vF(t)^{MO} = \vC^{\dagger}\vF(t)^{AO}\vC 
\end{equation}

where the $\vC$ matrix contains the reference MO expansion coefficients. The same 
coefficients satisfy a similar relation for $\vD(t)^{MO}$:

\begin{equation}
	\vD(t)^{AO} = \vC\vD(t)^{MO}\vC^{\dagger}	
\end{equation}
In a finite time interval, the solution of the Liouville-von Neumann equation
consists in the calculation of the Fock matrix at discrete time steps, and in propagating
the density matrix in time.

In the most general case, where the Fock operator depends on time even in absence of external fields, the 
time-evolution operator can be expressed by means of a Dyson-like series:
\begin{equation}
\begin{aligned}
&\vU(t,t_0)=\sum_{n=1}^{\infty} \frac{(-i)^n}{n!}\int_{t_0}^t d\tau_1 \int_{t_0}^t d\tau_2 \ldots \int_{t_0}^t d\tau_n \vF(\tau_1)\vF(\tau_2)\ldots \vF(\tau_n) \\
&\tau_1>\tau_2\ldots >\tau_n
\end{aligned}
\end{equation}
which in compact notation, using the the time ordering operator $\hat{\mathcal{T}}$, reads as:
\begin{equation}
	\vU(t,t_0) = \hat{\mathcal{T}} \exp \bigg(-i \int_{t_0}^{t} \vF(t') dt' \bigg)
\end{equation}
The time ordering is necessary since $\vF(t)$ at different times
do not necessarily commute ($[\vF(t),\vF(t')]\neq 0$).
Typically, this time-ordering problem is overcome by exploiting the composition property
of time-evolution operator ($\vU(t,t_0) = \vU(t,t_1)\vU(t_1,t_0)$)
and discretizing the time
using a small time step. It is clear that the exact time ordering can be achieved only in the limit of an infinitesimal time step.
Many different propagation schemes have been proposed\cite{castro}
in the context of rt-TDDFT.
Among others, we mention the Crank-Nicholson \cite{meng2008real}, Runge-Kutta \cite{numrecipes2007} or Magnus \cite{magnus,Casas_2006} methods.

The Magnus expansion has found the widest application, in particular, 
in those implementations that employ localized basis sets functions, 
for which matrix exponentiation can be performed exactly via matrix diagonalization.
Typically, the Magnus expansion is truncated to the first order 
 evaluating the integral over time using numerical quadrature, 
provided that the time interval $\Delta t$ is sufficiently short. 
Using the midpoint rule the propagator becomes
\begin{equation}\label{eq:u_op}
        \vU(t+\Delta t,t) \approx \exp\bigg[ - i \vF\Big(t+\frac{\Delta t}{2}\Big)\Delta t\bigg] .
\end{equation}
This approach, also referred as second-order midpoint Magnus propagator, is
unitary by construction, provided that $\vF$ is hermitian.
This scheme exhibits an error which is proportional to $(\Delta t)^3$.
The expression in Eq.\ref{eq:u_op} coincides with the
so-called modified-midpoint unitary transform time-propagation scheme originally 
 introduced by Schlegel et al.\cite{schlegel2005}.

The $\vF$ matrix at time $t+\Delta t/2$, where no density is available, can be obtained using an iterative
series of extrapolations and interpolations at
each time. Note that, if this predictor/corrector procedure is converged in a self-consistent
manner the
second-order midpoint Magnus propagator preserves the time reversal symmetry, which
is an exact property of the equation of motion in absence of magnetic field.
The predictor/corrector scheme is a key ingredient in preserving the
numerical stability of the propagation with a range of algorithms that can be applied in this context~\cite{propagators2}.
We have recently implemented a particularly stable  predictor/corrector scheme, originally proposed by
Repisky\textit{et al.}\cite{repisky2015excitation}, 
in the interactive quantum chemistry programming environment Psi4NumPy\cite{psi4numpy,pybertha,pyberthagit}.

The methodology that we have described above can be straightforwardly extended
to the subsystem density functional theory framework and in particular 
to FDE (FDE-rt-TDDFT)\cite{pavanello2015}.  In the present work
we consider one active subsystem and  keep frozen the density of the environment
along the time propagation (uncoupled scheme, to which we will refer as uFDE-rt-TDDFT).
Thus, a LvN type equation 
is solved in the space of the active subsystem. 
The only  modification to Eq.\ref{eq:lvn_eq}
is in the definition of the effective hamiltonian matrix representation  which
now refers to the active subsystem ($\vF^\text{I}(t) = \vT^\text{I} + \vv_\text{nuc}^\text{I}
+ \vV_\text{xc}[\rho^\text{I}(t)] + \vJ[\rho^\text{I}(t)] + \vv_\text{ext}(t)$) and to which
the matrix representation of the embedding potential
($\vV^\text{emb}(t)$) is
added to take into account the effect of the environment. 
The propagation scheme itself remains unaltered.

As in the case for the ground state, in which the change of the active subsystem 
density requires that $\vV^\text{emb}$ is updated at each SCF iteration, the time propagation of the
electron density will introduce a time dependence in  $\vV^\text{emb}$ even though 
the environment densities are kept frozen at their
ground state value (due to the use of the uncoupled scheme).

Thus, the $\vV^\text{emb}$ matrix needs to
be updated during the propagation. In the present implementation we use
atomic centered Gaussian function as basis set for the active subsystem
and evaluate  the $V^\text{emb}_{\mu\nu}$ matrix elements numerically~\cite{acetone_w_cluster}.  We will
show that the numerical noise associated with the construction of the embedding potential introduced by this scheme does
not affect the numerical stability of the density matrix propagation
in the linear and non-linear regimes. In the following sections we will
also demonstrate, for a specific application, that the updating frequency
of the embedding potential may be significantly reduced (much less than
one per time step used to solve the LvN equation) without jeopardising
the accuracy.

As usual, the key quantity in a real time simulation is the
time-dependent electric dipole moment ${\vec{\mu}}(t)$. 
Each Cartesian component $p$ (with $p = x,y,z$)
is given by
\begin{equation}\label{eq:dip}
{\mu_p}(t)=-\int \rho(t,{\bf r}) p\,d{\bf r} = \mathrm{Tr}(\vD(t)\vP_p) ,
\end{equation}
where $\vP_p$ is the matrix representation of the p-th component of the electric dipole moment
operator (see also Eq.~\ref{eq:dip}).
Since, in our uFDE-rt-TDDFT implementation, the time dependency response of the external field 
is due only from the active system, in the above expression (Eq.\ref{eq:dip}) all quantities refer to the active subsystem.
The vector ${\vec{\mu}}(t)$ defines the polarization response to all orders
and is easily computed by the electronic density at any time,
$t$.  From this quantity one can then compute both linear and non-linear
properties.

In the linear response regime, each component of electric dipole moment, $\mu_p(\omega)$,
with an external field $E_q$ in the direction $q$ (with $q =x,y,z$), is given in frequency space by
\begin{equation}
\mu_p(\omega)= \sum_q \alpha_{pq}(\omega) E_q(\omega).
\end{equation}
The components depend on the polarizability tensor ($\alpha_{pq}$) through  the
Fourier-transformation of the $q$-component of the applied field.
The dipole strength function $\mathrm{S}(\omega)$
is related to the imaginary
part of the frequency dependent linear polarizability by
\begin{equation}
        \mathrm{S}(\omega) = \frac{2\omega}{3\pi}\mathrm{Im
        \,Tr}[\alpha(\omega)]
        \label{eq:osc_str}
\end{equation} 
In our implementation\cite{pybertha} the perturbation
can be chosen to be either an impulsive kick or a continuous wave whose
amplitude is modulated by an analytic envelope function.  Different
explicit functional forms are available~\cite{pybertha,pyberthagit}.
In the case of an impulsive perturbation (${\bf E}(t)=k\delta(t)$${\bf
n }$, where ${\bf n }$ is a unit vector representing the orientation of
the field) we adopt the $\delta$-analytic representation as proposed in
Ref.~\cite{repisky2015excitation}.  One of the best-known examples of
non-linear optical phenomena is HHG in atoms and molecules.  HHG occurs
via photo-emission by the molecular system in a strong field and can
be also computed from ${\vec{\mu}}(t)$~\cite{Bandrauk:2009}.
In this work we calculate the HHG power spectrum for a particular
polarization direction as the Fourier transform of the laser-driven
induced dipole moment, 
\begin{equation}\label{eq:nlFT} P(\omega)
\propto \Bigg| \int_{t_1}^{t_2} \mu_z(t)\exp(- i \omega t) dt \Bigg|^2 .
\end{equation} 
Other suitable approaches have been investigated in the
literature\cite{Bandrauk:2009}, but in all cases the
key quantity is ${\vec{\mu}}(t)$.
\section{Computational Details and Implementation}
\label{sec:compdetails}
In this section we outline the computational strategy we adopted to 
implement the uFDE-rt-TDDFT scheme. We devised a multi-scale approach
where we take advantage of the real-time TDDFT reference procedure,
recently implemented within Psi4Numpy framework 
(i.e. Psi4-RT program)\cite{pybertha,pyberthagit}, while the FDE computational core relies on PyADF\cite{pyadf,pyadf-github-v096} 
and makes use of its PyEmbed module, which some of us have recently developed \cite{pyembed,schmittmonreal_frozen-density_2020}.
PyEmbed provides a Python implementation for computing the interaction energy (Eq.~\ref{eq:eint}) 
and embedding potential (Eq.~\ref{eq:vemb}) from FDE on user-defined integration grids, while using the XCFun
library\cite{ruud2010,xcfun:2019} to evaluate non-additive xc and kinetic energy contributions. With PyEmbed, quantum chemistry
codes require only minimal changes: functionality to provide electron densities and its derivatives, as well as the electrostatic potential, over the grid,
as well as to read in the embedding potential, and add it as a one-electron operator in the Fock matrix\cite{acetone_w_cluster}.
The PyADF scripting framework provides all the necessary tools to manage various computational tasks and
manipulate the relevant quantities for electronic-structure methods.  
The resulting Python code, referred as Psi4-rt-PyEmbed, is available under GPLv3 license at Ref.~\citenum{pyberthagitfdert}.
A data set collection of computational results, including numerical data and parameters used to obtain the absorption spectra of 
Sections 4.2, 4.3 and 4.5, is available in the Zenodo repository and can be freely accessed at Ref. ~\citenum{source_fig}.

\subsection{Rapid prototyping and implementation}

Psi4Numpy~\cite{psi4numpy,doi:10.1021/acs.jctc.7b00174} 
and PyADF~\cite{pyadf,pyadf-github-v096}, both  provide a Python 
interface, which greatly
simplifies the computational work-flow from input data to the results.
PyADF is a quantum chemistry scripting framework that
provides mechanisms for both controlling the execution of different
computational tasks and for managing the communication between these
tasks using Python object-oriented programming techniques. As we already
mentioned, its built-in classes permit to handle different aspects
involved in the work-flow as a single unit.  All the advantages coming
from object-oriented programming (i.e extensibility and inheritance) are
readily available and allow us to incorporate third-party scientific
code and directly manipulate quantities coming from different codes
(Psi4Numpy) in our case. 

The Python HLL (High-Level Language), among others, permits to formally express 
complex algorithms in comparatively few lines of codes. This
makes rather straightforward to let PyADF interact
with Psi4Numpy native Python API. For the sake of completeness we want
to finally mention that, to accomplish our goal, we 
firstly had to port some of the frameworks (specifically XCFun, PyADF and
PyEmbed) to the new Python 3.0 standard (i.e., we used a private branch
of the cited packages, available at Refs.~\citenum{pyadf-github-v096-python3,xcfun-python3}). 

As an explicit example of the interoperativity achieved 
between different codes we report in Algorithm~\ref{maincode} 
some basic directives used to compute those key quantities necessary 
for our uFDE-rt-TDDFT.  The electron density of an active system 
is obtained via Psi4Numpy while the electron density, the Coulomb potential
and non-additive terms of the environment  are managed 
using PyADF. These quantities 
can be easily mapped on a common numerical grid 
and used in PyEmbed to evaluate the relative non additive embedding
potential.
Thus, the geometry and basis set of the active system (in this
specific case a H$_2$O molecule) are parsed at Line 7 and the ground
state wavefunction object is returned by the \textit{psi4.energy()}
method. The corresponding electron density matrix is then obtained
as a NumPy array by the \textit{h2o\_wfn} object. The electron density is
mapped into a real-space grid representation using a preset numerical grid,
and used to populate a suitable object container (Line 14-20). A ground
state calculation of the environment molecule (that is a NH$_3$ molecule in this example)
is carried out using PyADF \textit{run()} method (Line 23). In this case
we use the \textit{adfsinglepointjob} method to execute the corresponding ADF calculation\cite{ADF2017authors}.
We mention here that PyADF, despite its name, is not specific to this program, but works with a number of 
different quantum chemistry codes.
The density and Coulomb potential resulting from this calculation, that are represented on
a common numerical grid, are obtained using \textit{get\_density()} and
\textit{get\_potential()} methods (Line 25,27) respectively.
The PyEmbed module has all the methods needed to manage the density of both the
reference system and environment to finally compute the non-additive embedding potential.
Indeed, the \textit{embed\_eval} object is instantiated  (Line 34) and the non-additive 
embedding potential is evaluated on the numerical grid 
using \textit{get\_nad\_pot} (Line 36), once
the density of both the active system and of the environment has been provided.

\begin{algorithm}
\begin{algorithmic}[1]
\STATE \textbf{import psi4}
\STATE \textbf{import pyadf}
\STATE \textbf{import pyadf.PyEmbed}
\STATE \textbf{from pyadf.Plot.GridFunctions import GridFunctionFactory}
\STATE \textbf{from pyadf.Plot.GridFunctions import GridFunctionContainer}
\STATE \textbf{...}
\STATE \textbf{geom,mol = fde\_util.set\_input('h2o.xyz',basis\_set)}
\STATE \# psi4 run
\STATE \textbf{ene, h2o\_wfn = psi4.energy(func,return\_wfn=True)}
\STATE \# get psi4 h2o density 
\STATE \textbf{D = np.array(h2o\_wfn.Da())} 
\STATE \textbf{...}
\STATE \# map h2o density matrix to the numerical grid
\STATE \textbf{temp = 2.0 * fde\_util.denstogrid( phi, D, S,ndocc)}
\STATE \textbf{rho = np.zeros((temp.shape[0],10),dtype=np.float\_)}
\STATE \textbf{rho[:,0] = temp}
\STATE \# fill in the container with density
\STATE \textbf{dens\_gf\:=\:GridFunctionFactory.newGridFunction(agrid,\
      np.ascontiguousarray(rho[:,0]),gf\_type="density")}
\STATE \textbf{...}
\STATE \textbf{density\_h2o = GridFunctionContainer([dens\_gf, densgrad, denshess])}
\STATE \textbf{m\_nh3 = pyadf.molecule(nh3.xyz)}
\STATE \# ADF run 
\STATE \textbf{run\_nh3\:=\:pyadf.adfsinglepointjob(m\_nh3,\:basis\_active,\:settings=adf\_settings,\
  options=['NOSYMFIT']).run()}
\STATE \# get nh3 density
\STATE \textbf{density\_nh3  = run\_nh3.get\_density(grid=agrid, fit=False, order=2)}
\STATE \# get nh3 coulomb potential
\STATE \textbf{nh3\_coul  = run\_nh3.get\_potential(grid=agrid, pot='coul')}
\STATE \textbf{...}
\STATE \textbf{}
\STATE \# PyEmbed run
\STATE \textbf{embed\_settings = pyadf.PyEmbed.EmbedXCFunSettings()}
\STATE \textbf{embed\_settings.set\_fun\_nad\_xc (\{'BeckeX': 1.0, 'LYPC': 1.0\})}
\STATE \textbf{embed\_settings.set\_fun\_nad\_kin(\{'pw91k' : 1.0\})}
\STATE \textbf{embed\_eval = pyadf.PyEmbed.EmbedXCFunEvaluator(settings=embed\_settings)}
\STATE \# compute non-additive part of the embedding potential
\STATE \textbf{nadpot\_h2o\:=\:embed\_eval.get\_nad\_pot(density\_h2o,\:density\_nh3)}
\STATE \textbf{nad\_val = nadpot\_h2o.get\_values()}
\end{algorithmic}
  \caption{Illustrative Python code to compute active system density (using the Psi4Numpy code), environment density and Coulomb potential (using the ADF code) and non-additive embedding potential via the PyEmbed module.}
\label{maincode}
\end{algorithm}

\begin{figure}[!h]
        \includegraphics[width=0.90\textwidth]{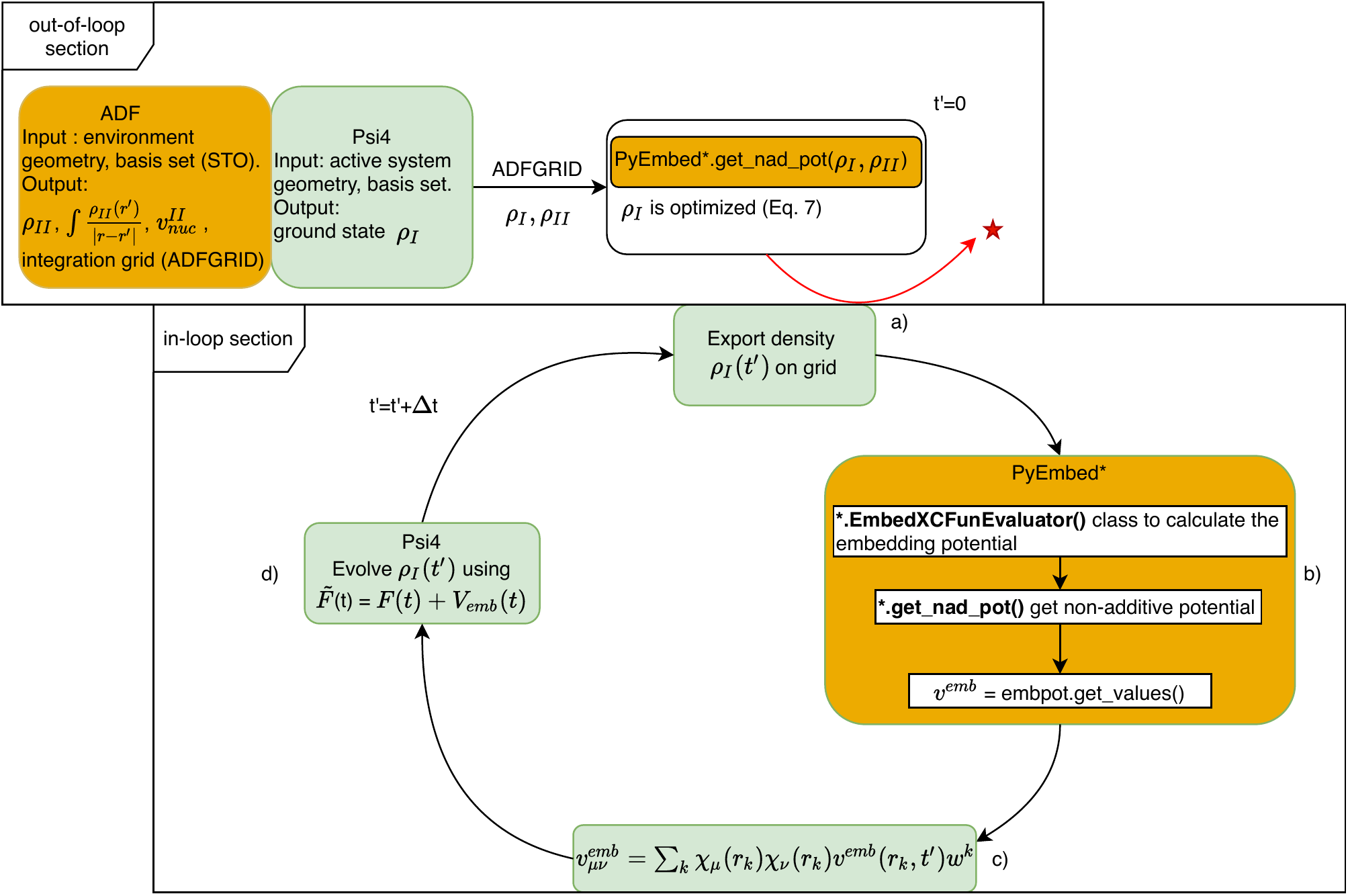}
	\caption{Working flowchart of the uFDE-RT-TDDFT. In the out-of-loop section the density and electrostatic potential of the environment are obtained as grid functions. The active system density matrix 
           is expressed as grid function object and used to calculate the embedding potential. The active system density is optimized self-consistently according to Eq 7.  
           The red star and the arrow pointing at it, symbolize that the out-of-loop blocks of tasks are involved only in the initial stage of the procedure. a) The relaxed active density matrix is exported as grid function.
         b) PyEmbed classes are used to calculate the embedding potential.
         c) The embedding potential is expressed on the finite basis set representation (GTO's). d) The active density matrix is evolved according to the real-time propagation scheme.}
        \label{fig:fde_cycle}
\end{figure}

Algorithm \ref{maincode} has well illustrated how we can utilize the classes 
provided by PyADF to obtain a very simple workflow
in which we are able to manipulate quantities coming from Psi4Numpy. 
Thus, we are now in a position to draw the main lines of our uFDE-rt-TDDFT implementation, the Psi4-RT-PyEmbed code\cite{pyberthagitfdert}. 
In Figure \ref{fig:fde_cycle} we present its pictorial  workflow.

We start describing the out-of-loop section. Firstly the geometry and basis set of the environment
are initialized (the orange left most block), thus the ADF package provides, 
through a standalone single point calculation, the electrostatic and nuclear 
potential of the environment and its 
density $\rho_\text{II}$ and a suitable integration grid for later use. At this stage all 
the basis sets and exchange-correlation functionals available in the ADF library 
can be used. In the next step, green block, the geometry and basis set of the active 
system are parsed from input 
and the ground state density $\rho_\text{I}$ is calculated using the Psi4Numpy related methods. 
The right pointing arrow, connecting the last block,
sketches the mapping of the density matrix onto the real-space grid representation. 
The evaluation of $\rho_\text{I}(\bm{r})$ on the numerical grid is efficiently accomplished using the 
molecular orbitals (MO), which requires 
the valuation the localized basis functions at the grid points.

Finally, the PyEmbed module comes into play 
(last block of the out-of-loop section), the real-space 
electron densities $\rho_\text{I}$ and $\rho_\text{II}$ serve as input for the \textit{get\_nad\_pot()} method. Thus
the non-additive kinetic and exchange potential are obtained. The embedding potential is then calculated from its constituents
(i.e. the environment electrostatic and nuclear potential and the non-additive contribution as 
detailed in Eq.~\ref{eq:vemb}) and evaluated at each grid point, $v^\text{emb}(\bm{r}_k)$.
The embedding potential matrix representation in the active subsystem basis set, $\vV^\text{emb}$,
is calculated numerically on the grid as 
\begin{equation}\label{eq:vembmatrix}
V_{\mu\nu}^\text{emb}= \sum_k \chi_\mu(r_k) \chi_\nu(r_k) v^\text{emb}(r_k) w_k
\end{equation}
where $\chi_\mu(\bm{r}_k)$ are the Gaussian-type basis set functions 
employed in the active systems (used in Psi4Numpy) evaluated at the grid 
point, $\bm{r}_k$. In the above expression, $w_k$ are specific integration  weights.

In the case of a FDE-rt-TDDFT calculation, the electron density of the
active system at the beginning of the propagation ($t_0 = 0$, initial
condition) is not the ground state density of the isolated  molecule,
rather a polarized ground state density. The latter is obtained through
a self-consistent-field calculation in the presence of the embedding
potential.  We adopt the so-called  split-scf scheme as described in
Ref.~\citenum{wesolowski09}.  It should be noted that the density matrix,
corresponding to the optimized $\rho_\text{I}$ electron density, is the input data
for the green block (block a) of the in-loop section.  The outgoing red arrow,
connecting the out- and the in-loop branch of the diagram, it means that
the former is only involved in the early step of the procedure and it
will no longer come into play during the time propagation.
As mentioned, the optimized density matrix of the active system as resulting from the SCF
procedure including the embedding potential, is the starting point for the real-time propagation. Whereupon, at each time step we determine the embedding potential
corresponding to the instantaneous active density ($v^\text{emb}[\rho_\text{I}(t), \rho_\text{II}]$).
Again we need its mapping onto the real-space grid as shown in the first green box (box a).
Then, we utilize the methods reported in the rectangular orange box (box b)
to calculate the non-additive part of the embedding potential at each grid point. 
Finally we add the non-additive (kinetic and exchange-correlation)
potential to the electrostatic potential of the environment calculated again at each 
grid point. It should be noted that 
because the density of the environment is frozen, thus the corresponding electrostatic 
potential remains constant during the time propagation. In the next phase 
its matrix representation in the localized Gaussian basis functions is obtained 
as in Eq.\ref{eq:vembmatrix}, (box c, in Figure).
The active system is evolved (box d) using an effective time-dependent Kohn-Sham matrix, which contains the usual implicit and explicit time-dependent terms, respectively (\textbf{J}$[\rho_\text{I}(t)]$+\textbf{V}$_{\mathrm{XC}}$$[\rho_\text{I}(t)]$) and
\textbf{v}$_{\mathrm{ext}}(t)$, plus the time-dependent 
embedding potential ($\vV^\text{emb}[\rho_\text{I}(t), \rho_\text{II}]$).

For the sake of completeness, the pseudo code needed to evolve the
density using the second-order midpoint Magnus propagator is reported in
SI and relies on the methodology illustrated in Section 2.2. We refer
the interested readers to our recent work on real-time propagation
for further details \cite{pybertha}. 
\section{Results and Discussion}
\label{sec:results}
In the present section we report a series of results mainly devoted 
to assess the correctness of the uFDE-rt-TDDFT scheme.
To the best of our knowledge, this implementation is 
the first available for localized basis sets. 
Since our implementation relies on the embedding strategies implemented in PyADF,
it appears natural and appropriate to choose as a useful reference the 
uncoupled FDE-TDDFT scheme,  based on the linear response\cite{jacs:2004,jacob2008flexible} and 
implemented in the ADF program package~\cite{te2001chemistry}.

\subsection{Initial validation and numerical stability}

Before going into the details of the numerical comparison between our implementation 
and the FDE-TDDFT scheme based on the linear response (ADF-LR) formalism, whether in 
combination with FDE (ADF-LR-FDE) or not, 
is important to first assess
the basis set dependence of the calculated excitation energies using the two different approaches.
This preliminary study is mandatory because Psi4Numpy (Gaussians) and ADF (Slaters) employ different 
types of atom-centered basis functions. Due to this difference, 
perfect numerical agreement between the two implementations can not be expected, 
but it is important to quantify the variability of our target 
observables (the excitation energies of a water molecule) with 
variations in the basis set. 

In order to simulate the linear response
regime within our Psi4-rt, 
the electronic ground-state of a water molecule,
calculated in absence of an external electric field, 
was perturbed by an analytic
$\delta$-function pulse with a strength of $\kappa =$ 1.0$\times10^{-5}$ a.u. along the three directions, $x,y,z$.
The induced dipole moment has been  collected for 9000 time steps with a length of 0.1 a.u. per time step,
corresponding to 21.7 fs of simulation. 
This time dependent  dipole moment is then Fourier transformed in order to obtain
the dipole strength function $S(w)$, accordingly to Eq.~\ref{eq:osc_str} and the transition energies.
The Fourier transform of the induced dipole moment has been carried out by means of Pad\'e approximants
\cite{bruner2016accelerated,goings2018real}.

As shown in Table~\ref{tab:wat_amm_conv}, convergence
can be observed with both Psi4-rt and ADF-LR, in particular
for the first low-lying transitions (additional excitation
energies are reported in the Supporting Information).  For some
of the higher energy transitions the
convergence is less prominent, pointing to deficiencies in the smaller basis sets.
We mention that the results obtained using our Psi4-rt implementation
perfectly agree with those obtained using the TDDFT implementation
based on linear response implemented in the NWChem code, which uses
the same Gaussian type basis set (see Table S1 in SI).
Thus, we conclude that most of the deviations from the ADF-LR values can
be ascribed to unavoidable basis set differences. A qualitatively similar 
pattern of  differences is to be expected when including the environment effect
within the FDE framework.

\begin{table}
	\caption{
Excitation energies (in eV) corresponding to the first five low-lying transitions of the isolated water molecule.
Data obtained using TDDFT based on linear response implemented in ADF (ADF-LR) and  
the our real-time TDDFT implemented (Psi4-rt).
The labels  (D, T, Q) correspond to data obtained using the 
 Gaussian-type basis sets aug-cc-pVXZ (X = D,T,Q)  and 
Slater-type basis sets AUG-X$'$ (X$'$ = DZP,TZ2P,QZ4P) which are used in the Psi4-rt and ADF-LR codes, respectively (see text for details).
}
\label{tab:wat_amm_conv}
\begin{tabular}{llllllllllll}
\multicolumn{8}{c}{Excitation energy (e.V)} \\ \cline{1-8}
\multicolumn{8}{c}{\qquad \qquad Psi4-rt  \qquad ADF-LR} \\ \cline{2-4} \cline{6-8}
& D & T & Q & & D     & T& Q \\
Root 1     & 6.2144      &6.2269 &6.2244         &  & 6.1610      &   6.1887 & 6.2868 \\ 
Root 2     & 7.5125      &7.4660 &7.4404         &  & 7.4540      &   7.4646 & 7.8841\\
Root 3     & 8.3626      &8.3516 &8.3436         &  & 8.3088      &   8.2881 & 8.4267\\
Root 4     &9.5357       &8.9526 &8.6506         &  & 8.8033      &   8.4825 & 8.6276\\
Root 5     &9.6436       &9.5721 &9.3056         &  &8.9446       &   8.8453 & 10.022\\
\end{tabular}
\end{table}

To assess differences in the presence of an environment, we next tested our uFDE-rt-TDDFT results against ADF-LR-FDE ones.
The target system is the water-ammonia adduct, in which the
water molecule is the active system that is bound to an ammonia molecule, which
plays the role of the embedding environment.   
In the Psi4-rt-PyEmbed case we
employed a contracted Gaussian aug-cc-pVXZ (X=D,T) 
basis set~\cite{dunning1989a,kendall1992a} for the active system  whereas the basis set used
in PyADF for the calculation of the environment frozen density (ammonia) and
the embedding potential is the AUG-X$'$ (X$'$=DZP,TZ2P) Slater-type set 
from the ADF library\cite{te2001chemistry}.
The ADF-LR-FDE employs the AUG-X$'$ (X$'$=DZP,TZ2P) basis sets from the same library.
For the real-time propagation of the active system (water), in both the isolated and the embedded case 
the BLYP~\cite{becke88,lyp1988} exchange-correlation functional is used, while the Thomas-Fermi and LDA functionals~\cite{vwn1980,slater1951} have been employed for the non-additive kinetic
and non-additive exchange-correlation potentential, respectively.
The numerical results are reported in Table~\ref{tab:wat_amm}.
Although, as expected, there is
no quantitative agreement on the absolute value of the
transitions, the shift $\Delta$ ($E_{iso.}-E_\text{emb}$) shows an acceptable agreement
for the lowest transitions
(see for additional excitation energies the Supporting Information).

\begin{table}
	\caption{Excitation energies (in eV) corresponding to the first five low-lying transitions of both the isolated  and  embedded calculations water molecule are reported. In the embedded water molecule, an ammonia molecule is used as 
environment.
Data have been obtained using our new Psi4-rt-PyEmbed implementation and reference ADF-LR-FDE implementation with (a) aug-cc-pVDZ and
AUG-DZP basis sets; and (b) aug-cc-pVTZ and AUG-TZ2P basis sets (see text for details).
The shift $\Delta$ ($E_{iso.}-E_\text{emb}$) in the transition energies due to
the embedding environment is also reported.
}
\label{tab:wat_amm}
\begin{tabular}{llllllllllll}
\multicolumn{8}{c}{Excitation energy (e.V)} \\ \cline{1-8}
\multicolumn{8}{c}{\qquad \qquad Psi4-rt-PyEmbed  \qquad ADF-LR-FDE} \\ \cline{2-4} \cline{6-8}
& isolated & emb. &$\Delta$& & isolated      & emb. &$\Delta$ \\
\hline
\multicolumn{8}{c}{(a) double-zeta calculations} \\
\hline
Root 1     & 6.2144      & 5.8167 &   0.398       & & 6.1610      &5.6871 &  0.474\\  
Root 2     & 7.5125      & 6.6940 &   0.818       & & 7.4540      &6.5779 &  0.876\\  
Root 3     & 8.3626      &7.8924  &   0.470       & & 8.3088      &7.7818 &  0.527\\  
Root 4     &9.5357       & 8.7677 &   0.768       &  & 8.8033     &8.3361 &  0.467\\  
Root 5      &9.6436       &9.1861 &   0.458        &  &8.9446     &8.4215&  0.523\\  
\hline
\multicolumn{8}{c}{(a) triple-zeta calculations} \\
\hline
Root 1     &6.2269       &5.7964 &   0.4305     & &   6.1887     & 5.6886    &   0.500 \\
Root 2     &7.4660       &6.5734 &   0.8926     & &   7.4646     & 6.5592     &  0.905 \\
Root 3     &8.3516       &7.8485 &   0.5031     & &   8.2881     & 7.7339     &  0.554 \\
Root 4     &8.9526       &8.5596 &   0.3930     & &   8.4825     & 7.9692     &  0.513 \\
Root 5     &9.5721       &8.6246 &   0.9475     & &   8.8453     & 8.3180     &  0.527 \\
\end{tabular}
\end{table}
From these results, we conclude that our implementation is both 
stable and numerically correct, with differences between the methods 
explainable by the  intrinsic basis set differences. 

\subsection{The water in water test case}

To provide a further test of our implementation, we also computed
the absorption spectra of a water molecule embedded in a water cluster of increasing size.
The geometries of the different water clusters are taken from Refs.~\cite{watercluster1,watercluster2}
which corresponds to one snapshot taken from a MD simulation.
Different cluster models were taken in consideration, by progressive addition of surrounding water molecules (from 1 to 5 molecules) to the single active water molecule.
For the active system water molecule propagation in Psi4-rt-PyEmbed we use the aug-cc-pVDZ basis set
while for the environment, computed using the ADF code,  we use the AUG-DZP basis set.
In both cases we use the BLYP~\cite{becke88,lyp1988} exchange-correlation functional while for the non-additive
kinetic and non-additive exchange-correlation terms in the generation of the embedding potential the Thomas-Fermi and LDA
functionals are used, respectively.
In each case, we use 9000 time steps of propagation which corresponds to a 
simulation of $\approx$ 22 fs (time step of 0.1 a.u.). 
The corresponding dipole strength functions ($S_z(w) = 2w/(3\pi)\mathrm{Im}[\alpha_{zz}(w)]$)
along the $z$-direction are reported in Fig.~\ref{fig:w_cluster_abs}.
Upon the increase of the cluster dimension, the lowest-lying transition
shifts within a range of about 1 eV and no spectra display cusps or 
irregular behavior.  These results give confidence in the numerical stability of the propagation 
when the number of molecules in the environment is increased. 

\begin{figure}[!h]
        \includegraphics[width=0.60\textwidth]{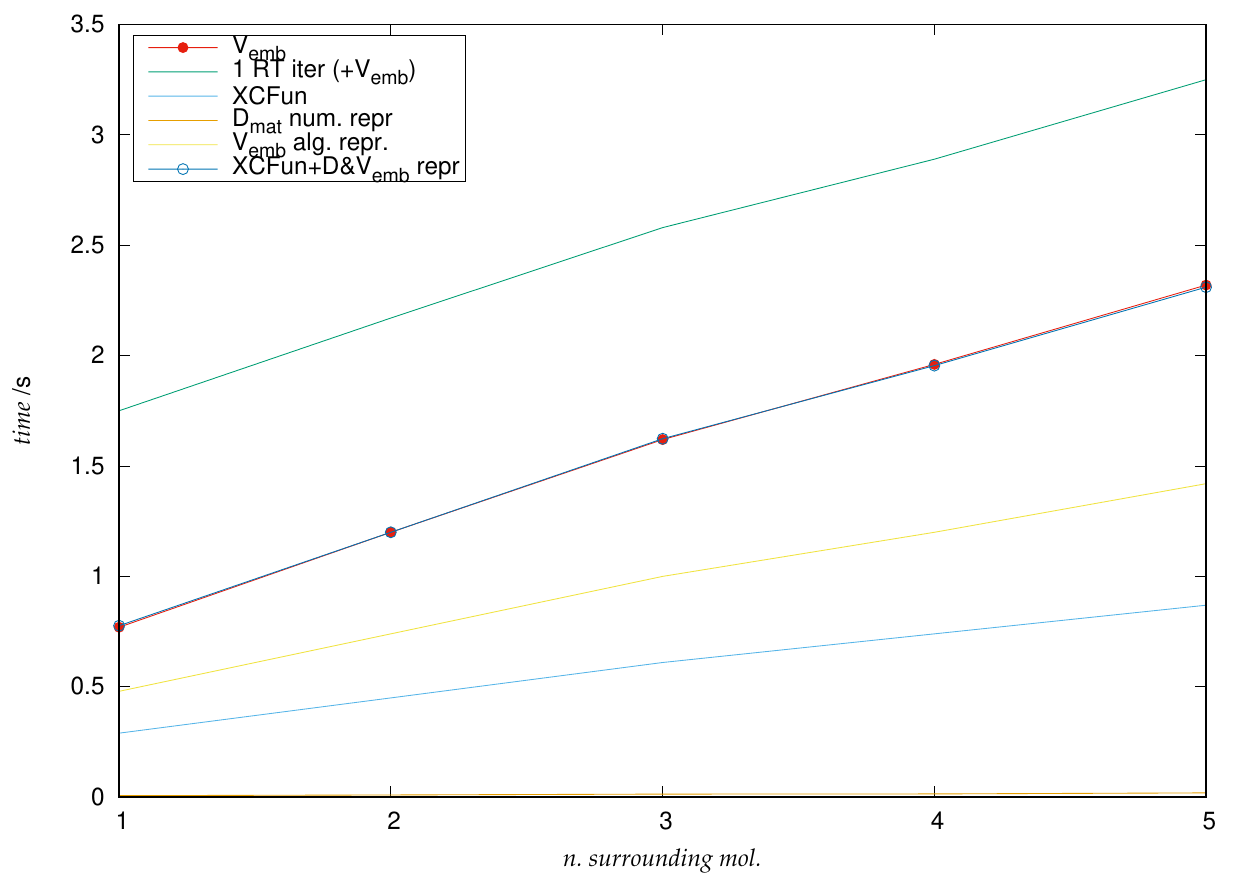}
        \caption{Time needed for different tasks vs number of surrounding molecules}
        \label{fig:w_cluster_time}
\end{figure}
\begin{figure}[!h]
        \includegraphics[width=0.55\textwidth]{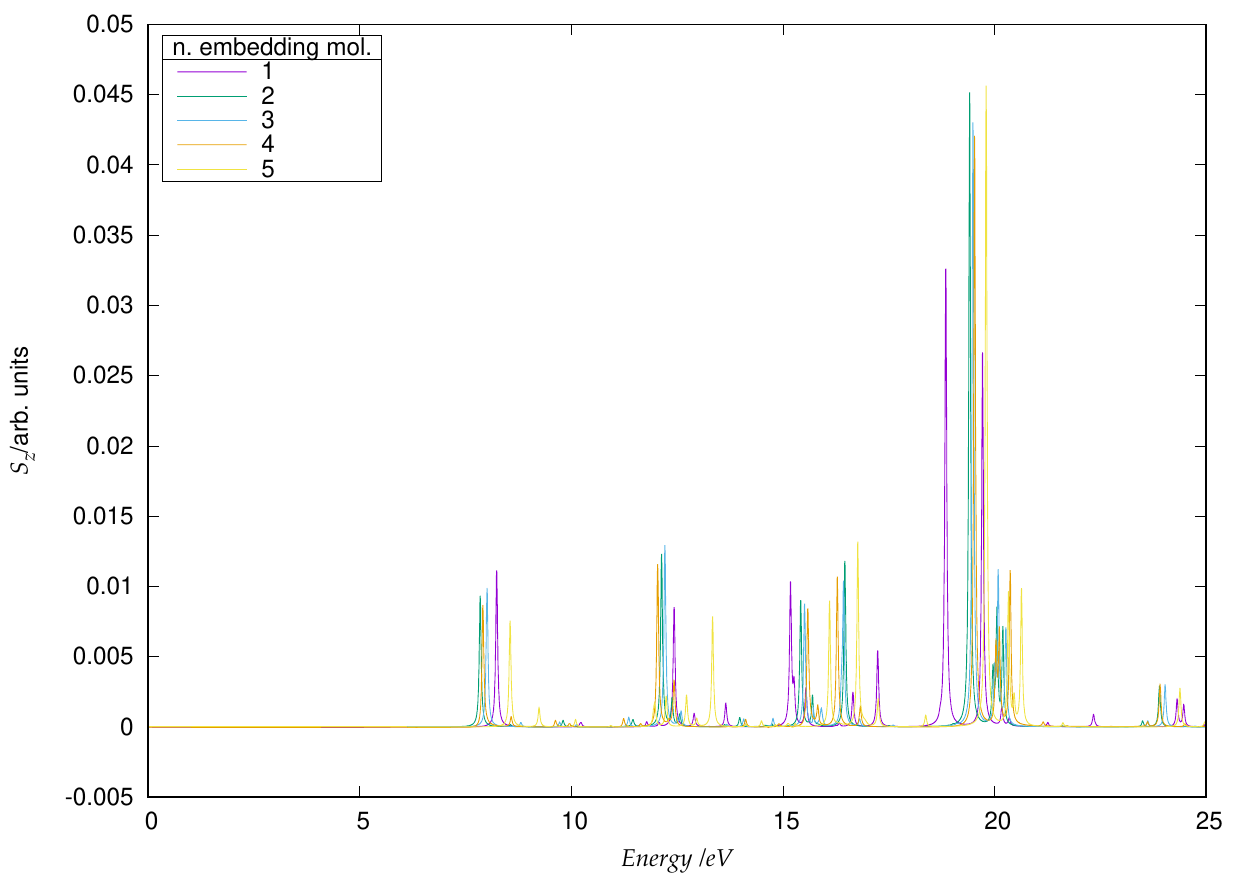}\includegraphics[width=0.55\textwidth]{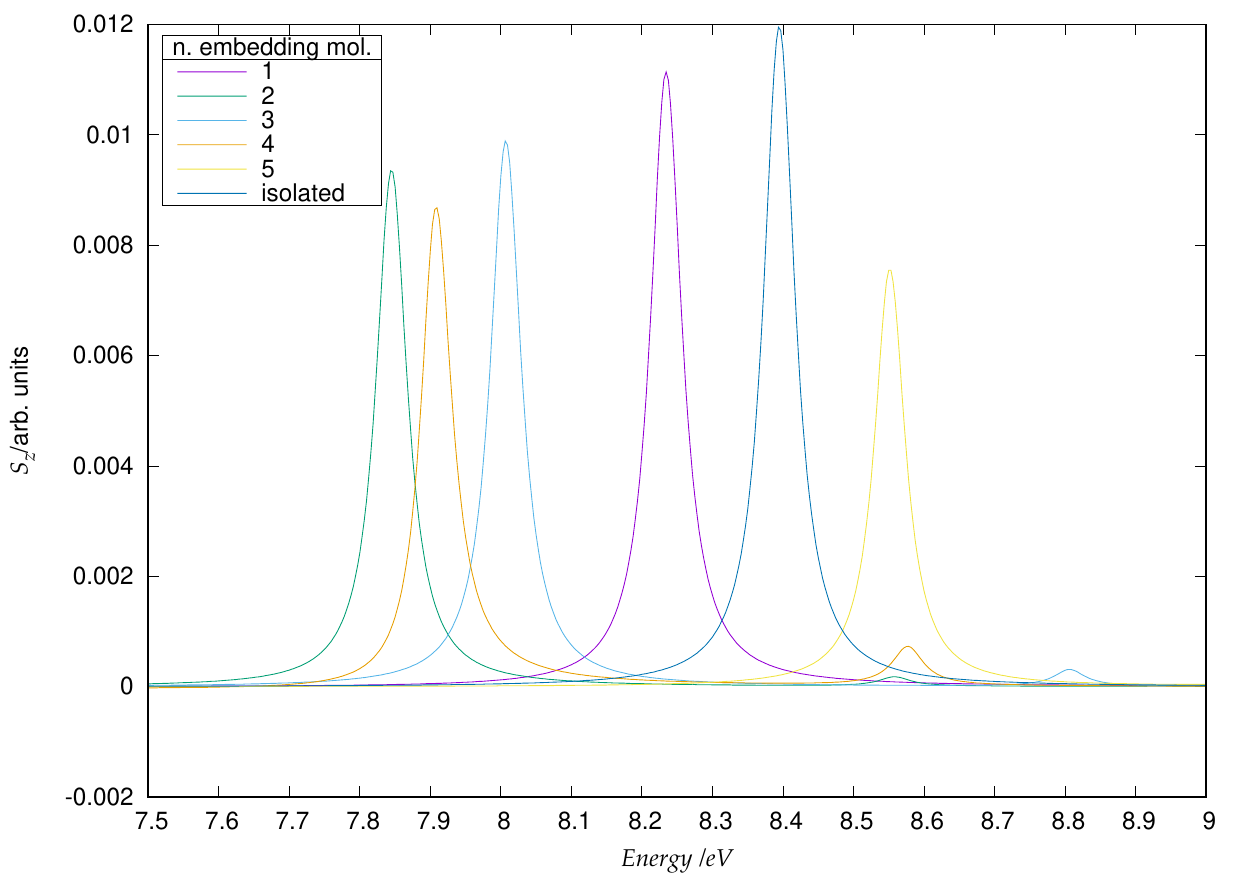}
        \caption{Dipole strength function $S_z$ of the water cluster as a function of the number of surrounding molecules (left panel).
        Right panel: detailed representation of the low lying transition. The peak corresponding to the isolated molecule is reported for comparison}
        \label{fig:w_cluster_abs}
\end{figure}

The systematic increase of the size of the environment makes it possible to also consider the actual computational scaling of the
Psi4-rt-PyEmbed code for this case. To show this scaling, we carried out a single time step of the real-time propagation and  broke down the 
computational cost into those of the different steps in the work-flow, as reported in  Fig.~\ref{fig:fde_cycle}.

\begin{table}
\caption{Time usage in seconds. $^a$ : Density on grid (through MOs). $^b$ : XCFun (non-additive potential calculation). 
	$^c$ : V$_\text{emb}$ projection onto the basis set $^d$ : Total time for V$_\text{emb}$ evaluation. $^{e}$ : Total time for a rt-iteration}
\label{tab:time_stats}
\begin{tabular}{llllllll}
\hline
	& t$^a$ &t$^b$ & t$^c$ &  t$^d$ & t$^e$   \\
\hline
1 &0.007 &0.29  &0.48  & 0.77  & 1.75 \\
2 &0.01  &0.45  &0.74  & 1.20  & 2.17 \\
3 &0.014 &0.61  &1.0   & 1.62  & 2.58 \\
4 &0.015 &0.74  &1.2   & 1.96  & 2.89 \\
5 & 0.02 &0.87  &1.42  & 2.32  & 3.25 \\
\end{tabular}
\end{table}

In Table~\ref{tab:time_stats}, and in Fig.\ref{fig:w_cluster_time},  we report how the time for the embedding 
potential calculation is distributed over the different tasks, when the number of surrounding water molecules increases
from one to five. It is interesting to note that the time needed to evaluate the embedding potential increases almost linearly,
for the limited number of water molecules considered here. The standard real-time iteration time
(corresponding to the isolated water molecule) takes 
less than 1 sec and shows up as a fixed cost in the increasing computation time, while the time spent in the embedding part
is dominated by the evaluation of the matrix representation for the active subsystem, e.g 
step \textit{c}) of Fig.\ref{fig:fde_cycle} (see for instance t$^c$ column of Table~\ref{tab:time_stats}). The time spent in this evaluation
depends on the number of numerical integration points used to represent the potential, and can be reduced by using special grids
for embedding purposes once the environment is large enough.

\subsection{The acetone in water test case}

\begin{figure}[!h]
        \includegraphics[width=0.50\textwidth]{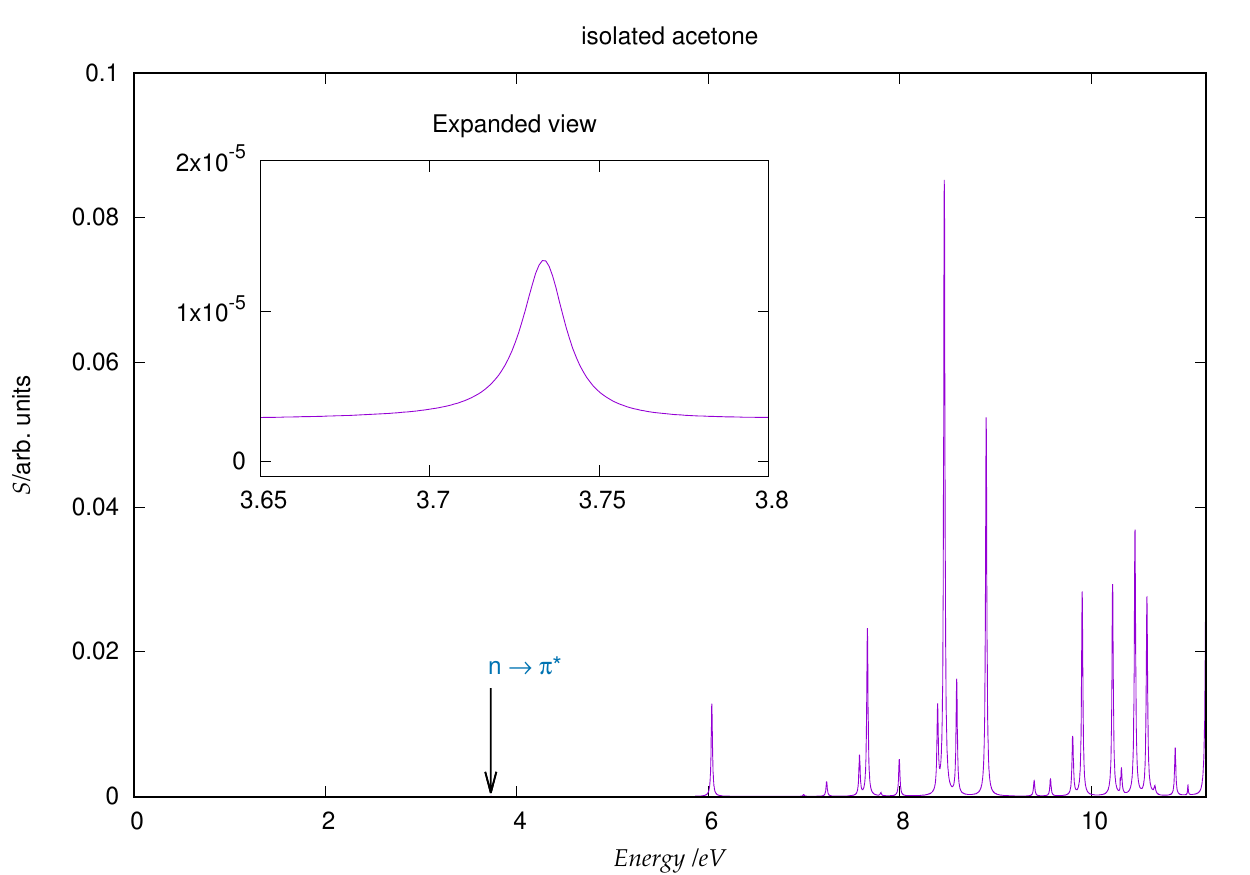}\includegraphics[width=0.50\textwidth]{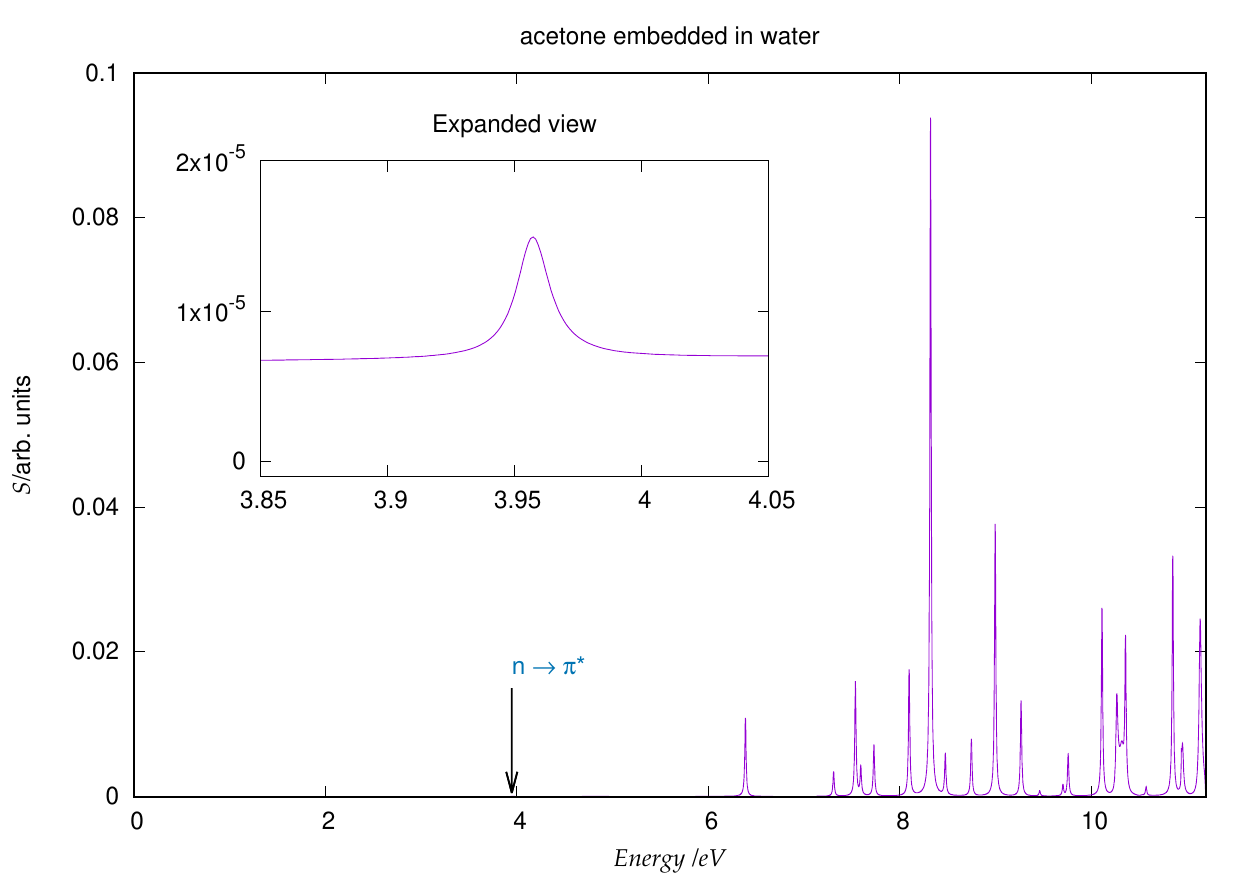}
        \caption{Absorption spectrum of isolated acetone (left panel) and embedded acetone in a water cluster (right panel)}
        \label{fig:acetone_emb}
\end{figure}

As a further test of the numerical stability of accuracy of the method, 
we investigated the $n \to \pi^*$  transition in the acetone molecule, both isolation and using an explicit water cluster to model
solvation.
In order to assess the shift due to the embedding potential, we calculate the absorption spectrum of the isolated molecule
at the same geometry it has in the cluster model.
The geometry for the solvated acetone system was taken
from Ref.~\citenum{acetone_w_cluster}, corresponding to one snapshot from a MD
simulation, where the acetone is surrounded by an environment consisting of 56 water molecules.
The uFDE-rt-TDDFT calculation has been obtained specifying in our Psi4-rt-PyEmbed framework 
all the computational details.
In particular, the frozen density of the environment is obtained from a ground
state calculation using ADF in combination with the PBE functional and DZP basis set,
while for the acetone we employ the BLYP functional and the Gaussian def2-svp basis set using the Psi4-rt code.
The non-additive kinetic and exchange-correlation terms of the embedding potential are calculated using
the Thomas-Fermi and LDA functionals respectively.
For the isolated acetone the  $n \to \pi^*$ transition is found at 3.73 eV whereas for the embedded molecule is 
located at 3.96 eV. The full absorption spectrum is reported in Fig. \ref{fig:acetone_emb}.

It is worth noting that, due to its low intensity, 
this transition is particularly challenging  for a real-time propagation framework.
To obtain a spectrum up to 11 eV, we carried out a simulation consisting of $20000$ time steps and lasting 
2000 a.u (48 fs). This relatively long simulation time demonstrates the 
numerical stability of the approach and its implementation.

\begin{table}
\begin{tabular}{llll}
\hline
	&iso./eV & emb./eV & $\Delta E$/eV \\
\hline
        Psi4-rt-PyEmbed & 3.7337 & 3.9583 & 0.225 \\
	ADF-LR-FDE      & 3.7928 & 3.9749 & 0.182\\
\end{tabular}
  \caption{Isolated and embedded in a water cluster acetone $n \to \pi^*$ transition, reported 
  for both ADF-LR-FDE and our Psi4-rt-PyEmbed code.}
  \label{tab:acetone_emb}
\end{table}

As an overall check of our implementation we compare the shift of the $n \to \pi^*$ transition 
observed between 
isolated and embedded in a water cluster acetone obtained using both our Psi4-rt-PyEmbed and the ADF-LR-FDE methods. 
The active system response was calculated at BLYP level of theory, while Thomas-Fermi and
LDA functionals were employed for the non-additive kinetic and exchange-correlation terms respectively of the embedding potential
in the ADF-LR-FDE calculation.
As one can observe by looking at the values reported in Table \ref{tab:acetone_emb},
we obtain a good agreement in the absolute values, both isolated and embedded acetone, and the computed shift
is likewise in rather good agreement.

\subsection{FDE-rt-TDDFT  in the non-linear regime}

A specificity in the the real-time approach is that the evolution of
the electron density can be driven by an real-valued electric field
whose shape can be explicitly  modulated. Realistic laser fields can
be modeled by a \textit{sine} function of $\omega_0$ frequency using any
physically meaningful enveloping function.  Using an explicit external 
field is a key tool in optical control theory, furthermore 
it is possible, employing high intensity field, to study phenomena beyond linear-response, i.e
hyperpolarizability coefficients and high harmonic generation in molecules.
The latter point will be detailed in the
following section.

In this section we demonstrate that the uFDE-rt-TDDFT scheme gives stable
numerical results not only in the perturbative regime, as shown above,
but also in the presence of intense fields.  Physically meaningful laser
fields are adequately represented by sinusoidal  pulse
 of the form $E(t) = f(t)\sin(\omega_0t)$ where $\omega_0$ is the carrier frequency.
In this work we employ,  a $\cos^2$  shape for the envelope
function\cite{luppi2012}:

\begin{flalign*}
f(t) = E_0\cos^2\Big(\frac{\pi}{2\sigma} (\tau -t)\Big)  \qquad for \quad |t-\tau| \le \tau \\
0 \qquad \text{elsewhere}
\end{flalign*} 

where $\tau$ is the width of the field envelope.  We have calculated
the response of H$_2$O embedded in a water cluster model made of five water
molecules (all the details about the geometry have been reported in the
previous section) to a $\cos^2$-shaped laser field with carrier frequency
$\omega_0 = $ 1.55 eV (analogously to a Ti:Sapphire laser), and intensity $I =
1.02\times10^{14} W cm^{-2}$ (which corresponds to a field $E = 0.054$
au) and a duration of 20 optical cycles. Each cycle lasts $2\pi/\omega_0$,
and the overall pulse spans over 2250.0 au (i.e. 54 fs).
The field has been chosen along the molecular symmetry axis ($z$)
and the 6-311++G** basis set and B3LYP functional  were used.
The propagation was carried 
out for a total time of 3500 a.u without any numerical instabilities.

\begin{figure}[!h]
        \includegraphics[width=0.80\textwidth]{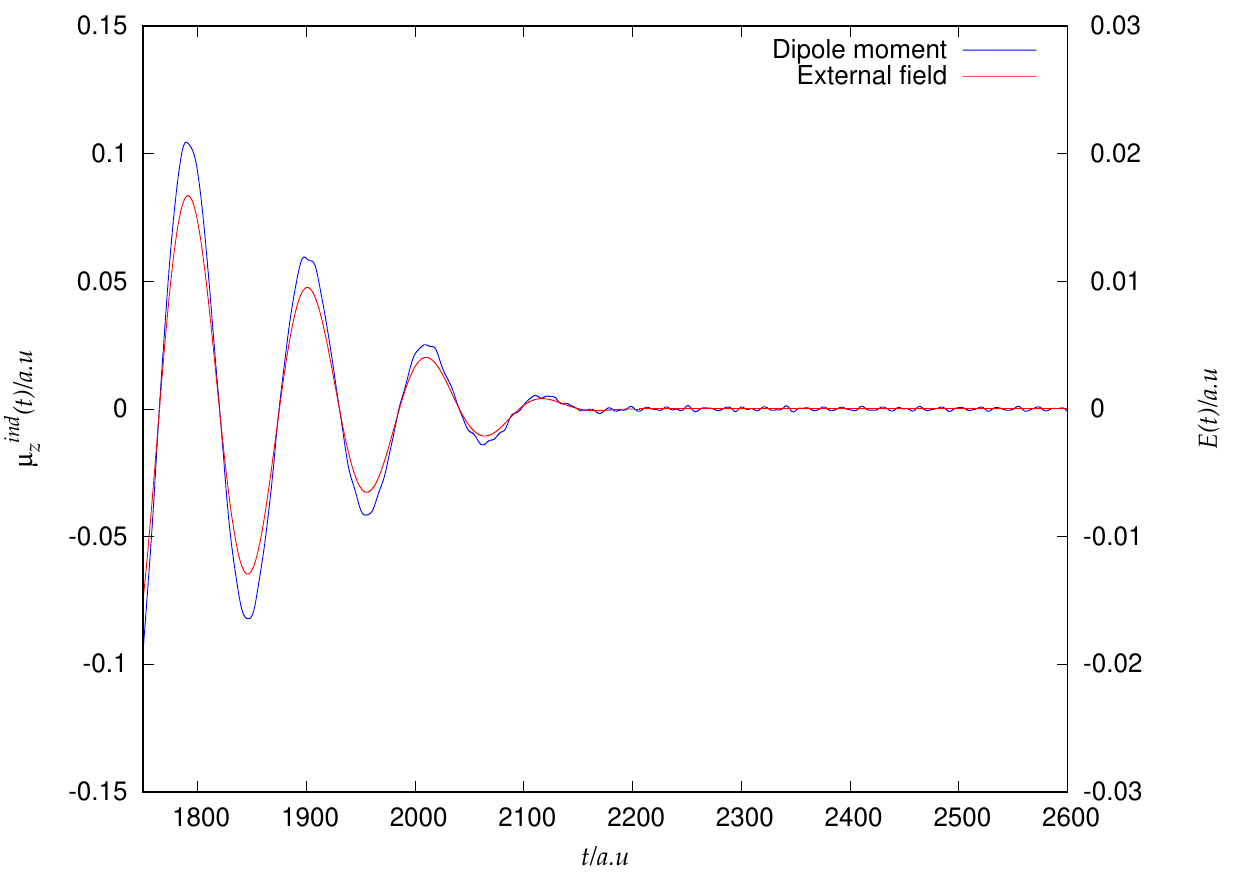}
	\caption{Induced dipole moment in H$_2$O molecule. The representation of the external field is also reported as
  a green line. }
        \label{fig:dipole_emb}
\end{figure}

As shown in Fig. \ref{fig:dipole_emb}, the induced dipole does not follow the applied field adiabatically 
when a strong field is applied, especially in the few last optical cycles, strong diabatic effects are clearly present.
These effects lead to the presence of a residual dipole oscillation.
Following a previous work on high harmonic generation (HHG) in H$_2$ molecule,~\cite{luppi2012}
we extract the high-order harmonic intensities via the Fourier transform of the laser-driven induced dipole moment
(neglecting the remaining part, i.e for $t$ larger than $\tau$, i.e 2250 au in the present simulation) as 

\begin{equation}\label{eq:hhg_eq}
P(\omega) \propto \Bigg| \int_{t_1}^{t_2} \mu_z(t)exp(-i \omega t) dt \Bigg|^2
\end{equation}

\begin{figure}[!h]
  \centering
  \centering
  \begin{subfigure}[t]{0.80\textwidth}
	  \includegraphics[width=\textwidth]{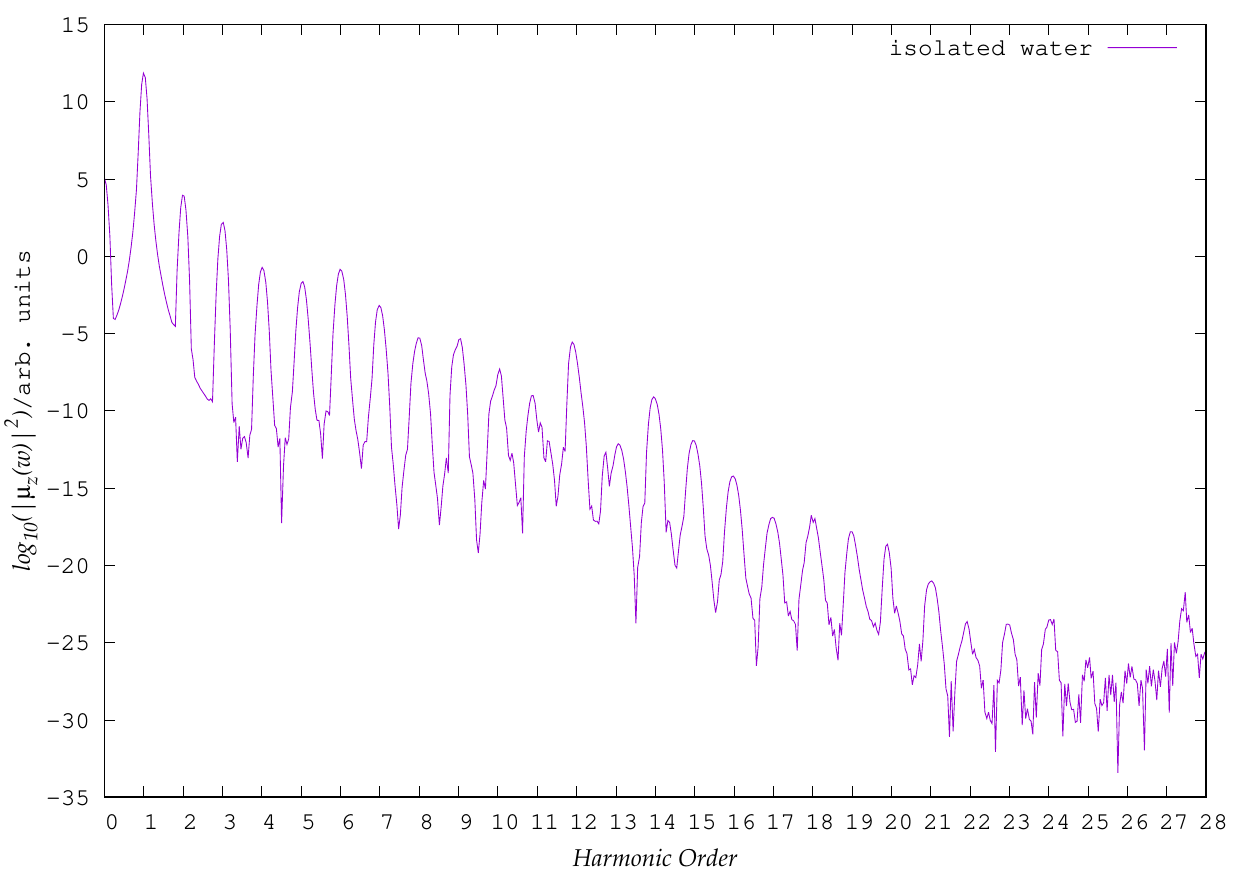}
  \end{subfigure}

  \centering
  \centering
  \begin{subfigure}[t]{0.80\textwidth}
    \includegraphics[width=\textwidth]{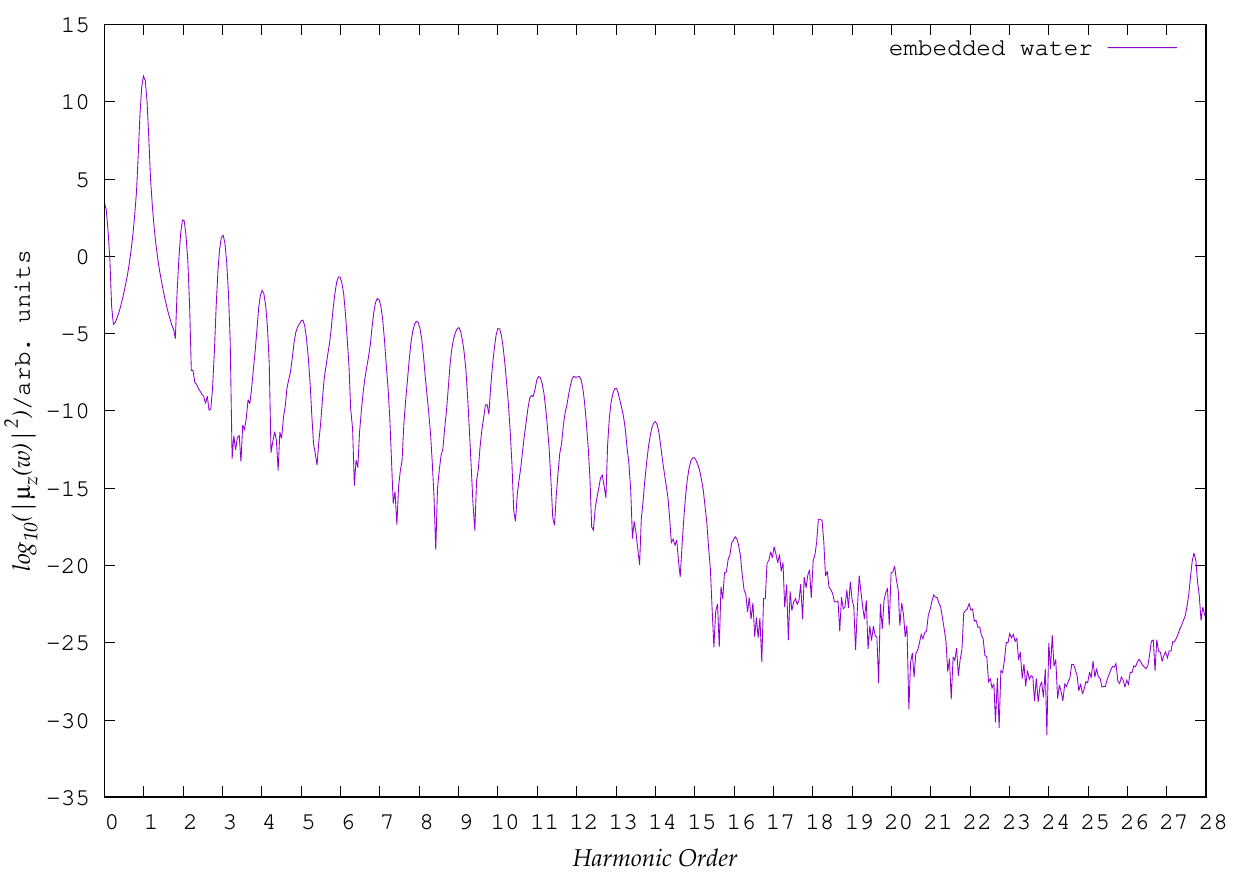}
  \end{subfigure}
	\caption{Upper panel: Emission spectrum of isolated water molecule. Lower panel: Emission spectrum of the same water molecule embedded in the (H$_2$O)$_5$ cluster}
  \label{fig:hhg_emission}
\end{figure}

In Fig.~\ref{fig:hhg_emission} we report the base-10 logarithm of the spectral intensity 
for the embedded water molecule and we compare it
to the HHG of the isolated water calculated that has the same geometry it has in the cluster model. 
In the case of the isolated water molecule we are able to observe relatively well defined peaks up to the 21th
harmonics. 
We mention that this finding qualitatively agrees with data obtained by Sun et al.~\cite{sun2007}
 (see Figure 3 of Ref.\cite{sun2007}).

An important parameter in the analysis of the HHG spectrum 
is the value of the energy cutoff ($E_{cutoff}$), which is related with  the maximum
number of high harmonics ($N_{max} \approx E_{cutoff}/\omega_0$).
In a semiclassical formulation~\cite{hhg_corkum}, which, among others
assumes that only a single electron is active for HHG, $E_{cutoff}\approx I_p + 3.17U_p$,
where $I_p$ is the ionization potential of the system  and $U_p$
($U_p=\frac{E^2}{4\omega_0^2}$) is the ponderomotive energy in
the laser field of strength $E$ and frequency $\omega_0$~\cite{hhg_corkum}.
In the case of molecular systems, the HHG spectra present
more complex features and the above formula it is not strictly valid.
With the laser parameters used here ($E=0.054$ a.u, $\omega_0=0.05696$ a.u.) and
the experimental ionization potential of H$_2O$ ($I_p=0.4637$ a.u.), the above 
formula predicts a $E_{cutoff}$ value of 1.17601 a.u. ($N_{max}$ at about the 21th harmonic),
which is remarkably consistent with HHG spectra we observed here.

For the water molecule embedded in the cluster the same boundary can be approximately found 
corresponding to the 16th harmonic. The peaks at higher energies have a very
small intensity and are much less resolved above the 16th harmonic.
The flattening of the HHG intensity 
pattern is therefore solely due the introduction of the embedding potential of the surrounding cluster. 
The latter is consistent with a shift towards lower ionization
energy passing from free water molecule to a small water cluster observed 
experimentally\cite{doi:10.1021/jp906113e}.

\subsection{Computational constraints}

Before concluding this work it may be interesting
to put forward some assessments in terms of time statistics,
to be used as a basis for optimizing the computation time and speed-up
any uFDE-rt-TDDFT calculations. 
We are using a water-ammonia complex as a general test-case, where the geometry of the adduct
has been taken from Ref.\citenum{geom_opts} and the water is the active subsystem.

In the real-time framework the embedding potential is, evidently, an implicit time-dependent 
quantity. Since in the uncoupled FDE framework the density of the environment is kept frozen, 
the embedding potential depends on time
only through the relatively small contributions given by the exchange-correlation and kinetic non-additive terms, 
which in turn it depends on time only through the density of the active subsystem. The electrostatic potential,
due to the frozen electron density and nuclear charges of the environment, is the leading term
in the overall potential. Thus, it may be reasonable to choose a
longer time step for the update of the embedding potential, which is weakly varying in time.

\begin{figure}
  \includegraphics[width=0.55\textwidth]{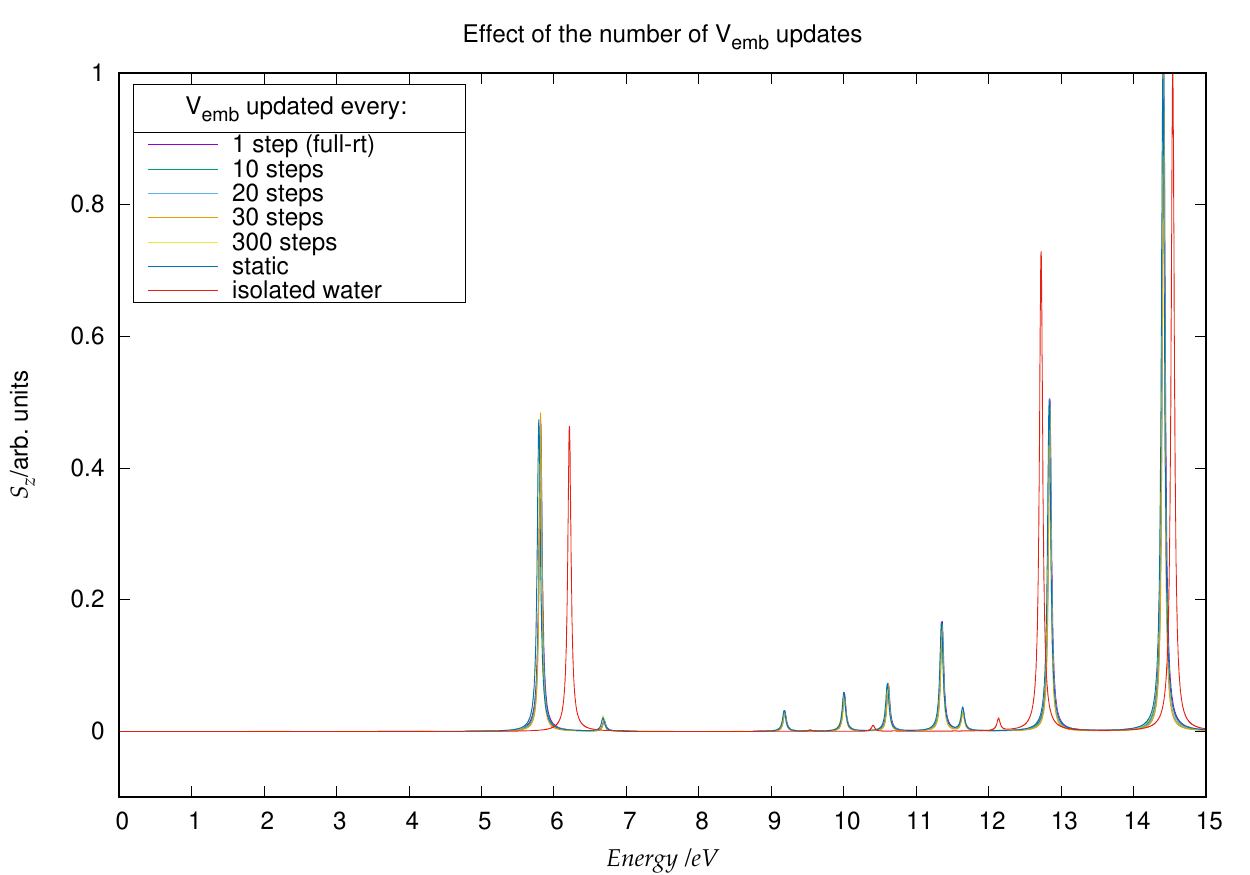}\includegraphics[width=0.55\textwidth]{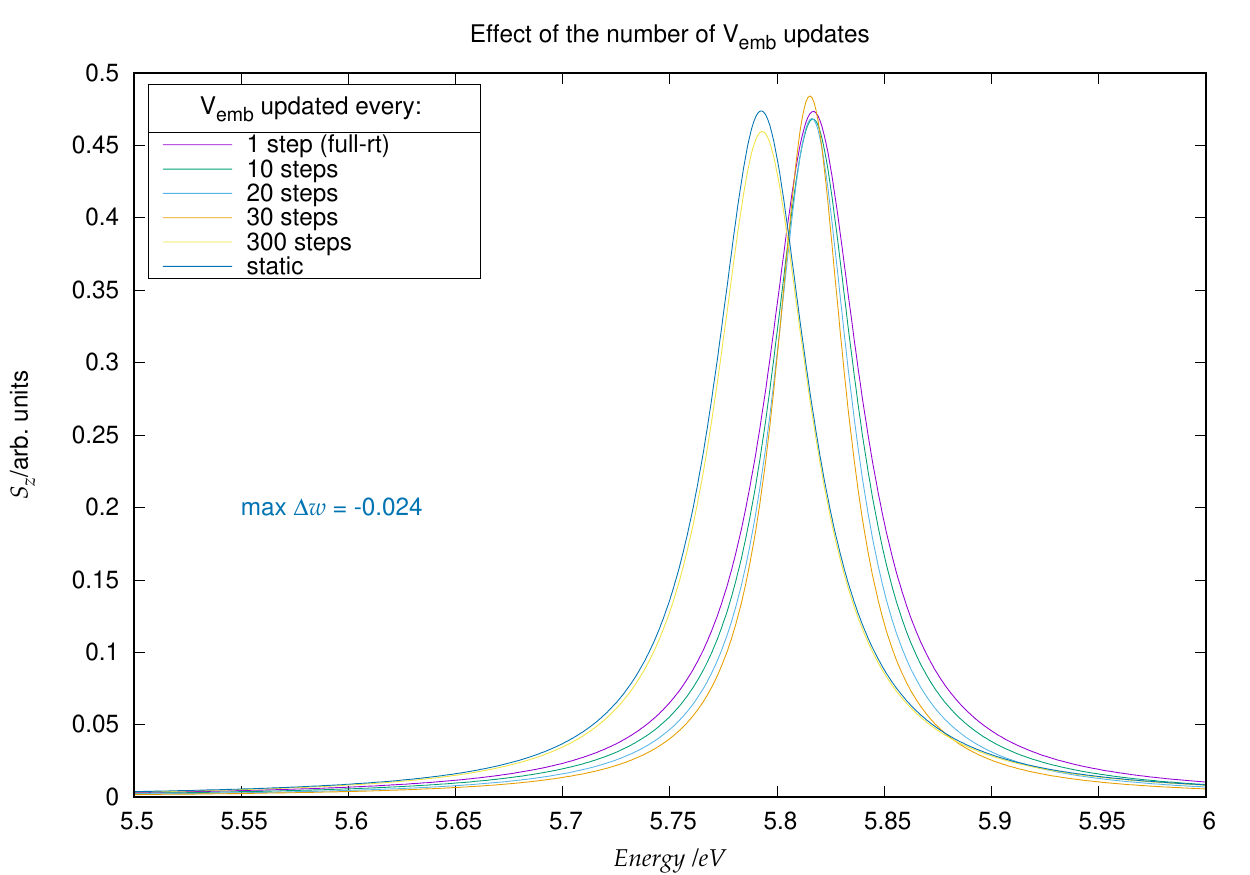}
  \caption{Left: Frequency shift in the S$_z$ function due to increasing rate of update of the embedding potential. The peaks corresponding
	to the isolated water molecule are also reported as red trace. Right: Expanded view of the homo-lumo transition}
  \label{fig:vemb_up}
\end{figure}

In order to investigate such a possible speed-up, we carried out different simulations in which the time interval of 
the embedding potential updating is progressively increased. 
The results are reported in Table~\ref{tab:time_stats_ammonia}.
Of course, as the number of time steps between consecutive updates is increased (i.e. the embedding potential is 
updated less often), the total time needed to perform the full simulation goes down, as the time spent in computing 
the embedding potential decreases.
The update rate of the embedding potential
during the propagation affects to some extent the position of the peaks in the absorption spectrum. As can be seen in Fig.~\ref{fig:vemb_up}
the different traces corresponding to dipole strength functions calculated with different update rates, do not differ significantly and tend to coalesce
as the number of time steps between consecutive updates decreases below 30 time steps. In particular, in the case of the lowest-energy transition,
the energy shift corresponding to a quite long update period (roughly 300 time steps) is of the order of 0.02 eV.

\begin{table}
\caption{Time in seconds as a function of the number n of time steps between consecutive updates of the 
embedding potential. $^f$ : time for V$_\text{emb}$ evaluation. $^g$ : Total time for V$_\text{emb}$ evaluation in 
the propagation. $^h$ : Total time needed for 100 real-time iterations}
\label{tab:time_stats_ammonia}
\begin{tabular}{llll}
\hline
	t$^f$ & t$^g$ & t$^h$ & $n$ \\
\hline
0.87 &- &94.84 &  inf (static)  \\
0.87 &2.59 & 97.52 &30\\
0.85 &4.32 &99.26 &  20\\
0.86 &8.56 & 103.31 &  10\\
0.86 &85.67 &180.97 & 1\\

\end{tabular}
\end{table}

\begin{figure}
  \includegraphics[width=0.55\textwidth]{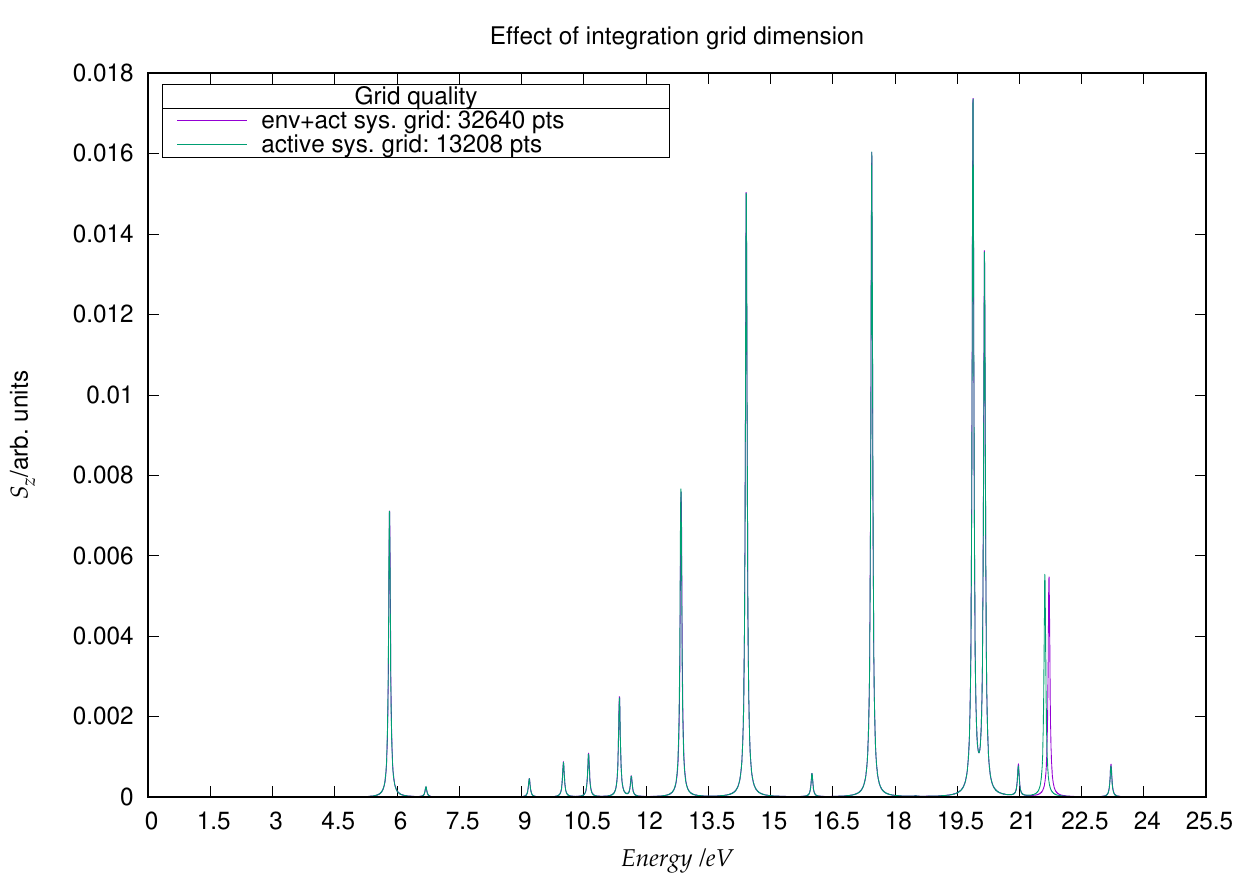}\includegraphics[width=0.55\textwidth]{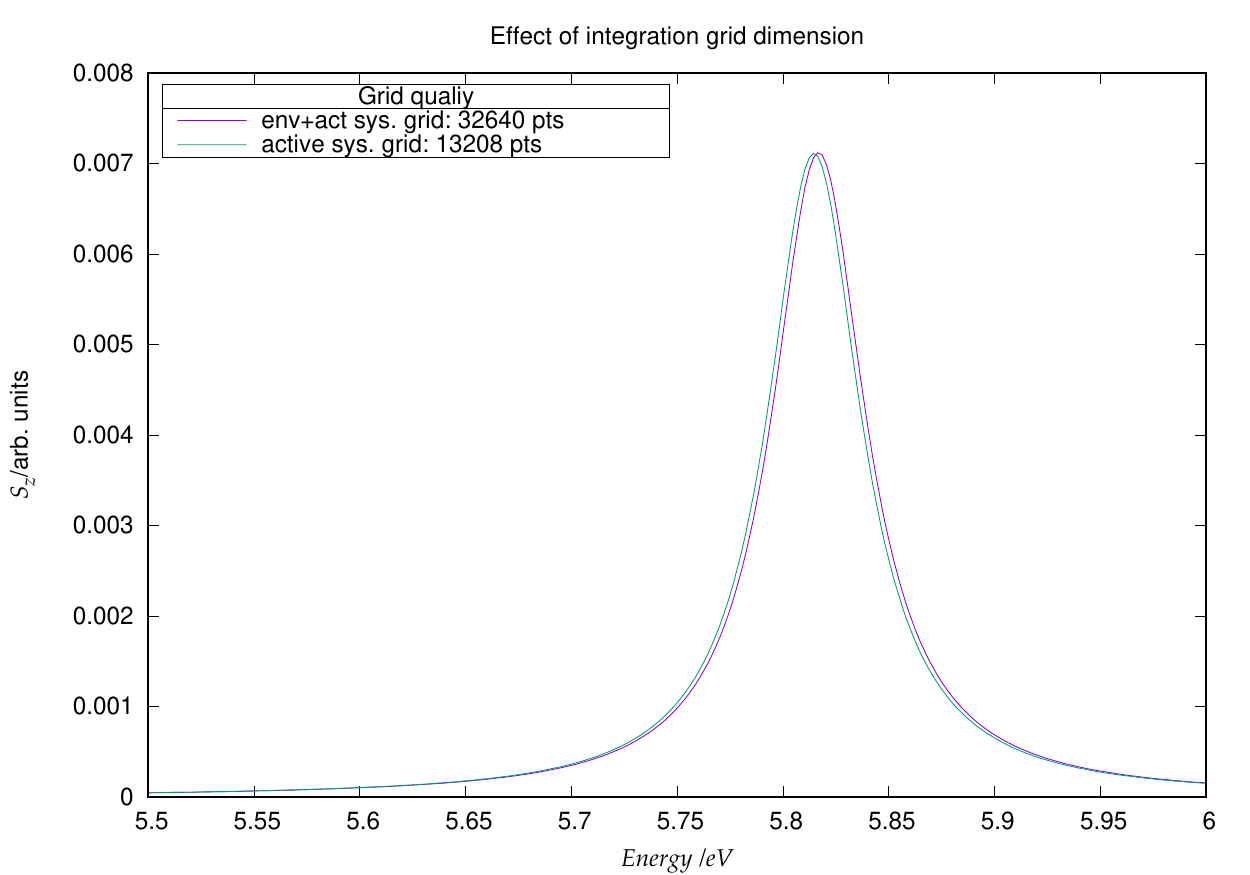}
  \caption{Comparison of $S_z$ dipole strength function obtained with two different integration grids. The violet trace corresponds
	to the full supramolecular integration grid, while the green trace to the active system grid. Right: Expanded view of the lowest-lying transition}
  \label{fig:grid_calc}
\end{figure}

We also reported the partition between different tasks of the time needed for the calculation of the embedding potential in Table~\ref{tab:time_part}.
As seen before, the calculation of the embedding potential is largely dominated by the projection to the basis set of the embedding potential from the 
numerical-grid representation.
Therefore, some preliminary tests in reducing the number of grid points were carried out, and the results are presented in Fig.~\ref{fig:grid_calc}.
It can be seen that there is no significant modification in the peak positions due
to the use of a coarser integration grid: the overall spectrum is essentially stable and no artifacts are introduced.

\begin{table}
\caption{Time usage in seconds. $^a$ : Density on grid (through MOs). $^b$ : XCFun (non-additive potential calculation). 
	$^c$ : V$_\text{emb}$ projection onto the basis set $^d$ : Total time for V$_\text{emb}$ evaluation.}
\begin{tabular}{llll}
\hline
	t$^a$ & t$^b$ & t$^c$ &  t$^d$  \\
\hline
	0.01 &  0.33   & 0.53 &0.87 \\
\end{tabular}
\label{tab:time_part}
\end{table}

We furthermore note the possibility to use small grid localized solely on the active system  by utilizing the fact that 
the embedding potential is projected on the localized basis set functions 
of the active system (see Eq.\ref{eq:vembmatrix}), which makes it possible to neglect points on which these functions
have a small value. 

\section{Conclusions and perspectives}
\label{sec:conclusions}
In this work we have focused on the implementation of the
 Frozen Density Embedding scheme in the real-time TDDFT. 
We have integrated the Psi4Numpy real-time module we 
recently developed within the PyADF framework. 
We have devised a real-time FDE scheme in which the active density is 
evolved under the presence of the embedding potential.
This implementation relies on a multiscale approach, since the
embedding potential is calculated by means of PyADF, while
the propagation is carried out by Psi4Numpy.
We tested the implementation on a simple water cluster showing
that the time needed for the propagation scales linearly with
 the cluster size.
We studied many low-lying transitions in the case of a water
molecule embedded in ammonia, and we showed that the shift
of excitation energies with respect to the isolated water molecule is in good 
agreement with the results obtained using linear response FDE TDDFT implemented in ADF.
Finally, we tackled a challenging case for rt-TDDFT, by computing the lowest-energy transition of acetone,
which features an extremely low intensity. The corresponding signal can be identified in the 
computed spectrum, and we evaluated the solvatochromic shift due to the presence of a surrounding
water cluster. We obtained a frequency shift of 0.225 eV, close to the reference value, 0.182 eV, 
from LR-FDE TDDFT as implemented in ADF. The scheme we developed has proven to be reliable also
in the case of propagation in the non-linear regime. As a demonstration, we perturbed with a strong electric field a water molecule
surrounded by five water molecules acting as frozen environment. Numerically stable induced dipole moment
and corresponding emission spectrum were obtained. 

Finally, we like to state that the present work provides an excellent 
framework for future developments. It is for instance possible and desirable to optimize the embedding potential
construction. In our implementation (i.e. Psi4-rt-PyEmbed) the projection onto the basis set
of the embedding potential from the numerical grid representation dominates the computational
burden. The change of the embedding potential matrix in time, (i.e the difference at two
consecutive time steps) depends on the relatively small contributions given by the
exchange-correlation and kinetic non-additive terms.
Significant improvement could be achieved by exploiting the sparsity of the matrix corresponding to 
that difference. Moreover, the use of smaller integration grid would probably further improve the 
procedure. Last but not least, the effect of relaxation of the environment has to be investigated.
In our uncoupled FDE-rt-TDDFT scheme we are able to study local transitions within a given subsystem, 
and particularly those of the active system under the influence of the embedding potential due to the frozen 
environment. Thus we neglect transitions involving the environment and those due to the couplings of 
the subsystems. Relaxing the environment can be crucial both in the linear-response framework, in order to recover
supramolecular excitations, and in the non-linear regime where a polarizable
environment could heavily affect the hyperpolarizabilities of the target system.
The limit of the uncoupled FDE scheme can be overcome by carrying out a simultaneous 
propagation of subsystems~\cite{pavanello2015} and the computational framework developed in
the present work represents an important step in that direction.

\section{Acknowledgements}
\label{sec:acknowledgements}
CRJ and ASPG acknowledge funding from the Franco-German project CompRIXS (Agence nationale de la recherche ANR-19-CE29-0019, Deutsche Forschungsgemeinschaft JA 2329/6-1).
ASPG further acknowledges support from the CNRS Institute of Physics (INP), PIA ANR project CaPPA (ANR-11-LABX-0005-01), I-SITE ULNE project OVERSEE (ANR-16-IDEX-0004), the French Ministry of Higher Education and Research, region Hauts de France council and European Regional Development Fund (ERDF) project CPER CLIMIBIO. CRJ acknowledges funding from the Deutsche Forschungsgemeinschaft for the development of PyADF (Project Suresoft, JA 2329/7-1).
\bibliography{shortj,biblio}

\providecommand{\latin}[1]{#1}
\makeatletter
\providecommand{\doi}
  {\begingroup\let\do\@makeother\dospecials
  \catcode`\{=1 \catcode`\}=2 \doi@aux}
\providecommand{\doi@aux}[1]{\endgroup\texttt{#1}}
\makeatother
\providecommand*\mcitethebibliography{\thebibliography}
\csname @ifundefined\endcsname{endmcitethebibliography}
  {\let\endmcitethebibliography\endthebibliography}{}
\begin{mcitethebibliography}{130}
\providecommand*\natexlab[1]{#1}
\providecommand*\mciteSetBstSublistMode[1]{}
\providecommand*\mciteSetBstMaxWidthForm[2]{}
\providecommand*\mciteBstWouldAddEndPuncttrue
  {\def\EndOfBibitem{\unskip.}}
\providecommand*\mciteBstWouldAddEndPunctfalse
  {\let\EndOfBibitem\relax}
\providecommand*\mciteSetBstMidEndSepPunct[3]{}
\providecommand*\mciteSetBstSublistLabelBeginEnd[3]{}
\providecommand*\EndOfBibitem{}
\mciteSetBstSublistMode{f}
\mciteSetBstMaxWidthForm{subitem}{(\alph{mcitesubitemcount})}
\mciteSetBstSublistLabelBeginEnd
  {\mcitemaxwidthsubitemform\space}
  {\relax}
  {\relax}

\bibitem[Hardin \latin{et~al.}(2009)Hardin, Hoke, Armstrong, Yum, Comte,
  Torres, Fr{\'e}chet, Nazeeruddin, Gr{\"a}tzel, and McGehee]{dssc1}
Hardin,~B.~E.; Hoke,~E.~T.; Armstrong,~P.~B.; Yum,~J.-H.; Comte,~P.;
  Torres,~T.; Fr{\'e}chet,~J.~M.; Nazeeruddin,~M.~K.; Gr{\"a}tzel,~M.;
  McGehee,~M.~D. Increased light harvesting in dye-sensitized solar cells with
  energy relay dyes. \emph{Nature Photonics} \textbf{2009}, \emph{3}, 406\relax
\mciteBstWouldAddEndPuncttrue
\mciteSetBstMidEndSepPunct{\mcitedefaultmidpunct}
{\mcitedefaultendpunct}{\mcitedefaultseppunct}\relax
\EndOfBibitem
\bibitem[Hagfeldt \latin{et~al.}(2010)Hagfeldt, Boschloo, Sun, Kloo, and
  Pettersson]{dssc2}
Hagfeldt,~A.; Boschloo,~G.; Sun,~L.; Kloo,~L.; Pettersson,~H. Dye-sensitized
  solar cells. \emph{Chem, Rev.} \textbf{2010}, \emph{110}, 6595--6663\relax
\mciteBstWouldAddEndPuncttrue
\mciteSetBstMidEndSepPunct{\mcitedefaultmidpunct}
{\mcitedefaultendpunct}{\mcitedefaultseppunct}\relax
\EndOfBibitem
\bibitem[Sali\`eres \latin{et~al.}(1999)Sali\`eres, Le~D\'eroff, Auguste,
  Monot, d'Oliveira, Campo, Hergott, Merdji, and Carr\'e]{prl_hhg}
Sali\`eres,~P.; Le~D\'eroff,~L.; Auguste,~T.; Monot,~P.; d'Oliveira,~P.;
  Campo,~D.; Hergott,~J.-F.; Merdji,~H.; Carr\'e,~B. Frequency-Domain
  Interferometry in the XUV with High-Order Harmonics. \emph{Phys. Rev. Lett.}
  \textbf{1999}, \emph{83}, 5483--5486\relax
\mciteBstWouldAddEndPuncttrue
\mciteSetBstMidEndSepPunct{\mcitedefaultmidpunct}
{\mcitedefaultendpunct}{\mcitedefaultseppunct}\relax
\EndOfBibitem
\bibitem[Paul \latin{et~al.}(2001)Paul, Toma, Breger, Mullot, Aug{\'e}, Balcou,
  Muller, and Agostini]{atto_hhg}
Paul,~P.~M.; Toma,~E.~S.; Breger,~P.; Mullot,~G.; Aug{\'e},~F.; Balcou,~P.;
  Muller,~H.~G.; Agostini,~P. Observation of a Train of Attosecond Pulses from
  High Harmonic Generation. \emph{Science} \textbf{2001}, \emph{292},
  1689--1692\relax
\mciteBstWouldAddEndPuncttrue
\mciteSetBstMidEndSepPunct{\mcitedefaultmidpunct}
{\mcitedefaultendpunct}{\mcitedefaultseppunct}\relax
\EndOfBibitem
\bibitem[Bass \latin{et~al.}(1962)Bass, Franken, Ward, and Weinreich]{prl_or}
Bass,~M.; Franken,~P.~A.; Ward,~J.~F.; Weinreich,~G. Optical Rectification.
  \emph{Phys. Rev. Lett.} \textbf{1962}, \emph{9}, 446--448\relax
\mciteBstWouldAddEndPuncttrue
\mciteSetBstMidEndSepPunct{\mcitedefaultmidpunct}
{\mcitedefaultendpunct}{\mcitedefaultseppunct}\relax
\EndOfBibitem
\bibitem[Kadlec \latin{et~al.}(2005)Kadlec, Ku\v{z}el, and Coutaz]{kadlec_or}
Kadlec,~F.; Ku\v{z}el,~P.; Coutaz,~J.-L. Study of terahertz radiation generated
  by optical rectification on thin gold films. \emph{Opt. Lett.} \textbf{2005},
  \emph{30}, 1402--1404\relax
\mciteBstWouldAddEndPuncttrue
\mciteSetBstMidEndSepPunct{\mcitedefaultmidpunct}
{\mcitedefaultendpunct}{\mcitedefaultseppunct}\relax
\EndOfBibitem
\bibitem[Keldysh(2017)]{Keldysh_2017}
Keldysh,~L.~V. Multiphoton ionization by a very short pulse.
  \emph{Physics-Uspekhi} \textbf{2017}, \emph{60}, 1187--1193\relax
\mciteBstWouldAddEndPuncttrue
\mciteSetBstMidEndSepPunct{\mcitedefaultmidpunct}
{\mcitedefaultendpunct}{\mcitedefaultseppunct}\relax
\EndOfBibitem
\bibitem[Eberly \latin{et~al.}(1991)Eberly, Javanainen, and Rza{\.z}ewski]{ati}
Eberly,~J.; Javanainen,~J.; Rza{\.z}ewski,~K. Above-threshold ionization.
  \emph{Physics Reports} \textbf{1991}, \emph{204}, 331 -- 383\relax
\mciteBstWouldAddEndPuncttrue
\mciteSetBstMidEndSepPunct{\mcitedefaultmidpunct}
{\mcitedefaultendpunct}{\mcitedefaultseppunct}\relax
\EndOfBibitem
\bibitem[Gallmann \latin{et~al.}(2012)Gallmann, Cirelli, and
  Keller]{attosecond_rev1}
Gallmann,~L.; Cirelli,~C.; Keller,~U. Attosecond Science: Recent Highlights and
  Future Trends. \emph{Annual Review of Physical Chemistry} \textbf{2012},
  \emph{63}, 447--469, PMID: 22404594\relax
\mciteBstWouldAddEndPuncttrue
\mciteSetBstMidEndSepPunct{\mcitedefaultmidpunct}
{\mcitedefaultendpunct}{\mcitedefaultseppunct}\relax
\EndOfBibitem
\bibitem[Ramasesha \latin{et~al.}(2016)Ramasesha, Leone, and
  Neumark]{attosecond_rev2}
Ramasesha,~K.; Leone,~S.~R.; Neumark,~D.~M. Real-Time Probing of Electron
  Dynamics Using Attosecond Time-Resolved Spectroscopy. \emph{Annual Review of
  Physical Chemistry} \textbf{2016}, \emph{67}, 41--63, PMID: 26980312\relax
\mciteBstWouldAddEndPuncttrue
\mciteSetBstMidEndSepPunct{\mcitedefaultmidpunct}
{\mcitedefaultendpunct}{\mcitedefaultseppunct}\relax
\EndOfBibitem
\bibitem[Attar \latin{et~al.}(2017)Attar, Bhattacherjee, Pemmaraju, Schnorr,
  Closser, Prendergast, and Leone]{Attar54}
Attar,~A.~R.; Bhattacherjee,~A.; Pemmaraju,~C.~D.; Schnorr,~K.; Closser,~K.~D.;
  Prendergast,~D.; Leone,~S.~R. Femtosecond x-ray spectroscopy of an
  electrocyclic ring-opening reaction. \emph{Science} \textbf{2017},
  \emph{356}, 54--59\relax
\mciteBstWouldAddEndPuncttrue
\mciteSetBstMidEndSepPunct{\mcitedefaultmidpunct}
{\mcitedefaultendpunct}{\mcitedefaultseppunct}\relax
\EndOfBibitem
\bibitem[Wolf \latin{et~al.}(2019)Wolf, Sanchez, Yang, Parrish, Nunes,
  Centurion, Coffee, Cryan, Ghr, Hegazy, Kirrander, Li, Ruddock, Shen,
  Vecchione, Weathersby, Weber, Wilkin, Yong, Zheng, Wang, Minitti, and
  Martinez]{Wolf2019}
Wolf,~T. J.~A.; Sanchez,~D.~M.; Yang,~J.; Parrish,~R.~M.; Nunes,~J. P.~F.;
  Centurion,~M.; Coffee,~R.; Cryan,~J.~P.; Ghr,~M.; Hegazy,~K.; Kirrander,~A.;
  Li,~R.~K.; Ruddock,~J.; Shen,~X.; Vecchione,~T.; Weathersby,~S.~P.;
  Weber,~P.~M.; Wilkin,~K.; Yong,~H.; Zheng,~Q.; Wang,~X.~J.; Minitti,~M.~P.;
  Martinez,~T.~J. The photochemical ring-opening of 1,3-cyclohexadiene imaged
  by ultrafast electron diffraction. \emph{Nat. Chem.} \textbf{2019},
  \emph{11}, 504--509\relax
\mciteBstWouldAddEndPuncttrue
\mciteSetBstMidEndSepPunct{\mcitedefaultmidpunct}
{\mcitedefaultendpunct}{\mcitedefaultseppunct}\relax
\EndOfBibitem
\bibitem[Ruddock \latin{et~al.}(2019)Ruddock, Zotev, Stankus, Yong, Bellshaw,
  Boutet, Lane, Liang, Carbajo, Du, Kirrander, Minitti, and Weber]{Adam2019}
Ruddock,~J.~M.; Zotev,~N.; Stankus,~B.; Yong,~H.; Bellshaw,~D.; Boutet,~S.;
  Lane,~T.~J.; Liang,~M.; Carbajo,~S.; Du,~W.; Kirrander,~A.; Minitti,~M.;
  Weber,~P.~M. Simplicity Beneath Complexity: Counting Molecular Electrons
  Reveals Transients and Kinetics of Photodissociation Reactions. \emph{Angew.
  Chem. Int. Ed.} \textbf{2019}, \emph{131}, 6437--6441\relax
\mciteBstWouldAddEndPuncttrue
\mciteSetBstMidEndSepPunct{\mcitedefaultmidpunct}
{\mcitedefaultendpunct}{\mcitedefaultseppunct}\relax
\EndOfBibitem
\bibitem[Kim \latin{et~al.}(2015)Kim, Kim, Nozawa, Sato, Oang, Kim, Ki, Jo,
  Park, Song, Sato, Ogawa, Togashi, Tono, Yabashi, Ishikawa, Kim, Ryoo, Kim,
  Ihee, and Adachi]{Kim2015}
Kim,~K.~H.; Kim,~J.~G.; Nozawa,~S.; Sato,~T.; Oang,~K.~Y.; Kim,~T.~W.; Ki,~H.;
  Jo,~J.; Park,~S.; Song,~C.; Sato,~T.; Ogawa,~K.; Togashi,~T.; Tono,~K.;
  Yabashi,~M.; Ishikawa,~T.; Kim,~J.; Ryoo,~R.; Kim,~J.; Ihee,~H.;
  Adachi,~S.-i. Direct observation of bond formation in solution with
  femtosecond X-ray scattering. \emph{Nature} \textbf{2015}, \emph{518},
  385--389\relax
\mciteBstWouldAddEndPuncttrue
\mciteSetBstMidEndSepPunct{\mcitedefaultmidpunct}
{\mcitedefaultendpunct}{\mcitedefaultseppunct}\relax
\EndOfBibitem
\bibitem[Sharifi \latin{et~al.}(2007)Sharifi, Kong, Chin, Mineo, Dyakov, Mebel,
  Chao, Hayashi, and Lin]{sharifi07ionization}
Sharifi,~M.; Kong,~F.; Chin,~S.~L.; Mineo,~H.; Dyakov,~Y.; Mebel,~A.~M.;
  Chao,~S.~D.; Hayashi,~M.; Lin,~S.~H. Experimental and Theoretical
  Investigation of High-Power Laser Ionization and Dissociation of Methane.
  \emph{J. Phys. Chem. A} \textbf{2007}, \emph{111}, 9405--9416\relax
\mciteBstWouldAddEndPuncttrue
\mciteSetBstMidEndSepPunct{\mcitedefaultmidpunct}
{\mcitedefaultendpunct}{\mcitedefaultseppunct}\relax
\EndOfBibitem
\bibitem[Zigo \latin{et~al.}(2017)Zigo, Le, Timilsina, and
  Trallero-Herrero]{zigo2017ionization}
Zigo,~S.; Le,~A.-T.; Timilsina,~P.; Trallero-Herrero,~C.~A. Ionization study of
  isomeric molecules in strong-field laser pulses. \emph{Scientific reports}
  \textbf{2017}, \emph{7}, 42149\relax
\mciteBstWouldAddEndPuncttrue
\mciteSetBstMidEndSepPunct{\mcitedefaultmidpunct}
{\mcitedefaultendpunct}{\mcitedefaultseppunct}\relax
\EndOfBibitem
\bibitem[Goings \latin{et~al.}(2018)Goings, Lestrange, and Li]{goings2018real}
Goings,~J.~J.; Lestrange,~P.~J.; Li,~X. Real-time time-dependent electronic
  structure theory. \emph{WIREs Computational Molecular Science} \textbf{2018},
  \emph{8}, e1341\relax
\mciteBstWouldAddEndPuncttrue
\mciteSetBstMidEndSepPunct{\mcitedefaultmidpunct}
{\mcitedefaultendpunct}{\mcitedefaultseppunct}\relax
\EndOfBibitem
\bibitem[Ekstr\"om \latin{et~al.}(2010)Ekstr\"om, Visscher, Bast, Thorvaldsen,
  and Ruud]{ruud2010}
Ekstr\"om,~U.; Visscher,~L.; Bast,~R.; Thorvaldsen,~A.~J.; Ruud,~K.
  Arbitrary-Order Density Functional Response Theory from Automatic
  Differentiation. \emph{J. Chem. Theory Comput.} \textbf{2010}, \emph{6},
  1971--1980\relax
\mciteBstWouldAddEndPuncttrue
\mciteSetBstMidEndSepPunct{\mcitedefaultmidpunct}
{\mcitedefaultendpunct}{\mcitedefaultseppunct}\relax
\EndOfBibitem
\bibitem[Rosa \latin{et~al.}(2019)Rosa, Gil, Corni, and Cammi]{cammi:2019}
Rosa,~M.; Gil,~G.; Corni,~S.; Cammi,~R. Quantum optimal control theory for
  solvated systems. \emph{J. Chem. Phys.} \textbf{2019}, \emph{151},
  194109\relax
\mciteBstWouldAddEndPuncttrue
\mciteSetBstMidEndSepPunct{\mcitedefaultmidpunct}
{\mcitedefaultendpunct}{\mcitedefaultseppunct}\relax
\EndOfBibitem
\bibitem[Sun \latin{et~al.}(2007)Sun, Song, Zhao, and Liang]{sun2007}
Sun,~J.; Song,~J.; Zhao,~Y.; Liang,~W.-Z. Real-time propagation of the reduced
  one-electron density matrix in atom-centered Gaussian orbitals: Application
  to absorption spectra of silicon clusters. \emph{J. Chem. Phys.}
  \textbf{2007}, \emph{127}, 234107\relax
\mciteBstWouldAddEndPuncttrue
\mciteSetBstMidEndSepPunct{\mcitedefaultmidpunct}
{\mcitedefaultendpunct}{\mcitedefaultseppunct}\relax
\EndOfBibitem
\bibitem[Li \latin{et~al.}(2005)Li, Smith, Markevitch, Romanov, Levis, and
  Schlegel]{schlegel2005}
Li,~X.; Smith,~S.~M.; Markevitch,~A.~N.; Romanov,~D.~A.; Levis,~R.~J.;
  Schlegel,~H.~B. A time-dependent Hartree-Fock approach for studying the
  electronic optical response of molecules in intense fields. \emph{Phys. Chem.
  Chem. Phys.} \textbf{2005}, \emph{7}, 233--239\relax
\mciteBstWouldAddEndPuncttrue
\mciteSetBstMidEndSepPunct{\mcitedefaultmidpunct}
{\mcitedefaultendpunct}{\mcitedefaultseppunct}\relax
\EndOfBibitem
\bibitem[Eshuis \latin{et~al.}(2008)Eshuis, Balint-Kurti, and
  Manby]{bkurti_manby2008}
Eshuis,~H.; Balint-Kurti,~G.~G.; Manby,~F.~R. Dynamics of molecules in strong
  oscillating electric fields using time-dependent Hartree-Fock theory.
  \emph{J. Chem. Phys.} \textbf{2008}, \emph{128}, 114113\relax
\mciteBstWouldAddEndPuncttrue
\mciteSetBstMidEndSepPunct{\mcitedefaultmidpunct}
{\mcitedefaultendpunct}{\mcitedefaultseppunct}\relax
\EndOfBibitem
\bibitem[Theilhaber(1992)]{theilhaber}
Theilhaber,~J. Ab initio simulations of sodium using time-dependent
  density-functional theory. \emph{Phys. Rev. B} \textbf{1992}, \emph{46},
  12990--13003\relax
\mciteBstWouldAddEndPuncttrue
\mciteSetBstMidEndSepPunct{\mcitedefaultmidpunct}
{\mcitedefaultendpunct}{\mcitedefaultseppunct}\relax
\EndOfBibitem
\bibitem[Yabana and Bertsch(1996)Yabana, and Bertsch]{yabana_bertsch}
Yabana,~K.; Bertsch,~G.~F. Time-dependent local-density approximation in real
  time. \emph{Phys. Rev. B} \textbf{1996}, \emph{54}, 4484--4487\relax
\mciteBstWouldAddEndPuncttrue
\mciteSetBstMidEndSepPunct{\mcitedefaultmidpunct}
{\mcitedefaultendpunct}{\mcitedefaultseppunct}\relax
\EndOfBibitem
\bibitem[Takimoto \latin{et~al.}(2007)Takimoto, Vila, and
  Rehr]{takimoto2007real}
Takimoto,~Y.; Vila,~F.; Rehr,~J. Real-time time-dependent density functional
  theory approach for frequency-dependent nonlinear optical response in
  photonic molecules. \emph{J. Chem. Phys.} \textbf{2007}, \emph{127},
  154114\relax
\mciteBstWouldAddEndPuncttrue
\mciteSetBstMidEndSepPunct{\mcitedefaultmidpunct}
{\mcitedefaultendpunct}{\mcitedefaultseppunct}\relax
\EndOfBibitem
\bibitem[Andrade \latin{et~al.}(2015)Andrade, Strubbe, De~Giovannini, Larsen,
  Oliveira, Alberdi-Rodriguez, Varas, Theophilou, Helbig, Verstraete, Stella,
  Nogueira, Aspuru-Guzik, Castro, Marques, and Rubio]{andrade2015real}
Andrade,~X.; Strubbe,~D.; De~Giovannini,~U.; Larsen,~A.~H.; Oliveira,~M.~J.;
  Alberdi-Rodriguez,~J.; Varas,~A.; Theophilou,~I.; Helbig,~N.;
  Verstraete,~M.~J.; Stella,~L.; Nogueira,~F.; Aspuru-Guzik,~A.; Castro,~A.;
  Marques,~M. A.~L.; Rubio,~A. Real-space grids and the Octopus code as tools
  for the development of new simulation approaches for electronic systems.
  \emph{Phys. Chem. Chem. Phys.} \textbf{2015}, \emph{17}, 31371--31396\relax
\mciteBstWouldAddEndPuncttrue
\mciteSetBstMidEndSepPunct{\mcitedefaultmidpunct}
{\mcitedefaultendpunct}{\mcitedefaultseppunct}\relax
\EndOfBibitem
\bibitem[Schleife \latin{et~al.}(2012)Schleife, Draeger, Kanai, and
  Correa]{schleife2012plane}
Schleife,~A.; Draeger,~E.~W.; Kanai,~Y.; Correa,~A.~A. Plane-wave
  pseudopotential implementation of explicit integrators for time-dependent
  Kohn-Sham equations in large-scale simulations. \emph{J. Chem. Phys.}
  \textbf{2012}, \emph{137}, 22A546\relax
\mciteBstWouldAddEndPuncttrue
\mciteSetBstMidEndSepPunct{\mcitedefaultmidpunct}
{\mcitedefaultendpunct}{\mcitedefaultseppunct}\relax
\EndOfBibitem
\bibitem[Giannozzi \latin{et~al.}(2020)Giannozzi, Baseggio, Bonf\`a, Brunato,
  Car, Carnimeo, Cavazzoni, de~Gironcoli, Delugas, Ferrari~Ruffino, Ferretti,
  Marzari, Timrov, Urru, and Baroni]{quantumespresso:2020}
Giannozzi,~P.; Baseggio,~O.; Bonf\`a,~P.; Brunato,~D.; Car,~R.; Carnimeo,~I.;
  Cavazzoni,~C.; de~Gironcoli,~S.; Delugas,~P.; Ferrari~Ruffino,~F.;
  Ferretti,~A.; Marzari,~N.; Timrov,~I.; Urru,~A.; Baroni,~S. Quantum ESPRESSO
  toward the exascale. \emph{J. Chem. Phys.} \textbf{2020}, \emph{152},
  154105\relax
\mciteBstWouldAddEndPuncttrue
\mciteSetBstMidEndSepPunct{\mcitedefaultmidpunct}
{\mcitedefaultendpunct}{\mcitedefaultseppunct}\relax
\EndOfBibitem
\bibitem[Genova \latin{et~al.}(2017)Genova, Ceresoli, Krishtal, Andreussi,
  DiStasio~Jr, and Pavanello]{Pavanello:2017}
Genova,~A.; Ceresoli,~D.; Krishtal,~A.; Andreussi,~O.; DiStasio~Jr,~R.~A.;
  Pavanello,~M. eQE: An open-source density functional embedding theory code
  for the condensed phase. \emph{Int. J. Quantum Chem.} \textbf{2017},
  \emph{117}, e25401\relax
\mciteBstWouldAddEndPuncttrue
\mciteSetBstMidEndSepPunct{\mcitedefaultmidpunct}
{\mcitedefaultendpunct}{\mcitedefaultseppunct}\relax
\EndOfBibitem
\bibitem[Liang \latin{et~al.}(2011)Liang, Chapman, and Li]{liang2011}
Liang,~W.; Chapman,~C.~T.; Li,~X. Efficient first-principles electronic
  dynamics. \emph{J. Chem. Phys.} \textbf{2011}, \emph{134}, 184102\relax
\mciteBstWouldAddEndPuncttrue
\mciteSetBstMidEndSepPunct{\mcitedefaultmidpunct}
{\mcitedefaultendpunct}{\mcitedefaultseppunct}\relax
\EndOfBibitem
\bibitem[Morzan \latin{et~al.}(2014)Morzan, Ram\'irez, Oviedo, S\'anchez,
  Scherlis, and Lebrero]{morzan2014}
Morzan,~U.~N.; Ram\'irez,~F.~F.; Oviedo,~M.~B.; S\'anchez,~C.~G.;
  Scherlis,~D.~A.; Lebrero,~M. C.~G. Electron dynamics in complex environments
  with real-time time dependent density functional theory in a QM-MM framework.
  \emph{J. Chem. Phys.} \textbf{2014}, \emph{140}, 164105\relax
\mciteBstWouldAddEndPuncttrue
\mciteSetBstMidEndSepPunct{\mcitedefaultmidpunct}
{\mcitedefaultendpunct}{\mcitedefaultseppunct}\relax
\EndOfBibitem
\bibitem[Lopata and Govind(2011)Lopata, and Govind]{nwchem}
Lopata,~K.; Govind,~N. Modeling Fast Electron Dynamics with Real-Time
  Time-Dependent Density Functional Theory: Application to Small Molecules and
  Chromophores. \emph{J. Chem. Theory Comput.} \textbf{2011}, \emph{7},
  1344--1355\relax
\mciteBstWouldAddEndPuncttrue
\mciteSetBstMidEndSepPunct{\mcitedefaultmidpunct}
{\mcitedefaultendpunct}{\mcitedefaultseppunct}\relax
\EndOfBibitem
\bibitem[Nguyen and Parkhill(2015)Nguyen, and Parkhill]{qchem1}
Nguyen,~T.~S.; Parkhill,~J. Nonadiabatic Dynamics for Electrons at
  Second-Order: Real-Time TDDFT and OSCF2. \emph{J. Chem. Theory Comput.}
  \textbf{2015}, \emph{11}, 2918--2924\relax
\mciteBstWouldAddEndPuncttrue
\mciteSetBstMidEndSepPunct{\mcitedefaultmidpunct}
{\mcitedefaultendpunct}{\mcitedefaultseppunct}\relax
\EndOfBibitem
\bibitem[Zhu and Herbert(2018)Zhu, and Herbert]{qchem2}
Zhu,~Y.; Herbert,~J.~M. Self-consistent predictor/corrector algorithms for
  stable and efficient integration of the time-dependent Kohn-Sham equation.
  \emph{J. Chem. Phys.} \textbf{2018}, \emph{148}, 044117\relax
\mciteBstWouldAddEndPuncttrue
\mciteSetBstMidEndSepPunct{\mcitedefaultmidpunct}
{\mcitedefaultendpunct}{\mcitedefaultseppunct}\relax
\EndOfBibitem
\bibitem[Repisky \latin{et~al.}(2015)Repisky, Konecny, Kadek, Komorovsky,
  Malkin, Malkin, and Ruud]{repisky2015excitation}
Repisky,~M.; Konecny,~L.; Kadek,~M.; Komorovsky,~S.; Malkin,~O.~L.;
  Malkin,~V.~G.; Ruud,~K. Excitation energies from real-time propagation of the
  four-component Dirac-Kohn-Sham equation. \emph{J. Chem. Theory Comput.}
  \textbf{2015}, \emph{11}, 980--991\relax
\mciteBstWouldAddEndPuncttrue
\mciteSetBstMidEndSepPunct{\mcitedefaultmidpunct}
{\mcitedefaultendpunct}{\mcitedefaultseppunct}\relax
\EndOfBibitem
\bibitem[Goings \latin{et~al.}(2016)Goings, Kasper, Egidi, Sun, and
  Li]{goings2016rtx2c}
Goings,~J.~J.; Kasper,~J.~M.; Egidi,~F.; Sun,~S.; Li,~X. Real time propagation
  of the exact two component time-dependent density functional theory. \emph{J.
  Chem. Phys.} \textbf{2016}, \emph{145}, 104107\relax
\mciteBstWouldAddEndPuncttrue
\mciteSetBstMidEndSepPunct{\mcitedefaultmidpunct}
{\mcitedefaultendpunct}{\mcitedefaultseppunct}\relax
\EndOfBibitem
\bibitem[De~Santis \latin{et~al.}(2020)De~Santis, Storchi, Belpassi, Quiney,
  and Tarantelli]{pybertha}
De~Santis,~M.; Storchi,~L.; Belpassi,~L.; Quiney,~H.~M.; Tarantelli,~F.
  PyBERTHART: A Relativistic Real-Time Four-Component TDDFT Implementation
  Using Prototyping Techniques Based on Python. \emph{J. Chem. Theory Comput.}
  \textbf{2020}, \emph{16}, 2410--2429\relax
\mciteBstWouldAddEndPuncttrue
\mciteSetBstMidEndSepPunct{\mcitedefaultmidpunct}
{\mcitedefaultendpunct}{\mcitedefaultseppunct}\relax
\EndOfBibitem
\bibitem[pyb()]{pyberthagit}
PyBertha project git URL: \url{https://github.com/lstorchi/pybertha} written
  by: L. Storchi, M. De Santis, L. Belpassi\relax
\mciteBstWouldAddEndPuncttrue
\mciteSetBstMidEndSepPunct{\mcitedefaultmidpunct}
{\mcitedefaultendpunct}{\mcitedefaultseppunct}\relax
\EndOfBibitem
\bibitem[Smith \latin{et~al.}(2018)Smith, Burns, Sirianni, Nascimento, Kumar,
  James, Schriber, Zhang, Zhang, Abbott, Berquist, Lechner, Cunha, Heide,
  Waldrop, Takeshita, Alenaizan, Neuhauser, King, Simmonett, Turney, Schaefer,
  Evangelista, DePrince, Crawford, Patkowski, and Sherrill]{psi4numpy}
Smith,~D. G.~A.; Burns,~L.~A.; Sirianni,~D.~A.; Nascimento,~D.~R.; Kumar,~A.;
  James,~A.~M.; Schriber,~J.~B.; Zhang,~T.; Zhang,~B.; Abbott,~A.~S.;
  Berquist,~E.~J.; Lechner,~M.~H.; Cunha,~L.~A.; Heide,~A.~G.; Waldrop,~J.~M.;
  Takeshita,~T.~Y.; Alenaizan,~A.; Neuhauser,~D.; King,~R.~A.;
  Simmonett,~A.~C.; Turney,~J.~M.; Schaefer,~H.~F.; Evangelista,~F.~A.;
  DePrince,~A.~E.; Crawford,~T.~D.; Patkowski,~K.; Sherrill,~C.~D. Psi4NumPy:
  An Interactive Quantum Chemistry Programming Environment for Reference
  Implementations and Rapid Development. \emph{J. Chem. Theory Comput.}
  \textbf{2018}, \emph{14}, 3504--3511\relax
\mciteBstWouldAddEndPuncttrue
\mciteSetBstMidEndSepPunct{\mcitedefaultmidpunct}
{\mcitedefaultendpunct}{\mcitedefaultseppunct}\relax
\EndOfBibitem
\bibitem[Belpassi \latin{et~al.}(2011)Belpassi, Storchi, Quiney, and
  Tarantelli]{belpassi11_12368}
Belpassi,~L.; Storchi,~L.; Quiney,~H.~M.; Tarantelli,~F. Recent Advances and
  Perspectives in Four-Component {D}irac-{K}ohn-{S}ham Calculations.
  \emph{Phys. Chem. Chem. Phys.} \textbf{2011}, \emph{13}, 12368--12394\relax
\mciteBstWouldAddEndPuncttrue
\mciteSetBstMidEndSepPunct{\mcitedefaultmidpunct}
{\mcitedefaultendpunct}{\mcitedefaultseppunct}\relax
\EndOfBibitem
\bibitem[Belpassi \latin{et~al.}(2006)Belpassi, Tarantelli, Sgamellotti, and
  Quiney]{belpassi06_124104}
Belpassi,~L.; Tarantelli,~F.; Sgamellotti,~A.; Quiney,~H.~M. Electron density
  fitting for the Coulomb problem in relativistic density-functional theory.
  \emph{The Journal of Chemical Physics} \textbf{2006}, \emph{124},
  124104\relax
\mciteBstWouldAddEndPuncttrue
\mciteSetBstMidEndSepPunct{\mcitedefaultmidpunct}
{\mcitedefaultendpunct}{\mcitedefaultseppunct}\relax
\EndOfBibitem
\bibitem[Storchi \latin{et~al.}(2010)Storchi, Belpassi, Tarantelli,
  Sgamellotti, and Quiney]{storchi10_384}
Storchi,~L.; Belpassi,~L.; Tarantelli,~F.; Sgamellotti,~A.; Quiney,~H.~M. An
  Efficient Parallel All-Electron Four-Component {D}irac-{K}ohn-{S}ham Program
  Using a Distributed Matrix Approach. \emph{J. Chem. Theory Comput.}
  \textbf{2010}, \emph{6}, 384--394\relax
\mciteBstWouldAddEndPuncttrue
\mciteSetBstMidEndSepPunct{\mcitedefaultmidpunct}
{\mcitedefaultendpunct}{\mcitedefaultseppunct}\relax
\EndOfBibitem
\bibitem[Belpassi \latin{et~al.}(2020)Belpassi, De~Santis, Quiney, Tarantelli,
  and Storchi]{paperelectronic}
Belpassi,~L.; De~Santis,~M.; Quiney,~H.~M.; Tarantelli,~F.; Storchi,~L. BERTHA:
  Implementation of a four-component Dirac-Kohn-Sham relativistic framework.
  \emph{The Journal of Chemical Physics} \textbf{2020}, \emph{152},
  164118\relax
\mciteBstWouldAddEndPuncttrue
\mciteSetBstMidEndSepPunct{\mcitedefaultmidpunct}
{\mcitedefaultendpunct}{\mcitedefaultseppunct}\relax
\EndOfBibitem
\bibitem[Storchi \latin{et~al.}(2019)Storchi, De~Santis, and
  Belpassi]{parcopaper}
Storchi,~L.; De~Santis,~M.; Belpassi,~L. {BERTHA} and PyBERTHA: State of the
  Art for Full Four-Component Dirac-Kohn-Sham Calculations. Parallel Computing:
  Technology Trends, Proceedings of the International Conference on Parallel
  Computing, {PARCO} 2019, Prague, Czech Republic, September 10-13, 2019. 2019;
  pp 354--363\relax
\mciteBstWouldAddEndPuncttrue
\mciteSetBstMidEndSepPunct{\mcitedefaultmidpunct}
{\mcitedefaultendpunct}{\mcitedefaultseppunct}\relax
\EndOfBibitem
\bibitem[Lopata \latin{et~al.}(2012)Lopata, Van~Kuiken, Khalil, and
  Govind]{lopata2012}
Lopata,~K.; Van~Kuiken,~B.~E.; Khalil,~M.; Govind,~N. Linear-Response and
  Real-Time Time-Dependent Density Functional Theory Studies of Core-Level
  Near-Edge X-Ray Absorption. \emph{J. Chem. Theory Comput.} \textbf{2012},
  \emph{8}, 3284--3292\relax
\mciteBstWouldAddEndPuncttrue
\mciteSetBstMidEndSepPunct{\mcitedefaultmidpunct}
{\mcitedefaultendpunct}{\mcitedefaultseppunct}\relax
\EndOfBibitem
\bibitem[Ding \latin{et~al.}(2013)Ding, Van~Kuiken, Eichinger, and Li]{ding}
Ding,~F.; Van~Kuiken,~B.~E.; Eichinger,~B.~E.; Li,~X. An efficient method for
  calculating dynamical hyperpolarizabilities using real-time time-dependent
  density functional theory. \emph{J. Chem. Phys.} \textbf{2013}, \emph{138},
  064104\relax
\mciteBstWouldAddEndPuncttrue
\mciteSetBstMidEndSepPunct{\mcitedefaultmidpunct}
{\mcitedefaultendpunct}{\mcitedefaultseppunct}\relax
\EndOfBibitem
\bibitem[Cheng \latin{et~al.}(2006)Cheng, Evans, and Van~Voorhis]{voohris}
Cheng,~C.-L.; Evans,~J.~S.; Van~Voorhis,~T. Simulating molecular conductance
  using real-time density functional theory. \emph{Phys. Rev. B} \textbf{2006},
  \emph{74}, 155112\relax
\mciteBstWouldAddEndPuncttrue
\mciteSetBstMidEndSepPunct{\mcitedefaultmidpunct}
{\mcitedefaultendpunct}{\mcitedefaultseppunct}\relax
\EndOfBibitem
\bibitem[Isborn and Li(2009)Isborn, and Li]{isborn}
Isborn,~C.~M.; Li,~X. Singlet-Triplet Transitions in Real-Time Time-Dependent
  Hartree-Fock/Density Functional Theory. \emph{J. Chem. Theory Comput.}
  \textbf{2009}, \emph{5}, 2415--2419\relax
\mciteBstWouldAddEndPuncttrue
\mciteSetBstMidEndSepPunct{\mcitedefaultmidpunct}
{\mcitedefaultendpunct}{\mcitedefaultseppunct}\relax
\EndOfBibitem
\bibitem[Goings and Li(2016)Goings, and Li]{gauss_jjgoings}
Goings,~J.~J.; Li,~X. An atomic orbital based real-time time-dependent density
  functional theory for computing electronic circular dichroism band spectra.
  \emph{J. Chem. Phys.} \textbf{2016}, \emph{144}, 234102\relax
\mciteBstWouldAddEndPuncttrue
\mciteSetBstMidEndSepPunct{\mcitedefaultmidpunct}
{\mcitedefaultendpunct}{\mcitedefaultseppunct}\relax
\EndOfBibitem
\bibitem[Peralta \latin{et~al.}(2015)Peralta, Hod, and Scuseria]{mag_dyn}
Peralta,~J.~E.; Hod,~O.; Scuseria,~G.~E. Magnetization Dynamics from
  Time-Dependent Noncollinear Spin Density Functional Theory Calculations.
  \emph{J. Chem. Theory Comput.} \textbf{2015}, \emph{11}, 3661--3668\relax
\mciteBstWouldAddEndPuncttrue
\mciteSetBstMidEndSepPunct{\mcitedefaultmidpunct}
{\mcitedefaultendpunct}{\mcitedefaultseppunct}\relax
\EndOfBibitem
\bibitem[Li \latin{et~al.}(2005)Li, Tully, Schlegel, and
  Frisch]{rt_ehrenfestdyn}
Li,~X.; Tully,~J.~C.; Schlegel,~H.~B.; Frisch,~M.~J. Ab initio Ehrenfest
  dynamics. \emph{J. Chem. Phys.} \textbf{2005}, \emph{123}, 084106\relax
\mciteBstWouldAddEndPuncttrue
\mciteSetBstMidEndSepPunct{\mcitedefaultmidpunct}
{\mcitedefaultendpunct}{\mcitedefaultseppunct}\relax
\EndOfBibitem
\bibitem[Kolesov \latin{et~al.}(2016)Kolesov, Gr{\aa}n\"as, Hoyt, Vinichenko,
  and Kaxiras]{grigory2016}
Kolesov,~G.; Gr{\aa}n\"as,~O.; Hoyt,~R.; Vinichenko,~D.; Kaxiras,~E. Real-Time
  TD-DFT with Classical Ion Dynamics: Methodology and Applications. \emph{J.
  Chem. Theory Comput.} \textbf{2016}, \emph{12}, 466--476\relax
\mciteBstWouldAddEndPuncttrue
\mciteSetBstMidEndSepPunct{\mcitedefaultmidpunct}
{\mcitedefaultendpunct}{\mcitedefaultseppunct}\relax
\EndOfBibitem
\bibitem[Kadek \latin{et~al.}(2015)Kadek, Konecny, Gao, Repisky, and
  Ruud]{kadek2015}
Kadek,~M.; Konecny,~L.; Gao,~B.; Repisky,~M.; Ruud,~K. X-ray absorption
  resonances near L2{,}3-edges from real-time propagation of the
  Dirac-Kohn-Sham density matrix. \emph{Phys. Chem. Chem. Phys.} \textbf{2015},
  \emph{17}, 22566--22570\relax
\mciteBstWouldAddEndPuncttrue
\mciteSetBstMidEndSepPunct{\mcitedefaultmidpunct}
{\mcitedefaultendpunct}{\mcitedefaultseppunct}\relax
\EndOfBibitem
\bibitem[Konecny \latin{et~al.}(2016)Konecny, Kadek, Komorovsky, Malkina, Ruud,
  and Repisky]{konecny2016nlo}
Konecny,~L.; Kadek,~M.; Komorovsky,~S.; Malkina,~O.~L.; Ruud,~K.; Repisky,~M.
  Acceleration of Relativistic Electron Dynamics by Means of X2C
  Transformation: Application to the Calculation of Nonlinear Optical
  Properties. \emph{J. Chem. Theory Comput.} \textbf{2016}, \emph{12},
  5823--5833\relax
\mciteBstWouldAddEndPuncttrue
\mciteSetBstMidEndSepPunct{\mcitedefaultmidpunct}
{\mcitedefaultendpunct}{\mcitedefaultseppunct}\relax
\EndOfBibitem
\bibitem[Konecny \latin{et~al.}(2018)Konecny, Kadek, Komorovsky, Ruud, and
  Repisky]{konecnyresolution-of-identity}
Konecny,~L.; Kadek,~M.; Komorovsky,~S.; Ruud,~K.; Repisky,~M.
  Resolution-of-identity accelerated relativistic two- and four-component
  electron dynamics approach to chiroptical spectroscopies. \emph{J. Chem.
  Phys.} \textbf{2018}, \emph{149}, 204104\relax
\mciteBstWouldAddEndPuncttrue
\mciteSetBstMidEndSepPunct{\mcitedefaultmidpunct}
{\mcitedefaultendpunct}{\mcitedefaultseppunct}\relax
\EndOfBibitem
\bibitem[Marques \latin{et~al.}(2003)Marques, L\'opez, Varsano, Castro, and
  Rubio]{PhysRevLett.90.258101}
Marques,~M. A.~L.; L\'opez,~X.; Varsano,~D.; Castro,~A.; Rubio,~A.
  Time-Dependent Density-Functional Approach for Biological Chromophores: The
  Case of the Green Fluorescent Protein. \emph{Phys. Rev. Lett.} \textbf{2003},
  \emph{90}, 258101\relax
\mciteBstWouldAddEndPuncttrue
\mciteSetBstMidEndSepPunct{\mcitedefaultmidpunct}
{\mcitedefaultendpunct}{\mcitedefaultseppunct}\relax
\EndOfBibitem
\bibitem[Liang \latin{et~al.}(2012)Liang, Chapman, Ding, and Li]{Li:2012}
Liang,~W.; Chapman,~C.~T.; Ding,~F.; Li,~X. Modeling Ultrafast Solvated
  Electronic Dynamics Using Time-Dependent Density Functional Theory and
  Polarizable Continuum Model. \emph{J. Phys. Chem. A} \textbf{2012},
  \emph{116}, 1884--1890\relax
\mciteBstWouldAddEndPuncttrue
\mciteSetBstMidEndSepPunct{\mcitedefaultmidpunct}
{\mcitedefaultendpunct}{\mcitedefaultseppunct}\relax
\EndOfBibitem
\bibitem[Nguyen \latin{et~al.}(2012)Nguyen, Ding, Fischer, Liang, and
  Li]{Li_b:2012}
Nguyen,~P.~D.; Ding,~F.; Fischer,~S.~A.; Liang,~W.; Li,~X. Solvated
  First-Principles Excited-State Charge-Transfer Dynamics with Time-Dependent
  Polarizable Continuum Model and Solvent Dielectric Relaxation. \emph{J. Phys.
  Chem. Lett.} \textbf{2012}, \emph{3}, 2898--2904\relax
\mciteBstWouldAddEndPuncttrue
\mciteSetBstMidEndSepPunct{\mcitedefaultmidpunct}
{\mcitedefaultendpunct}{\mcitedefaultseppunct}\relax
\EndOfBibitem
\bibitem[Pipolo \latin{et~al.}(2014)Pipolo, Corni, and Cammi]{PIPOLO2014112}
Pipolo,~S.; Corni,~S.; Cammi,~R. The cavity electromagnetic field within the
  polarizable continuum model of solvation: An application to the real-time
  time dependent density functional theory. \emph{Computational and Theoretical
  Chemistry} \textbf{2014}, \emph{1040-1041}, 112 -- 119, Excited states: From
  isolated molecules to complex environments\relax
\mciteBstWouldAddEndPuncttrue
\mciteSetBstMidEndSepPunct{\mcitedefaultmidpunct}
{\mcitedefaultendpunct}{\mcitedefaultseppunct}\relax
\EndOfBibitem
\bibitem[Corni \latin{et~al.}(2015)Corni, Pipolo, and Cammi]{Corni_jpca:2015}
Corni,~S.; Pipolo,~S.; Cammi,~R. Equation of Motion for the Solvent
  Polarization Apparent Charges in the Polarizable Continuum Model: Application
  to Real-Time TDDFT. \emph{J. Phys. Chem. A} \textbf{2015}, \emph{119},
  5405--5416\relax
\mciteBstWouldAddEndPuncttrue
\mciteSetBstMidEndSepPunct{\mcitedefaultmidpunct}
{\mcitedefaultendpunct}{\mcitedefaultseppunct}\relax
\EndOfBibitem
\bibitem[Ding \latin{et~al.}(2015)Ding, Lingerfelt, Mennucci, and
  Li]{mennucci2015}
Ding,~F.; Lingerfelt,~D.~B.; Mennucci,~B.; Li,~X. Time-dependent
  non-equilibrium dielectric response in QM/continuum approaches. \emph{J.
  Chem. Phys.} \textbf{2015}, \emph{142}, 034120\relax
\mciteBstWouldAddEndPuncttrue
\mciteSetBstMidEndSepPunct{\mcitedefaultmidpunct}
{\mcitedefaultendpunct}{\mcitedefaultseppunct}\relax
\EndOfBibitem
\bibitem[Donati \latin{et~al.}(2017)Donati, Wildman, Caprasecca, Lingerfelt,
  Lipparini, Mennucci, and Li]{mennucci2017}
Donati,~G.; Wildman,~A.; Caprasecca,~S.; Lingerfelt,~D.~B.; Lipparini,~F.;
  Mennucci,~B.; Li,~X. Coupling Real-Time Time-Dependent Density Functional
  Theory with Polarizable Force Field. \emph{J. Phys. Chem. Lett.}
  \textbf{2017}, \emph{8}, 5283--5289\relax
\mciteBstWouldAddEndPuncttrue
\mciteSetBstMidEndSepPunct{\mcitedefaultmidpunct}
{\mcitedefaultendpunct}{\mcitedefaultseppunct}\relax
\EndOfBibitem
\bibitem[Wu \latin{et~al.}(2017)Wu, Teuler, Cailliez, Clavagu\'era, Salahub,
  and de~la Lande]{demon2krt}
Wu,~X.; Teuler,~J.-M.; Cailliez,~F.; Clavagu\'era,~C.; Salahub,~D.~R.; de~la
  Lande,~A. Simulating Electron Dynamics in Polarizable Environments. \emph{J.
  Chem. Theory Comput.} \textbf{2017}, \emph{13}, 3985--4002\relax
\mciteBstWouldAddEndPuncttrue
\mciteSetBstMidEndSepPunct{\mcitedefaultmidpunct}
{\mcitedefaultendpunct}{\mcitedefaultseppunct}\relax
\EndOfBibitem
\bibitem[Gil \latin{et~al.}(2019)Gil, Pipolo, Delgado, Rozzi, and
  Corni]{Corni:2019}
Gil,~G.; Pipolo,~S.; Delgado,~A.; Rozzi,~C.~A.; Corni,~S. Nonequilibrium
  Solvent Polarization Effects in Real-Time Electronic Dynamics of Solute
  Molecules Subject to Time-Dependent Electric Fields: A New Feature of the
  Polarizable Continuum Model. \emph{J. Chem. Theory Comput.} \textbf{2019},
  \emph{15}, 2306--2319\relax
\mciteBstWouldAddEndPuncttrue
\mciteSetBstMidEndSepPunct{\mcitedefaultmidpunct}
{\mcitedefaultendpunct}{\mcitedefaultseppunct}\relax
\EndOfBibitem
\bibitem[Koh \latin{et~al.}(2017)Koh, Nguyen-Beck, and Parkhill]{Koh:2017}
Koh,~K.~J.; Nguyen-Beck,~T.~S.; Parkhill,~J. Accelerating Realtime TDDFT with
  Block-Orthogonalized Manby-Miller Embedding Theory. \emph{J. Chem. Theory
  Comput.} \textbf{2017}, \emph{13}, 4173--4178\relax
\mciteBstWouldAddEndPuncttrue
\mciteSetBstMidEndSepPunct{\mcitedefaultmidpunct}
{\mcitedefaultendpunct}{\mcitedefaultseppunct}\relax
\EndOfBibitem
\bibitem[Lee \latin{et~al.}(2019)Lee, Welborn, Manby, and
  Miller]{lee_projection-based_2019}
Lee,~S. J.~R.; Welborn,~M.; Manby,~F.~R.; Miller,~T.~F. {P}rojection-{B}ased
  {W}avefunction-in-{D}{F}{T} {E}mbedding. \emph{Acc. Chem. Res.}
  \textbf{2019}, \emph{52}, 1359--1368\relax
\mciteBstWouldAddEndPuncttrue
\mciteSetBstMidEndSepPunct{\mcitedefaultmidpunct}
{\mcitedefaultendpunct}{\mcitedefaultseppunct}\relax
\EndOfBibitem
\bibitem[Gomes and Jacob(2012)Gomes, and Jacob]{gomes_quantum-chemical_2012}
Gomes,~A. S.~P.; Jacob,~{\relax Ch}.~R. {Q}uantum-chemical embedding methods
  for treating local electronic excitations in complex chemical systems.
  \emph{Annu. Rep. Prog. Chem., Sect. C} \textbf{2012}, \emph{108}, 222\relax
\mciteBstWouldAddEndPuncttrue
\mciteSetBstMidEndSepPunct{\mcitedefaultmidpunct}
{\mcitedefaultendpunct}{\mcitedefaultseppunct}\relax
\EndOfBibitem
\bibitem[Jacob and Neugebauer(2014)Jacob, and Neugebauer]{jacob_subsystem_2014}
Jacob,~{\relax Ch}.~R.; Neugebauer,~J. {S}ubsystem density-functional theory.
  \emph{WIREs Comput. Mol. Sci.} \textbf{2014}, \emph{4}, 325--362\relax
\mciteBstWouldAddEndPuncttrue
\mciteSetBstMidEndSepPunct{\mcitedefaultmidpunct}
{\mcitedefaultendpunct}{\mcitedefaultseppunct}\relax
\EndOfBibitem
\bibitem[Wesolowski \latin{et~al.}(2015)Wesolowski, Shedge, and
  Zhou]{wesolowski_frozen-density_2015}
Wesolowski,~T.~A.; Shedge,~S.; Zhou,~X. {F}rozen-{D}ensity {E}mbedding
  {S}trategy for {M}ultilevel {S}imulations of {E}lectronic {S}tructure.
  \emph{Chem. Rev.} \textbf{2015}, \emph{115}, 5891--5928\relax
\mciteBstWouldAddEndPuncttrue
\mciteSetBstMidEndSepPunct{\mcitedefaultmidpunct}
{\mcitedefaultendpunct}{\mcitedefaultseppunct}\relax
\EndOfBibitem
\bibitem[Krishtal \latin{et~al.}(2015)Krishtal, Ceresoli, and
  Pavanello]{pavanello2015}
Krishtal,~A.; Ceresoli,~D.; Pavanello,~M. Subsystem real-time time dependent
  density functional theory. \emph{J. Chem. Phys.} \textbf{2015}, \emph{142},
  154116\relax
\mciteBstWouldAddEndPuncttrue
\mciteSetBstMidEndSepPunct{\mcitedefaultmidpunct}
{\mcitedefaultendpunct}{\mcitedefaultseppunct}\relax
\EndOfBibitem
\bibitem[Wesolowski and Warshel(1993)Wesolowski, and Warshel]{wesolowski93}
Wesolowski,~T.~A.; Warshel,~A. Frozen density functional approach for ab initio
  calculations of solvated molecules. \emph{J. Phys. Chem.} \textbf{1993},
  \emph{97}, 8050--8053\relax
\mciteBstWouldAddEndPuncttrue
\mciteSetBstMidEndSepPunct{\mcitedefaultmidpunct}
{\mcitedefaultendpunct}{\mcitedefaultseppunct}\relax
\EndOfBibitem
\bibitem[Senatore and Subbaswamy(1986)Senatore, and Subbaswamy]{Senatore86}
Senatore,~G.; Subbaswamy,~K.~R. Density dependence of the dielectric constant
  of rare-gas crystals. \emph{Phys. Rev. B} \textbf{1986}, \emph{34},
  5754--5757\relax
\mciteBstWouldAddEndPuncttrue
\mciteSetBstMidEndSepPunct{\mcitedefaultmidpunct}
{\mcitedefaultendpunct}{\mcitedefaultseppunct}\relax
\EndOfBibitem
\bibitem[Cortona(1992)]{Cortona92}
Cortona,~P. Direct determination of self-consistent total energies and charge
  densities of solids: A study of the cohesive properties of the alkali
  halides. \emph{Phys. Rev. B} \textbf{1992}, \emph{46}, 2008--2014\relax
\mciteBstWouldAddEndPuncttrue
\mciteSetBstMidEndSepPunct{\mcitedefaultmidpunct}
{\mcitedefaultendpunct}{\mcitedefaultseppunct}\relax
\EndOfBibitem
\bibitem[Iannuzzi \latin{et~al.}(2006)Iannuzzi, Kirchner, and
  Hutter]{iannuzzi2006}
Iannuzzi,~M.; Kirchner,~B.; Hutter,~J. Density functional embedding for
  molecular systems. \emph{Chem. Phys. Lett.} \textbf{2006}, \emph{421}, 16 --
  20\relax
\mciteBstWouldAddEndPuncttrue
\mciteSetBstMidEndSepPunct{\mcitedefaultmidpunct}
{\mcitedefaultendpunct}{\mcitedefaultseppunct}\relax
\EndOfBibitem
\bibitem[Jacob \latin{et~al.}(2008)Jacob, Neugebauer, and
  Visscher]{jacob2008flexible}
Jacob,~C.~R.; Neugebauer,~J.; Visscher,~L. A flexible implementation of
  frozen-density embedding for use in multilevel simulations. \emph{J. Comput.
  Chem.} \textbf{2008}, \emph{29}, 1011--1018\relax
\mciteBstWouldAddEndPuncttrue
\mciteSetBstMidEndSepPunct{\mcitedefaultmidpunct}
{\mcitedefaultendpunct}{\mcitedefaultseppunct}\relax
\EndOfBibitem
\bibitem[Casida and Wesolowski(2004)Casida, and
  Wesolowski]{doi:10.1002/qua.10744}
Casida,~M.~E.; Wesolowski,~T.~A. Generalization of the Kohn-Sham equations with
  constrained electron density formalism and its time-dependent response theory
  formulation. \emph{Int. J. Quantum Chem.} \textbf{2004}, \emph{96},
  577--588\relax
\mciteBstWouldAddEndPuncttrue
\mciteSetBstMidEndSepPunct{\mcitedefaultmidpunct}
{\mcitedefaultendpunct}{\mcitedefaultseppunct}\relax
\EndOfBibitem
\bibitem[Neugebauer(2007)]{coupl_neugebauer2007}
Neugebauer,~J. Couplings between electronic transitions in a subsystem
  formulation of time-dependent density functional theory. \emph{J. Chem.
  Phys.} \textbf{2007}, \emph{126}, 134116\relax
\mciteBstWouldAddEndPuncttrue
\mciteSetBstMidEndSepPunct{\mcitedefaultmidpunct}
{\mcitedefaultendpunct}{\mcitedefaultseppunct}\relax
\EndOfBibitem
\bibitem[Neugebauer(2009)]{neugebauer2009}
Neugebauer,~J. On the calculation of general response properties in subsystem
  density functional theory. \emph{J. Chem. Phys.} \textbf{2009}, \emph{131},
  084104\relax
\mciteBstWouldAddEndPuncttrue
\mciteSetBstMidEndSepPunct{\mcitedefaultmidpunct}
{\mcitedefaultendpunct}{\mcitedefaultseppunct}\relax
\EndOfBibitem
\bibitem[T\"olle \latin{et~al.}(2019)T\"olle, B\"ockers, and
  Neugebauer]{toelle_1}
T\"olle,~J.; B\"ockers,~M.; Neugebauer,~J. Exact subsystem time-dependent
  density-functional theory. \emph{J. Chem. Phys.} \textbf{2019}, \emph{150},
  181101\relax
\mciteBstWouldAddEndPuncttrue
\mciteSetBstMidEndSepPunct{\mcitedefaultmidpunct}
{\mcitedefaultendpunct}{\mcitedefaultseppunct}\relax
\EndOfBibitem
\bibitem[T\"olle \latin{et~al.}(2019)T\"olle, B\"ockers, Niemeyer, and
  Neugebauer]{toelle_2}
T\"olle,~J.; B\"ockers,~M.; Niemeyer,~N.; Neugebauer,~J. Inter-subsystem
  charge-transfer excitations in exact subsystem time-dependent
  density-functional theory. \emph{J. Chem. Phys.} \textbf{2019}, \emph{151},
  174109\relax
\mciteBstWouldAddEndPuncttrue
\mciteSetBstMidEndSepPunct{\mcitedefaultmidpunct}
{\mcitedefaultendpunct}{\mcitedefaultseppunct}\relax
\EndOfBibitem
\bibitem[Fux \latin{et~al.}(2010)Fux, Jacob, Neugebauer, Visscher, and
  Reiher]{fux_exemb}
Fux,~S.; Jacob,~C.~R.; Neugebauer,~J.; Visscher,~L.; Reiher,~M. Accurate
  frozen-density embedding potentials as a first step towards a subsystem
  description of covalent bonds. \emph{J. Chem. Phys.} \textbf{2010},
  \emph{132}, 164101\relax
\mciteBstWouldAddEndPuncttrue
\mciteSetBstMidEndSepPunct{\mcitedefaultmidpunct}
{\mcitedefaultendpunct}{\mcitedefaultseppunct}\relax
\EndOfBibitem
\bibitem[Goodpaster \latin{et~al.}(2010)Goodpaster, Ananth, Manby, and
  Miller]{goodpaster_exemb}
Goodpaster,~J.~D.; Ananth,~N.; Manby,~F.~R.; Miller,~T.~F. Exact nonadditive
  kinetic potentials for embedded density functional theory. \emph{J. Chem.
  Phys.} \textbf{2010}, \emph{133}, 084103\relax
\mciteBstWouldAddEndPuncttrue
\mciteSetBstMidEndSepPunct{\mcitedefaultmidpunct}
{\mcitedefaultendpunct}{\mcitedefaultseppunct}\relax
\EndOfBibitem
\bibitem[Goodpaster \latin{et~al.}(2011)Goodpaster, Barnes, and
  Miller]{goodpaster_exemb2}
Goodpaster,~J.~D.; Barnes,~T.~A.; Miller,~T.~F. Embedded density functional
  theory for covalently bonded and strongly interacting subsystems. \emph{J.
  Chem. Phys.} \textbf{2011}, \emph{134}, 164108\relax
\mciteBstWouldAddEndPuncttrue
\mciteSetBstMidEndSepPunct{\mcitedefaultmidpunct}
{\mcitedefaultendpunct}{\mcitedefaultseppunct}\relax
\EndOfBibitem
\bibitem[Huang \latin{et~al.}(2011)Huang, Pavone, and Carter]{huang_exemb}
Huang,~C.; Pavone,~M.; Carter,~E.~A. Quantum mechanical embedding theory based
  on a unique embedding potential. \emph{J. Chem. Phys.} \textbf{2011},
  \emph{134}, 154110\relax
\mciteBstWouldAddEndPuncttrue
\mciteSetBstMidEndSepPunct{\mcitedefaultmidpunct}
{\mcitedefaultendpunct}{\mcitedefaultseppunct}\relax
\EndOfBibitem
\bibitem[Nafziger \latin{et~al.}(2011)Nafziger, Wu, and
  Wasserman]{nafziger_exemb}
Nafziger,~J.; Wu,~Q.; Wasserman,~A. Molecular binding energies from partition
  density functional theory. \emph{J. Chem. Phys.} \textbf{2011}, \emph{135},
  234101\relax
\mciteBstWouldAddEndPuncttrue
\mciteSetBstMidEndSepPunct{\mcitedefaultmidpunct}
{\mcitedefaultendpunct}{\mcitedefaultseppunct}\relax
\EndOfBibitem
\bibitem[Jacob \latin{et~al.}(2011)Jacob, Beyhan, Bulo, Gomes, G\"otz,
  Kiewisch, Sikkema, and Visscher]{pyadf}
Jacob,~C.~R.; Beyhan,~S.~M.; Bulo,~R.~E.; Gomes,~A. S.~P.; G\"otz,~A.~W.;
  Kiewisch,~K.; Sikkema,~J.; Visscher,~L. PyADF -- A scripting framework for
  multiscale quantum chemistry. \emph{J. Comput. Chem.} \textbf{2011},
  \emph{32}, 2328--2338\relax
\mciteBstWouldAddEndPuncttrue
\mciteSetBstMidEndSepPunct{\mcitedefaultmidpunct}
{\mcitedefaultendpunct}{\mcitedefaultseppunct}\relax
\EndOfBibitem
\bibitem[Ekstr\"{o}m(2019)]{xcfun:2019}
Ekstr\"{o}m,~U. XCFun: Exchange-Correlation functionals with arbitrary order
  derivatives. \url{ https://github.com/dftlibs/xcfun}, 2019\relax
\mciteBstWouldAddEndPuncttrue
\mciteSetBstMidEndSepPunct{\mcitedefaultmidpunct}
{\mcitedefaultendpunct}{\mcitedefaultseppunct}\relax
\EndOfBibitem
\bibitem[Thomas(1927)]{thomas_1927}
Thomas,~L.~H. The calculation of atomic fields. \emph{Mathematical Proceedings
  of the Cambridge Philosophical Society} \textbf{1927}, \emph{23},
  542--548\relax
\mciteBstWouldAddEndPuncttrue
\mciteSetBstMidEndSepPunct{\mcitedefaultmidpunct}
{\mcitedefaultendpunct}{\mcitedefaultseppunct}\relax
\EndOfBibitem
\bibitem[Lembarki and Chermette(1994)Lembarki, and Chermette]{PhysRevA.50.5328}
Lembarki,~A.; Chermette,~H. Obtaining a gradient-corrected kinetic-energy
  functional from the Perdew-Wang exchange functional. \emph{Phys. Rev. A}
  \textbf{1994}, \emph{50}, 5328--5331\relax
\mciteBstWouldAddEndPuncttrue
\mciteSetBstMidEndSepPunct{\mcitedefaultmidpunct}
{\mcitedefaultendpunct}{\mcitedefaultseppunct}\relax
\EndOfBibitem
\bibitem[Mi and Pavanello(2020)Mi, and Pavanello]{Pavanello:2020}
Mi,~W.; Pavanello,~M. Nonlocal Subsystem Density Functional Theory. \emph{J.
  Phys. Chem. Lett.} \textbf{2020}, \emph{11}, 272--279\relax
\mciteBstWouldAddEndPuncttrue
\mciteSetBstMidEndSepPunct{\mcitedefaultmidpunct}
{\mcitedefaultendpunct}{\mcitedefaultseppunct}\relax
\EndOfBibitem
\bibitem[Gomes \latin{et~al.}(2008)Gomes, Jacob, and
  Visscher]{acetone_w_cluster}
Gomes,~A. S.~P.; Jacob,~C.~R.; Visscher,~L. Calculation of local excitations in
  large systems by embedding wave-function theory in density-functional theory.
  \emph{Phys. Chem. Chem. Phys.} \textbf{2008}, \emph{10}, 5353--5362\relax
\mciteBstWouldAddEndPuncttrue
\mciteSetBstMidEndSepPunct{\mcitedefaultmidpunct}
{\mcitedefaultendpunct}{\mcitedefaultseppunct}\relax
\EndOfBibitem
\bibitem[Bouchafra \latin{et~al.}(2018)Bouchafra, Shee, R{\'e}al, Vallet, and
  Gomes]{Bouchafra:2019}
Bouchafra,~Y.; Shee,~A.; R{\'e}al,~F.; Vallet,~V.; Gomes,~A. S.~P. Predictive
  simulations of ionization energies of solvated halide ions with relativistic
  embedded Equation of Motion Coupled Cluster Theory. \emph{Phys. Rev. Lett.}
  \textbf{2018}, \emph{121}, 266001\relax
\mciteBstWouldAddEndPuncttrue
\mciteSetBstMidEndSepPunct{\mcitedefaultmidpunct}
{\mcitedefaultendpunct}{\mcitedefaultseppunct}\relax
\EndOfBibitem
\bibitem[Halbert \latin{et~al.}(2019)Halbert, Olejniczak, Vallet, and
  Gomes]{Halbert:2020}
Halbert,~L.; Olejniczak,~M.; Vallet,~V.; Gomes,~A. S.~P. {Investigating solvent
  effects on the magnetic properties of molybdate ions (MoO$_4^{2-}$) with
  relativistic embedding}. \emph{Int. J. Quantum Chem.} \textbf{2019}, \relax
\mciteBstWouldAddEndPunctfalse
\mciteSetBstMidEndSepPunct{\mcitedefaultmidpunct}
{}{\mcitedefaultseppunct}\relax
\EndOfBibitem
\bibitem[H\"ofener \latin{et~al.}(2012)H\"ofener, Severo Pereira~Gomes, and
  Visscher]{visscher12}
H\"ofener,~S.; Severo Pereira~Gomes,~A.; Visscher,~L. Molecular properties via
  a subsystem density functional theory formulation: A common framework for
  electronic embedding. \emph{J. Chem. Phys.} \textbf{2012}, \emph{136},
  044104\relax
\mciteBstWouldAddEndPuncttrue
\mciteSetBstMidEndSepPunct{\mcitedefaultmidpunct}
{\mcitedefaultendpunct}{\mcitedefaultseppunct}\relax
\EndOfBibitem
\bibitem[Olejniczak \latin{et~al.}(2017)Olejniczak, Bast, and
  Gomes]{Olejniczak:2017}
Olejniczak,~M.; Bast,~R.; Gomes,~A. S.~P. On the calculation of second-order
  magnetic properties using subsystem approaches in a relativistic framework.
  \emph{Phys. Chem. Chem. Phys.} \textbf{2017}, \emph{19}, 8400--8415\relax
\mciteBstWouldAddEndPuncttrue
\mciteSetBstMidEndSepPunct{\mcitedefaultmidpunct}
{\mcitedefaultendpunct}{\mcitedefaultseppunct}\relax
\EndOfBibitem
\bibitem[Neugebauer \latin{et~al.}(2005)Neugebauer, Jacob, Wesolowski, and
  Baerends]{Johannes:2005}
Neugebauer,~J.; Jacob,~C.~R.; Wesolowski,~T.~A.; Baerends,~E.~J. An Explicit
  Quantum Chemical Method for Modeling Large Solvation Shells Applied to
  Aminocoumarin C151. \emph{J. Phys. Chem. A} \textbf{2005}, \emph{109},
  7805--7814\relax
\mciteBstWouldAddEndPuncttrue
\mciteSetBstMidEndSepPunct{\mcitedefaultmidpunct}
{\mcitedefaultendpunct}{\mcitedefaultseppunct}\relax
\EndOfBibitem
\bibitem[Bulo \latin{et~al.}(2008)Bulo, Jacob, and Visscher]{Bulo:2008}
Bulo,~R.~E.; Jacob,~C.~R.; Visscher,~L. NMR Solvent Shifts of Acetonitrile from
  Frozen Density Embedding Calculations. \emph{The Journal of Physical
  Chemistry A} \textbf{2008}, \emph{112}, 2640--2647\relax
\mciteBstWouldAddEndPuncttrue
\mciteSetBstMidEndSepPunct{\mcitedefaultmidpunct}
{\mcitedefaultendpunct}{\mcitedefaultseppunct}\relax
\EndOfBibitem
\bibitem[Castro \latin{et~al.}(2004)Castro, Marques, and Rubio]{castro}
Castro,~A.; Marques,~M. A.~L.; Rubio,~A. Propagators for the time-dependent
  Kohni-Sham equations. \emph{J. Chem. Phys.} \textbf{2004}, \emph{121},
  3425--3433\relax
\mciteBstWouldAddEndPuncttrue
\mciteSetBstMidEndSepPunct{\mcitedefaultmidpunct}
{\mcitedefaultendpunct}{\mcitedefaultseppunct}\relax
\EndOfBibitem
\bibitem[Meng and Kaxiras(2008)Meng, and Kaxiras]{meng2008real}
Meng,~S.; Kaxiras,~E. Real-time, local basis-set implementation of
  time-dependent density functional theory for excited state dynamics
  simulations. \emph{J. Chem. Phys.} \textbf{2008}, \emph{129}, 054110\relax
\mciteBstWouldAddEndPuncttrue
\mciteSetBstMidEndSepPunct{\mcitedefaultmidpunct}
{\mcitedefaultendpunct}{\mcitedefaultseppunct}\relax
\EndOfBibitem
\bibitem[Press \latin{et~al.}(2007)Press, Teukolsky, Vetterling, and
  Flannery]{numrecipes2007}
Press,~W.~H.; Teukolsky,~S.~A.; Vetterling,~W.~T.; Flannery,~B.~P.
  \emph{Numerical recipes 3rd edition: The art of scientific computing};
  Cambridge university press, 2007\relax
\mciteBstWouldAddEndPuncttrue
\mciteSetBstMidEndSepPunct{\mcitedefaultmidpunct}
{\mcitedefaultendpunct}{\mcitedefaultseppunct}\relax
\EndOfBibitem
\bibitem[Magnus(1954)]{magnus}
Magnus,~W. On the exponential solution of differential equations for a linear
  operator. \emph{Communications on Pure and Applied Mathematics}
  \textbf{1954}, \emph{7}, 649--673\relax
\mciteBstWouldAddEndPuncttrue
\mciteSetBstMidEndSepPunct{\mcitedefaultmidpunct}
{\mcitedefaultendpunct}{\mcitedefaultseppunct}\relax
\EndOfBibitem
\bibitem[Casas and Iserles(2006)Casas, and Iserles]{Casas_2006}
Casas,~F.; Iserles,~A. Explicit Magnus expansions for nonlinear equations.
  \emph{Journal of Physics A: Mathematical and General} \textbf{2006},
  \emph{39}, 5445--5461\relax
\mciteBstWouldAddEndPuncttrue
\mciteSetBstMidEndSepPunct{\mcitedefaultmidpunct}
{\mcitedefaultendpunct}{\mcitedefaultseppunct}\relax
\EndOfBibitem
\bibitem[Zhu and Herbert(2018)Zhu, and Herbert]{propagators2}
Zhu,~Y.; Herbert,~J.~M. Self-consistent predictor/corrector algorithms for
  stable and efficient integration of the time-dependent Kohn-Sham equation.
  \emph{J. Chem. Phys.} \textbf{2018}, \emph{148}, 044117\relax
\mciteBstWouldAddEndPuncttrue
\mciteSetBstMidEndSepPunct{\mcitedefaultmidpunct}
{\mcitedefaultendpunct}{\mcitedefaultseppunct}\relax
\EndOfBibitem
\bibitem[Bandrauk \latin{et~al.}(2009)Bandrauk, Chelkowski, Diestler, Manz, and
  Yuan]{Bandrauk:2009}
Bandrauk,~A.~D.; Chelkowski,~S.; Diestler,~D.~J.; Manz,~J.; Yuan,~K.-J. Quantum
  simulation of high-order harmonic spectra of the hydrogen atom. \emph{Phys.
  Rev. A} \textbf{2009}, \emph{79}, 023403\relax
\mciteBstWouldAddEndPuncttrue
\mciteSetBstMidEndSepPunct{\mcitedefaultmidpunct}
{\mcitedefaultendpunct}{\mcitedefaultseppunct}\relax
\EndOfBibitem
\bibitem[Jacob \latin{et~al.}(2020)Jacob, Beyhan, Bulo, Gomes, Goetz, Handzlik,
  Kiewisch, Klammler, Sikkema, and Visscher]{pyadf-github-v096}
Jacob,~{\relax Ch}.~R.; Beyhan,~S.~M.; Bulo,~R.~E.; Gomes,~A. S.~P.; Goetz,~A.;
  Handzlik,~M.; Kiewisch,~K.; Klammler,~M.; Sikkema,~J.; Visscher,~L. {PyADF}
  --- {A} {S}cripting {F}ramework for {M}ultiscale {Q}uantum {C}hemistry:
  {V}ersion 0.96. 2020; URL: https://github.com/chjacob-tubs/pyadf-releases,
  DOI: 10.5281/zenodo.3834283\relax
\mciteBstWouldAddEndPuncttrue
\mciteSetBstMidEndSepPunct{\mcitedefaultmidpunct}
{\mcitedefaultendpunct}{\mcitedefaultseppunct}\relax
\EndOfBibitem
\bibitem[Gomes and Jacob(2020)Gomes, and Jacob]{pyembed}
Gomes,~A. S.~P.; Jacob,~{\relax Ch}.~R. {PyEmbed} --- {A} {F}rozen-{D}ensity
  {E}mbedding {M}odule for {PyADF}. 2020; available at DOI:
  10.5281/zenodo.3834283\relax
\mciteBstWouldAddEndPuncttrue
\mciteSetBstMidEndSepPunct{\mcitedefaultmidpunct}
{\mcitedefaultendpunct}{\mcitedefaultseppunct}\relax
\EndOfBibitem
\bibitem[Schmitt-Monreal and Jacob(2020)Schmitt-Monreal, and
  Jacob]{schmittmonreal_frozen-density_2020}
Schmitt-Monreal,~D.; Jacob,~{\relax Ch}.~R. {F}rozen-density embedding-based
  many-body expansions. \emph{Int. J. Quantum Chem.} \textbf{2020}, \emph{n/a},
  e26228, in press, {DOI}: 10.1002/qua.26228\relax
\mciteBstWouldAddEndPuncttrue
\mciteSetBstMidEndSepPunct{\mcitedefaultmidpunct}
{\mcitedefaultendpunct}{\mcitedefaultseppunct}\relax
\EndOfBibitem
\bibitem[pyb()]{pyberthagitfdert}
M. De Santis, git URL:
  \url{https://github.com/lstorchi/pybertha/tree/master/psi4embedrt} within the
  PyBertha project: \url{https://github.com/lstorchi/pybertha} written by: L.
  Storchi, M. De Santis, L. Belpassi\relax
\mciteBstWouldAddEndPuncttrue
\mciteSetBstMidEndSepPunct{\mcitedefaultmidpunct}
{\mcitedefaultendpunct}{\mcitedefaultseppunct}\relax
\EndOfBibitem
\bibitem[De~Santis(2020)]{source_fig}
De~Santis,~M. Numerical data and post-processing material. 2020; available at:
  DOI: 10.5281/zenodo.3885610\relax
\mciteBstWouldAddEndPuncttrue
\mciteSetBstMidEndSepPunct{\mcitedefaultmidpunct}
{\mcitedefaultendpunct}{\mcitedefaultseppunct}\relax
\EndOfBibitem
\bibitem[Parrish \latin{et~al.}(2017)Parrish, Burns, Smith, Simmonett,
  DePrince, Hohenstein, Bozkaya, Sokolov, Di~Remigio, Richard, Gonthier, James,
  McAlexander, Kumar, Saitow, Wang, Pritchard, Verma, Schaefer, Patkowski,
  King, Valeev, Evangelista, Turney, Crawford, and
  Sherrill]{doi:10.1021/acs.jctc.7b00174}
Parrish,~R.~M.; Burns,~L.~A.; Smith,~D. G.~A.; Simmonett,~A.~C.;
  DePrince,~A.~E.; Hohenstein,~E.~G.; Bozkaya,~U.; Sokolov,~A.~Y.;
  Di~Remigio,~R.; Richard,~R.~M.; Gonthier,~J.~F.; James,~A.~M.;
  McAlexander,~H.~R.; Kumar,~A.; Saitow,~M.; Wang,~X.; Pritchard,~B.~P.;
  Verma,~P.; Schaefer,~H.~F.; Patkowski,~K.; King,~R.~A.; Valeev,~E.~F.;
  Evangelista,~F.~A.; Turney,~J.~M.; Crawford,~T.~D.; Sherrill,~C.~D. Psi4 1.1:
  An Open-Source Electronic Structure Program Emphasizing Automation, Advanced
  Libraries, and Interoperability. \emph{J. Chem. Theory Comput.}
  \textbf{2017}, \emph{13}, 3185--3197\relax
\mciteBstWouldAddEndPuncttrue
\mciteSetBstMidEndSepPunct{\mcitedefaultmidpunct}
{\mcitedefaultendpunct}{\mcitedefaultseppunct}\relax
\EndOfBibitem
\bibitem[Storchi(2020)]{pyadf-github-v096-python3}
Storchi,~L. {P}ython 3 port of {PyADF} v0.96. 2020; DOI:
  10.5281/zenodo.3834286\relax
\mciteBstWouldAddEndPuncttrue
\mciteSetBstMidEndSepPunct{\mcitedefaultmidpunct}
{\mcitedefaultendpunct}{\mcitedefaultseppunct}\relax
\EndOfBibitem
\bibitem[Storchi(2020)]{xcfun-python3}
Storchi,~L. {P}ython 3 port of {XcFun} a486a3f148. 2020; URL:
  https://github.com/lstorchi/xcfun\relax
\mciteBstWouldAddEndPuncttrue
\mciteSetBstMidEndSepPunct{\mcitedefaultmidpunct}
{\mcitedefaultendpunct}{\mcitedefaultseppunct}\relax
\EndOfBibitem
\bibitem[Baerends \latin{et~al.}()Baerends, Ziegler, Atkins, Autschbach,
  Bashford, Baseggio, B{\'{e}}rces, Bickelhaupt, Bo, Boerritger, Cavallo, Daul,
  Chong, Chulhai, Deng, Dickson, Dieterich, Ellis, van Faassen, Ghysels,
  Giammona, van Gisbergen, Goez, G{\"{o}}tz, Gusarov, Harris, van~den Hoek, Hu,
  Jacob, Jacobsen, Jensen, Joubert, Kaminski, van Kessel, K{\"{o}}nig,
  Kootstra, Kovalenko, Krykunov, van Lenthe, McCormack, Michalak, Mitoraj,
  Morton, Neugebauer, Nicu, Noodleman, Osinga, Patchkovskii, Pavanello,
  Peeples, Philipsen, Post, Pye, Ramanantoanina, Ramos, Ravenek,
  Rodr{\'{i}}guez, Ros, R{\"{u}}ger, Schipper, Schl{\"{u}}ns, van Schoot,
  Schreckenbach, Seldenthuis, Seth, Snijders, Sol{\`{a}}, M., Swart, Swerhone,
  te~Velde, Tognetti, Vernooijs, Versluis, Visscher, Visser, Wang, Wesolowski,
  van Wezenbeek, Wiesenekker, Wolff, Woo, and Yakovlev]{ADF2017authors}
Baerends,~E.~J.; Ziegler,~T.; Atkins,~A.~J.; Autschbach,~J.; Bashford,~D.;
  Baseggio,~O.; B{\'{e}}rces,~A.; Bickelhaupt,~F.~M.; Bo,~C.;
  Boerritger,~P.~M.; Cavallo,~L.; Daul,~C.; Chong,~D.~P.; Chulhai,~D.~V.;
  Deng,~L.; Dickson,~R.~M.; Dieterich,~J.~M.; Ellis,~D.~E.; van Faassen,~M.;
  Ghysels,~A.; Giammona,~A.; van Gisbergen,~S. J.~A.; Goez,~A.;
  G{\"{o}}tz,~A.~W.; Gusarov,~S.; Harris,~F.~E.; van~den Hoek,~P.; Hu,~Z.;
  Jacob,~C.~R.; Jacobsen,~H.; Jensen,~L.; Joubert,~L.; Kaminski,~J.~W.; van
  Kessel,~G.; K{\"{o}}nig,~C.; Kootstra,~F.; Kovalenko,~A.; Krykunov,~M.; van
  Lenthe,~E.; McCormack,~D.~A.; Michalak,~A.; Mitoraj,~M.; Morton,~S.~M.;
  Neugebauer,~J.; Nicu,~V.~P.; Noodleman,~L.; Osinga,~V.~P.; Patchkovskii,~S.;
  Pavanello,~M.; Peeples,~C.~A.; Philipsen,~P. H.~T.; Post,~D.; Pye,~C.~C.;
  Ramanantoanina,~H.; Ramos,~P.; Ravenek,~W.; Rodr{\'{i}}guez,~J.~I.; Ros,~P.;
  R{\"{u}}ger,~R.; Schipper,~P. R.~T.; Schl{\"{u}}ns,~D.; van Schoot,~H.;
  Schreckenbach,~G.; Seldenthuis,~J.~S.; Seth,~M.; Snijders,~J.~G.;
  Sol{\`{a}},~M.; M.,~S.; Swart,~M.; Swerhone,~D.; te~Velde,~G.; Tognetti,~V.;
  Vernooijs,~P.; Versluis,~L.; Visscher,~L.; Visser,~O.; Wang,~F.;
  Wesolowski,~T.~A.; van Wezenbeek,~E.~M.; Wiesenekker,~G.; Wolff,~S.~K.;
  Woo,~T.~K.; Yakovlev,~A.~L. {ADF2017, SCM, Theoretical Chemistry, Vrije
  Universiteit, Amsterdam, The Netherlands, https://www.scm.com}\relax
\mciteBstWouldAddEndPuncttrue
\mciteSetBstMidEndSepPunct{\mcitedefaultmidpunct}
{\mcitedefaultendpunct}{\mcitedefaultseppunct}\relax
\EndOfBibitem
\bibitem[Du\l{}ak \latin{et~al.}(2009)Du\l{}ak, Kami\'nski, and
  Wesolowski]{wesolowski09}
Du\l{}ak,~M.; Kami\'nski,~J.~W.; Wesolowski,~T.~A. Linearized orbital-free
  embedding potential in self-consistent calculations. \emph{Int. J. Quantum
  Chem.} \textbf{2009}, \emph{109}, 1886--1897\relax
\mciteBstWouldAddEndPuncttrue
\mciteSetBstMidEndSepPunct{\mcitedefaultmidpunct}
{\mcitedefaultendpunct}{\mcitedefaultseppunct}\relax
\EndOfBibitem
\bibitem[Wesolowski(2004)]{jacs:2004}
Wesolowski,~T.~A. Hydrogen-Bonding-Induced Shifts of the Excitation Energies in
  Nucleic Acid Bases: An Interplay between Electrostatic and Electron Density
  Overlap Effects. \emph{J. Am. Chem. Soc.} \textbf{2004}, \emph{126},
  11444--11445\relax
\mciteBstWouldAddEndPuncttrue
\mciteSetBstMidEndSepPunct{\mcitedefaultmidpunct}
{\mcitedefaultendpunct}{\mcitedefaultseppunct}\relax
\EndOfBibitem
\bibitem[Te~Velde \latin{et~al.}(2001)Te~Velde, Bickelhaupt, Baerends,
  Fonseca~Guerra, van Gisbergen, Snijders, and Ziegler]{te2001chemistry}
Te~Velde,~G.~t.; Bickelhaupt,~F.~M.; Baerends,~E.~J.; Fonseca~Guerra,~C.; van
  Gisbergen,~S.~J.; Snijders,~J.~G.; Ziegler,~T. Chemistry with ADF. \emph{J.
  Comput. Chem.} \textbf{2001}, \emph{22}, 931--967\relax
\mciteBstWouldAddEndPuncttrue
\mciteSetBstMidEndSepPunct{\mcitedefaultmidpunct}
{\mcitedefaultendpunct}{\mcitedefaultseppunct}\relax
\EndOfBibitem
\bibitem[Bruner \latin{et~al.}(2016)Bruner, LaMaster, and
  Lopata]{bruner2016accelerated}
Bruner,~A.; LaMaster,~D.; Lopata,~K. Accelerated broadband spectra using
  transition dipole decomposition and Pad{\'e} approximants. \emph{J. Chem.
  Theory Comput.} \textbf{2016}, \emph{12}, 3741--3750\relax
\mciteBstWouldAddEndPuncttrue
\mciteSetBstMidEndSepPunct{\mcitedefaultmidpunct}
{\mcitedefaultendpunct}{\mcitedefaultseppunct}\relax
\EndOfBibitem
\bibitem[Dunning(1989)]{dunning1989a}
Dunning,~T.~H. Gaussian basis sets for use in correlated molecular
  calculations. I. The atoms boron through neon and hydrogen. \emph{J. Chem.
  Phys.} \textbf{1989}, \emph{90}, 1007--1023\relax
\mciteBstWouldAddEndPuncttrue
\mciteSetBstMidEndSepPunct{\mcitedefaultmidpunct}
{\mcitedefaultendpunct}{\mcitedefaultseppunct}\relax
\EndOfBibitem
\bibitem[Kendall \latin{et~al.}(1992)Kendall, Dunning, and
  Harrison]{kendall1992a}
Kendall,~R.~A.; Dunning,~T.~H.; Harrison,~R.~J. Electron affinities of the
  first‐row atoms revisited. Systematic basis sets and wave functions.
  \emph{J. Chem. Phys.} \textbf{1992}, \emph{96}, 6796--6806\relax
\mciteBstWouldAddEndPuncttrue
\mciteSetBstMidEndSepPunct{\mcitedefaultmidpunct}
{\mcitedefaultendpunct}{\mcitedefaultseppunct}\relax
\EndOfBibitem
\bibitem[Becke(1988)]{becke88}
Becke,~A.~D. Density-functional exchange-energy approximation with correct
  asymptotic behavior. \emph{Phys. Rev. A} \textbf{1988}, \emph{38},
  3098--3100\relax
\mciteBstWouldAddEndPuncttrue
\mciteSetBstMidEndSepPunct{\mcitedefaultmidpunct}
{\mcitedefaultendpunct}{\mcitedefaultseppunct}\relax
\EndOfBibitem
\bibitem[Lee \latin{et~al.}(1988)Lee, Yang, and Parr]{lyp1988}
Lee,~C.; Yang,~W.; Parr,~R.~G. Development of the Colle-Salvetti
  correlation-energy formula into a functional of the electron density.
  \emph{Phys. Rev. B} \textbf{1988}, \emph{37}, 785--789\relax
\mciteBstWouldAddEndPuncttrue
\mciteSetBstMidEndSepPunct{\mcitedefaultmidpunct}
{\mcitedefaultendpunct}{\mcitedefaultseppunct}\relax
\EndOfBibitem
\bibitem[Vosko \latin{et~al.}(1980)Vosko, Wilk, and Nusair]{vwn1980}
Vosko,~S.~H.; Wilk,~L.; Nusair,~M. Accurate spin-dependent electron liquid
  correlation energies for local spin density calculations: a critical
  analysis. \emph{Canadian Journal of Physics} \textbf{1980}, \emph{58},
  1200--1211\relax
\mciteBstWouldAddEndPuncttrue
\mciteSetBstMidEndSepPunct{\mcitedefaultmidpunct}
{\mcitedefaultendpunct}{\mcitedefaultseppunct}\relax
\EndOfBibitem
\bibitem[Slater(1951)]{slater1951}
Slater,~J.~C. A Simplification of the Hartree-Fock Method. \emph{Phys. Rev.}
  \textbf{1951}, \emph{81}, 385--390\relax
\mciteBstWouldAddEndPuncttrue
\mciteSetBstMidEndSepPunct{\mcitedefaultmidpunct}
{\mcitedefaultendpunct}{\mcitedefaultseppunct}\relax
\EndOfBibitem
\bibitem[Jacob \latin{et~al.}(2006)Jacob, Neugebauer, Jensen, and
  Visscher]{watercluster1}
Jacob,~C.~R.; Neugebauer,~J.; Jensen,~L.; Visscher,~L. Comparison of
  frozen-density embedding and discrete reaction field solvent models for
  molecular properties. \emph{Phys. Chem. Chem. Phys.} \textbf{2006}, \emph{8},
  2349--2359\relax
\mciteBstWouldAddEndPuncttrue
\mciteSetBstMidEndSepPunct{\mcitedefaultmidpunct}
{\mcitedefaultendpunct}{\mcitedefaultseppunct}\relax
\EndOfBibitem
\bibitem[H\"ofener \latin{et~al.}(2013)H\"ofener, Gomes, and
  Visscher]{watercluster2}
H\"ofener,~S.; Gomes,~A. S.~P.; Visscher,~L. Solvatochromic shifts from
  coupled-cluster theory embedded in density functional theory. \emph{J. Chem.
  Phys.} \textbf{2013}, \emph{139}, 104106\relax
\mciteBstWouldAddEndPuncttrue
\mciteSetBstMidEndSepPunct{\mcitedefaultmidpunct}
{\mcitedefaultendpunct}{\mcitedefaultseppunct}\relax
\EndOfBibitem
\bibitem[Luppi and Head-Gordon(2012)Luppi, and Head-Gordon]{luppi2012}
Luppi,~E.; Head-Gordon,~M. Computation of high-harmonic generation spectra of
  H2 And N2 in intense laser pulses using quantum chemistry methods and
  time-dependent density functional theory. \emph{Mol. Phys.} \textbf{2012},
  \emph{110}, 909--923\relax
\mciteBstWouldAddEndPuncttrue
\mciteSetBstMidEndSepPunct{\mcitedefaultmidpunct}
{\mcitedefaultendpunct}{\mcitedefaultseppunct}\relax
\EndOfBibitem
\bibitem[Lewenstein \latin{et~al.}(1994)Lewenstein, Balcou, Ivanov, L'Huillier,
  and Corkum]{hhg_corkum}
Lewenstein,~M.; Balcou,~P.; Ivanov,~M.~Y.; L'Huillier,~A.; Corkum,~P.~B. Theory
  of high-harmonic generation by low-frequency laser fields. \emph{Phys. Rev.
  A} \textbf{1994}, \emph{49}, 2117--2132\relax
\mciteBstWouldAddEndPuncttrue
\mciteSetBstMidEndSepPunct{\mcitedefaultmidpunct}
{\mcitedefaultendpunct}{\mcitedefaultseppunct}\relax
\EndOfBibitem
\bibitem[Barth \latin{et~al.}(2009)Barth, Oncak, Ulrich, Mucke, Lischke,
  Slavicek, and Hergenhahn]{doi:10.1021/jp906113e}
Barth,~S.; Oncak,~M.; Ulrich,~V.; Mucke,~M.; Lischke,~T.; Slavicek,~P.;
  Hergenhahn,~U. Valence Ionization of Water Clusters: From Isolated Molecules
  to Bulk. \emph{J. Phys. Chem. A} \textbf{2009}, \emph{113},
  13519--13527\relax
\mciteBstWouldAddEndPuncttrue
\mciteSetBstMidEndSepPunct{\mcitedefaultmidpunct}
{\mcitedefaultendpunct}{\mcitedefaultseppunct}\relax
\EndOfBibitem
\bibitem[Klahr \latin{et~al.}(2018)Klahr, Schl\"uns, and Neugebauer]{geom_opts}
Klahr,~K.; Schl\"uns,~D.; Neugebauer,~J. Geometry Optimizations in a Subsystem
  Density Functional Theory Formalism: A Benchmark Study. \emph{J. Chem. Theory
  Comput.} \textbf{2018}, \emph{14}, 5631--5644\relax
\mciteBstWouldAddEndPuncttrue
\mciteSetBstMidEndSepPunct{\mcitedefaultmidpunct}
{\mcitedefaultendpunct}{\mcitedefaultseppunct}\relax
\EndOfBibitem
\end{mcitethebibliography}

\end{document}